\documentclass[11pt]{article}
\usepackage{graphicx}
\usepackage{amsmath,amssymb,enumerate}
\usepackage{color}

\newtheorem{Prop}{Prop.}
\newtheorem{Def}{Definition}[section]
\newtheorem{Ex}{Example}

\newcommand{\QED}[0]{\hskip 14cm Q.E.D.}
\newcommand{\ket}[1]{|#1 \rangle}

\newcommand{\bra}[1]{\langle #1 |}

\newcommand{\ave}[1]{\langle #1 \rangle}

\newcommand{\rk}[1]{\noindent{\bf Remark\ #1}\\}
\newcommand{\reso}[1]{r_{\cal A}\left( #1 \right)}
\newcommand{\spect}[1]{\sigma_{\cal A}\left( #1 \right)}
\newcommand{\integ}[1]{\int d#1 \hspace*{1mm}}
\newcommand{\mat}[2]{\Bigg( \begin{array}{cc} #1 \\ #2 \end{array} \Bigg)}

\begin{document}
\baselineskip 5mm

\vspace*{1cm}
\centerline{\LARGE \bf Quantum statistical mechanics} 

\centerline{\LARGE \bf in infinitely extended systems}
\vspace*{2mm}
\centerline{\Large --- $C^*$ algebraic approach ---}
\vspace*{2mm}

\begin{flushright}
{\large 
$^*$ Department of Applied Physics, Waseda University 
\\ 
$^\dag$ Depto. de F\'isica, Facultad de ciencias F\'isicas y Matem\'aticas,  Universidad de Chile}\\
{\large Shuichi Tasaki$^*$, Shigeru Ajisaka$^\dag$\footnote{E-mail: g00k0056@suou.jp}, and
Felipe Barra$^\dag$}

\end{flushright}

\vspace{5mm}
This is a lecture note of Professor Shuichi Tasaki.
In the last part of this manuscript, SA and FB demonstrate some recent results as well.
As far as SA's knowledge, regular courses were given every two years for graduate students at Waseda university, and SA attended the course twice (probably in 2005 and 2007 or in 2006 and 2008).
Some topics were added in the lectures of the later year, probably Professor Tasaki modified his lectures as he was discovering new results.
A similar but intensive short course was given once as  at Kyoto university.

SA thinks Professor Tasaki had a unique philosophy of physics, which enabled him to give very attractive 
lectures that unfortunately it is not possible to attend any more.\footnote{SA joined Professor Tasaki's courses on dynamics (1st year), thermodynamics (2nd year), statistical mechanics (3rd year - as a master student), and two courses for graduate students, one about $C^*$ algebra and dynamical systems, the other about renormalization group}
In Professor Tasaki's lectures, both beautiful mathematical views and clear physical intuition were always given together, which SA believe, is quite difficult.
As many people know Professor Tasaki was very strong in mathematics, but he never forgot to think of physical interpretations.
Professor Tasaki was writing a lecture note of this course based on the note taken by a student of Kyoto university.
Unfortunately, Prof. Tasaki passed away before he finished writing the manuscript.
This lecture note is based on the incomplete Japanese note\cite{Bussei_Tasaki} and 
a note taken by SA at Waseda university.
It seems that main stories of the lectures are the same, but Professor Tasaki 
gave different materials in detail, and use different notations.
This manuscript mainly follows the note of SA.

%%%%% end SA

SA and FB decided to write this manuscript (1) to complete a note of Professor Tasaki's lectures, and (2) to give an access for non-Japanese speaker to his lecture note.
However, they are slow writer, and they could not finish it before the deadline of this volume.
They could finish only parts before nonequilibrium states (only some of the results will be shown in this version), and some recent results about Landauer formula and sufficient condition on interaction for the existence of unique steady states.
They hope to complete the lecture note part, and upload it on Arxiv in the future.

\section{introduction}
One of the main problems in a mathematical formulation of statistical mechanics might be the treatment of infinite systems because some fundamental quantities that usually appear in the equations have formal meaning but are ill-defined.
In addition, states in quantum system also do not have asymptotic limit since $e^{-iHt}$ oscillate very frequently.

%$C^*$ algebra is a set of elements with finite norm, and 

The method of $C^*$ algebra was first introduced on the purpose to axiomatically study quantum system~\cite{Seg47}, and it has been applied to the study of statistical mechanics of infinitely extended systems~\cite{Cornfeld00,Bunimovich82,Rue69,Sinai82,BR02}.
$C^*$ algebra is constructed as a set of elements with finite norm, and thus, the problem of divergence does not exist by definition.

Recently, the $C^*$ algebra approach to quantum statistical mechanics found
 important applications in the study of nonequilibrium systems
 because we can rigorously consider reservoirs in contrast to usual
 formulation where the infinite size of them prevents a rigorous
 analysis.
Starting 
from Ruelle's work\cite{Rue00,Rue01} on scattering-theoretical characterizations of NESS and Jak\v si\'c-Pillet's 
investigation\cite{JP01,JP02a,JP02b} on entropy production, the algebraic approach to NESS has been extensively developed 
(see Ref.\cite{LecMath1880,FMSU03,TM03,TT05,TM06,TT06a,TT06b,TT06c,Tasaki06} and references therein). Currently, in 
addition to various aspects\cite{LecMath1880,JP07,Merkli07,Merkli08,Zagrebnov09}, linear response theories\cite{JOP06a,JOP06b,JOP06c,JOP07},
thermodynamics properties\cite{TM06,Tasaki06,SalemFrohlich07,Salem07}, Landauer-B\"uttiker formula\cite{TT05,TT06a,Tasaki01,FMU06,AJPP07,Nenciu07}, nonequilibrium phase transition\cite{AjiTa09}, and quantum dissipative structure\cite{AjiTa11} are investigated.

Since quantum time evolution is unitary, the consideration of the reservoirs is inevitable for discussion of many phenomena such as dissipation and decoherence, and thus, it had been important to establish mathematically rigorous theory to discuss these issues.
Several tools exist to study those systems, like kinetic theory\cite{Resibois77}, steady state thermodynamics\cite{ST06}, linear response theory\cite{Kubo88}, etc, but their foundation has been the subject of long debates. 
For application, $C^*$ algebra has been only applied to quasi-free systems, though under this circumstances, the rigorous study with $C^*$ algebra for nonequilibrium systems has significant values.

In systems connected to two infinitely extended reservoirs with different sets of thermodynamic quantities (temperatures $T$ and chemical potentials $\mu$), one might expect that systems reach steady states after sufficiently long time.
However, it is not always the case, and the conditions to reach steady state are not well understood.
With the method of $C^*$ algebra, it was shown that\cite{Rue00}, if time-evolution satisfies $L^1$ asymptotic abelian properties
and some other conditions, there exists a stable unique steady state.

Part of $C^*$ algebraic works is highly mathematically oriented, and physical systems are typically studied with evolution equations for reduced density matrix\cite{Koss76,Lindblad76,Breuer02} (super operators obtained by tracing out reservoir's freedom) or with the Keldysh formalism\cite{Kamenev04}.

Maybe one of the achievements of Professor Tasaki was that he studied physically interesting systems rigorously and presented the analysis in a language accessible to most physicists. 
%The main purpose of this note is to give a knowledge for beginners to read papers of this subject.

 In the first part of this article we have presented in great detail the general
 framework of $C^*$ algebra approach to quantum mechanics and to
 equilibrium
 quantum statistical mechanics as though by S. Tasaki in Kyoto and Waseda University. 
The extension to nonequilibrium situations is presented without detailed profs, which are left for a
 second part of the lecture
 notes, however, in the second part of this article, namely in Sec.\ref{sec_Landauer},
 we include a detailed derivation of Landauer formula,
 an important formula in the study of transport properties of nonequilibrium systems.
 Consider a system of non-interacting fermions and two reservoirs,
 one called the left reservoir and the second the right reservoir, each
 characterized by a given temperature and chemical potential.
 Then the system is put in contact with the left reservoir and the
 right reservoir such that a current of particles is established.
 Landauer formula, connects the (particle) current through the system
 in the nonequilibrium steady state with the Fermi distribution
 characterizing the left and right reservoirs. Landauer formula is very
 appealing because it has a simple physical interpretation linking
 current to transmission properties of the system, and in fact it can be
 derived using simple and reasonable physical assumptions \cite{Imry99}. 
Here we present one derivation that uses the results of $C^*$ algebra approach without  invoking these assumptions. In particular, there is no need to assume that the reservoirs remain
in Fermi distribution once the nonequilibrium steady state (NESS) is established but still
NESS current is determined by the Fermi distribution that reservoirs had at least in the infinite past.

The manuscript is organized as follows.
In \S 2, we review the basics of second quantization,  Fock Space and
operators acting on that space. In \S 3, 
we define $*$ and $C^*$ algebras, analyze their spectral properties,
an analyze special properties of some members of the algebra know as
selfadjoint elements. In \S 4, we introduce the notion of time evolution
by considering the action of a one parameter group on the algebra. The
analogy with usual time evolution in quantum mechanics is explained.
In \S 5, we introduce states as linear functionals from the $C^*$ algebra to
the complex numbers. We show that for finite system this is
equivalent to the notion of density matrix. Then we consider the so
called GNS representation of the $C^*$ algebra, which provides a
useful tool of analysis and also of physical interpretation in terms
of usual quantum mechanics. At the end of this section and in \S 6,
statistical mechanics is developed in the $C^*$ approach to quantum
mechanics. \S 6 ends with the introduction on nonequilibrium
steady states. In \S 7, we include a result of our own research and
analyze a particular problem which is the validity of Landauer formula and explicit form of tunneling probability for systems described by a quadratic hamiltonian.
Sufficient conditions for the existence of a unique steady state are also
derived in that section. We end with a few conclusions in section \S
8.

\section{Second Quantization}
In this section, we briefly review the second quantization of fermions.
Let us start from 2 body wave function.
\begin{eqnarray}
\psi(x_1,x_2)=-\psi(x_2,x_1)
\label{fermi}
\end{eqnarray}
Let $\{\phi_n(x)\}$ be a complete orthonormal system (CONS), then, $\psi(x_1,x_2)$ can be expanded as follows.
\begin{eqnarray*}
\psi(x_1,x_2) &=& \sum_n \widetilde{C}_n (x_1) \phi_n (x_2)
\\
&=&
\sum_{n,m} C_{nm}\phi_m(x_1) \phi_n(x_2),
\end{eqnarray*}
where $\widetilde{C}_n (x_1)$ and $C_{nm}$ are defined by
\begin{eqnarray*}
\widetilde{C}_n (x_1) &\equiv& \int \phi_n^*(x_2) \psi(x_1,x_2) dx_2
\\
C_{nm} &\equiv& \int dx_1 dx_2\ \phi_m^*(x_1) \phi_n^*(x_2) \psi(x_1,x_2)
\ .
\end{eqnarray*}
In this expansion, fermionic condition~(\ref{fermi}) reads
\begin{eqnarray*}
\sum C_{nm}\phi_m(x_1)\phi_n(x_2)
=
-\sum C_{nm}\phi_m(x_2)\phi_n(x_1)
\Rightarrow\ \ 
C_{nm}=-C_{mn}
\end{eqnarray*}
Therefore, the wave function is rewritten by
\begin{eqnarray*}
\psi(x_1,x_2) &=& \sum_{n>m} C_{nm}\phi_m(x_1) \phi_n(x_2)
+
\sum_{n<m} C_{nm}\phi_m(x_1) \phi_n(x_2)
\\&=&
\sum_{n>m} C_{nm}\left[
\phi_m(x_1) \phi_n(x_2)-
\phi_n(x_1) \phi_m(x_2)\right]
\\&=&
\sum_{n>m} \sqrt{2}C_{nm}
\frac{1}{\sqrt{2}}
\left| \begin{array}{cc}
\phi_m(x_1) & \phi_m(x_2) \\
\phi_n(x_1) & \phi_n(x_2)
\end{array} \right|,
\end{eqnarray*}
where the matrix
\begin{eqnarray*}
\frac{1}{\sqrt{2}}
\left| \begin{array}{cc}
\phi_m(x_1) & \phi_m(x_2) \\
\phi_n(x_1) & \phi_n(x_2)
\end{array} \right|
\end{eqnarray*}
is called the Slater matrix. 
By interpreting $n$ and $m$ as states, 
$\sqrt{2}C_{nm}$ is a probability amplitude of which state $(n,m)$ are occupied by particles.

\noindent Note:
In general, wave function of $N$ particle systems are expressed as
\begin{eqnarray*}
\phi(x_1,\cdots,x_N)=\sum_{ \{n_j\} }
C_{ \{n_j\}}
\frac{1}{\sqrt{N!}}
	\left| \begin{array}{ccc}
	\phi_{n_1} (x_1) & \cdots & \phi_{n_1}(x_N) \\
	\vdots & \ddots & \vdots \\
	\phi_{n_N} (x_1) & \cdots & \phi_{n_N}(x_N) 
	\end{array} \right|
\ \ .
\end{eqnarray*}

Next we are going to study the action of operators in this space and show that is useful to introduce the notion
of Fock space.
Let $\hat{h}$ be an operator from $L^2$ to $L^2$, then, $\hat{h}f(x)$ 
can be expanded as 
%\begin{eqnarray}
%\hat{h}\phi_m(x)
%&=& \sum_n \phi_n(x) \int dx'\ \phi_n^*(x')\hat{h}\phi_m(x')
%\\ &\equiv& \sum_n \bra{n}h \ket{m}\phi_n(x)
%\label{13}
%\end{eqnarray}
\begin{eqnarray*}
\hat{h}f(x)
&=& \sum_n \phi_n(x) \int dx'\ \phi_n^*(x')\hat{h}f(x')
\\ &\equiv& \sum_n \bra{n}h \ket{f}\phi_n(x).
\end{eqnarray*}
Later we will use this formula when $f$ is an element of the base $\phi_m$ or $\phi_\alpha$. In that case we
use the notation $ \ket{f}\to \ket{m}$ or $ \ket{f}\to \ket{\alpha}.$

Suppose, $\hat{h}_k\ (k=1,2)$ acts on $x_k$. 
Then, ${\displaystyle \sum_{i=1}^2 \hat{h}_i\psi(x_1,x_2)}$ reads, 
\begin{eqnarray}
&& {\!\!\!\!\!\!\!\!\!\!\!\!\!\!}
\sum_{i=1}^2 \hat{h}_i \psi(x_1,x_2)  \nonumber \\
&&= \sum_{n>m} C_{nm}(\hat{h}_1 + \hat{h}_2)
\left[\phi_m(x_1)\phi_n(x_2)-\phi_m(x_2)\phi_n(x_1)\right]
\nonumber \\ 
&&= \sum_{n>m} \sum_k C_{nm}
\left[\bra{k}h\ket{m} \phi_k(x_1)\phi_n(x_2)-\bra{k}h\ket{n} \phi_m(x_2)\phi_k(x_1)\right]
\nonumber \\ 
&& \quad + \sum_{n>m} \sum_k  C_{nm}
\left[\bra{k}h\ket{n} \phi_m(x_1)\phi_k(x_2)-\bra{k}h\ket{m} \phi_k(x_2)\phi_n(x_1)\right]
\nonumber \\ 
&&=\sum_{n>m}\sum_{k}\sqrt{2}C_{nm}\left\{
\frac{\bra{k}h\ket{m}}{\sqrt{2}}
\left| \begin{array}{cc}
\phi_k(x_1) & \phi_k(x_2) \\
\phi_n(x_1) & \phi_n(x_2)
\end{array} \right|
-
\frac{\bra{k}h\ket{n}}{\sqrt{2}}
\left| \begin{array}{cc}
\phi_k(x_1) & \phi_k(x_2) \\
\phi_m(x_1) & \phi_m(x_2)
\end{array} \right|
\right\}\ .
\nonumber\\
\label{1particle}
\end{eqnarray}
One can interpret (\ref{1particle}) as $\sum_i \hat{h}_i$ transforming
the state $(n,m)$ into $(n,k)$ and $(k,m)$.
It is know that it is convenient to separate this transformation into 2 steps.
Namely, we first annihilate one particle, and, then, create one particle.
For that purpose, we introduce a space which contains states with different particle number (Fock space).

\subsection{Fock Space}

Quantum state of the 1 particle system is described by the Hilbert space ${\cal H}$ as a space of $L^2$ functions. On the other hand, 
Quantum state of the $N$ particle system is described by $\psi(x_1,\cdots,x_n)\in {\cal H}^{\otimes N}$ (we will write $({\cal H}^N)_S$ for symmetric case 
and $({\cal H}^N)_A$ for anti-symmetric case).
To treat states with different particle numbers, the Fock space is introduced as follows.
\begin{eqnarray*}
{\cal F}_{A/S} = {\bf C} \oplus (H)_{A/S}\oplus (H^2)_{A/S}\oplus \cdots\oplus,
(H^N)_{A/S}\oplus\cdots
\end{eqnarray*}
where {\bf C} corresponds to vacuum.
Now wave function is described as
\begin{eqnarray*}
\psi =
	\begin{pmatrix}
	C_1
	\\
	\psi^{(1)}(x)
	\\
	\psi^{(2)}(x_1,x_2)
	\\
	\vdots
	\end{pmatrix}
\in{\cal F}_{A/S},\qquad \psi^{(n)}(x_1,\cdots,x_n)
\in ({\cal H}^n)_{A/S}
\ .
\end{eqnarray*}
One can prove that the Fock space equipped with the following inner product
is a Hilbert space.
\begin{eqnarray*}
&&(\psi, \phi)\equiv C_1^* C_2+\int dx \psi^{(1)*}(x) \phi^{(1)}(x)
+\int dx_1 dx_2 \psi^{(2)*}(x_1,x_2) \phi^{(2)}(x_1,x_2)+\cdots
\\
&&\psi =
	\begin{pmatrix}
	C_1
	\\
	\psi^{(1)}(x)
	\\
	\psi^{(2)}(x_1,x_2)
	\\
	\vdots
	\end{pmatrix}
,\ \ 
\phi =
	\begin{pmatrix}
	C_2
	\\
	\phi^{(1)}(x)
	\\
	\phi^{(2)}(x_1,x_2)
	\\
	\vdots
	\end{pmatrix}
\end{eqnarray*}

\subsection{creation and annihilation operator}
Let ${\cal H}$ be a separable Hilbert space of $L^2$ function, and $\{\phi_\alpha(x)\}$ be a CONS of ${\cal H}$. Then, let us define annihilation ($a_\alpha$) and creation operators ($b_\alpha$):
\begin{eqnarray}
(a_\alpha \psi)^{(n)}(x_1,\cdots,x_n) \mskip -10 mu &\equiv&  \mskip -10 mu 
\sqrt{n+1}\int dx \phi_\alpha (x)^* \psi^{(n+1)}(x,x_1,\cdots,x_n)\label{migui1}\\
(b_\alpha \psi)^{(n)}(x_1,\cdots,x_n) \mskip -10 mu &\equiv&  \mskip -10 mu 
\left\{\begin{array}{cc}
\frac{1}{\sqrt{n}}\sum_{i=1}^n(\pm1)^{i-1}
\phi_\alpha(x_i)\psi^{(n-1)}(x_1,\cdots,x_{i-1},x_{i+1},\cdots,x_n) & n\neq 0 \\0 & n=0
\end{array}\right.
\label{migui2}
\end{eqnarray}
where plus sign corresponds to boson ($S$) and minus sign corresponds to fermion ($A$).
Then, one can easily prove the following propositions.

\begin{Prop}
\begin{eqnarray*}
[a_\alpha,b_\beta]_{\pm}=\delta_{\alpha,\beta}{\bf 1}
,\ [a_\alpha,a_\beta]_{\pm}=[b_\alpha,b_\beta]_{\pm}=0
,\ (a_\alpha)^\dag=b_\alpha\ \ ,
\end{eqnarray*}
where $[\cdot,\cdot]_+$ and $[\cdot,\cdot]_-$ represent 
anti-commutation relation and commutation relation respectively, and $\dag$ 
represent Hermite conjugate in Fock space.
Since $b_\alpha$ is the Hermite conjugate of $a_\alpha$, 
we will denote $b_\alpha$ as $a_\alpha^\dag$.
\end{Prop}
For instance, suppose 
\begin{eqnarray*}
\Omega=
	\begin{pmatrix}
	1
	\\
	0
	\\
	0
	\\
	\vdots
	\end{pmatrix}\in {\cal F}\ \ ,
\end{eqnarray*}
then, we have
\begin{eqnarray*}
a^\dag_\alpha \Omega=	
	\begin{pmatrix}
	0
	\\
	\phi_\alpha(x_1)
	\\
	0
	\\
	\vdots
	\end{pmatrix}
,\ \ 
a^\dag_\beta a^\dag_\alpha \Omega=	
	\begin{pmatrix}
	0
	\\
	0
	\\
	\frac{1}{\sqrt{2}}
	\left\{ \phi_\beta(x_1)\phi_\alpha(x_2)
	\pm
	\phi_\beta(x_2)\phi_\alpha(x_1) \right\}
	\\
	0
	\\
	\vdots
	\end{pmatrix}
	,\cdots\ \ .
\end{eqnarray*}
In general, 
for operator $A$ on ${\cal H}$, 
a one particle operator $\hat{A}:\ {\cal F} \to {\cal F}$ 
is defined by
\begin{eqnarray*}
	\hat{A}\psi=
	\begin{pmatrix}
	0
	\\
	A\psi^{(1)}(x_1)
	\\
	\vdots
	\\
	\sum_{j=1}^n A_j \psi^{(n)}(x_1,\cdots,x_n)
	\\
	\vdots
	\end{pmatrix}
	,\ \ 
%%%
	\psi=
	\begin{pmatrix}
	0
	\\
	\psi^{(1)}(x_1)
	\\
	\vdots
	\\
	\psi^{(n)}(x_1,\cdots,x_n)
	\\
	\vdots
	\end{pmatrix}
\ .
\end{eqnarray*}
We will express it in a very useful way. For this first note that due to 
Eq.(\ref{expansion}),  we have
\begin{eqnarray*}
A_j \psi^{(n)}(x_1,\cdots,x_n)&=&\sum_\alpha \bra{\alpha}A_j \ket{\psi^{(n)}}\phi_\alpha(x_j)
\nonumber
\\&=&
\sum_{\alpha,\beta} \bra{\alpha}A_j \ket{\beta}\int d\bar{x}_j\phi_\beta(\bar{x}_j)\psi^{(n)}(x_1,\cdots,\bar{x}_j,\cdots,x_\alpha)\phi_\alpha(x_j)
\ .
\end{eqnarray*}
With the aids of 
$\psi^{(n)}(x_1,\cdots,\bar{x}_j,\cdots,x_n)=(\pm 1)^{j-1}\psi^{(n)}(\bar{x}_j,x_1,\cdots,x_{j-1},x_{j+1}\cdots,x_n)$, 
one recognizes (see Eqs.(\ref{migui1},\ref{migui2})) the component $(a_\alpha^\dag a_\beta \psi)^{(n)}$ in the second line of the previous equation.
Then, the one particle operator  $\hat{A}$ reads
\begin{eqnarray}
	\hat{A}=\sum_{\alpha,\beta} 
	\bra{\alpha} A \ket{\beta}a_\alpha^\dag a_\beta
,\qquad	\bra{\alpha}A\ket{\beta}
	=\int dx\ \phi_\alpha (x)^* A \phi_\beta (x)
	\ \ .
\end{eqnarray}
In a similar way, two particle operators can be defined. 
Let $V(y,y')$ be a symmetric operator, i.e., $V(y,y')=V(y',y)$, then, two particle operators on Fock space are defined by
\begin{eqnarray*}
\hat{V}\psi =
	\begin{pmatrix}
	0 \\ 0 \\ 
	V(y_1,y_2)\psi^{(2)}(y_1,y_2) \\ \vdots \\
	\sum_{i<j} V(y_i,y_j)\psi^{(n)}(y_1,\cdots,y_n) \\
	\vdots
	\end{pmatrix}
\ .
\end{eqnarray*}
This operator can be rewritten in terms of creation and annihilation operators:
\begin{eqnarray*}
	\hat{V} &=& \sum_{\alpha,\beta,\alpha',\beta'} 
	\bra{\alpha,\beta} V \ket{\alpha',\beta'}
	a_\alpha^\dag a_\beta^\dag
	a_{\beta'} a_{\alpha'}
\\
	\bra{\alpha,\beta} V \ket{\alpha',\beta'}
	&=& \int dx dx'\ \phi_\alpha (x)^*  \phi_{\alpha'} (x) V(x,x') 
	 \phi_\beta (x')^*  \phi_{\beta'} (x')
\end{eqnarray*}
Note:\\
Let 
\begin{eqnarray*}
\hat{\psi}(y)=\sum_\alpha \phi_\alpha (y) a_\alpha\ \ ,
\end{eqnarray*}
then, $\hat{A}$ and $\hat{V}$ reads
\begin{eqnarray*}
\hat{A} &=& \int dy \hat{\psi}(y)^\dag A \hat{\psi}(y)
\\
\hat{V} &=& \frac{1}{2}
	\int dy\ dy'\ \hat{\psi}(y)^\dag \hat{\psi}(y')^\dag V(y,y') 
	\hat{\psi}(y') \hat{\psi}(y)
\ .
\end{eqnarray*}

%\newpage
\section{$C^*$ algebra}
A satisfactory theory for nonequilibrium systems should include treatments of reservoir, sine system might be strongly influenced by them.

The most common approaches to nonequilibrium systems involve those with Keldysh Green function and reduced density operator.
Applicabilities of those methods are remarkable; however those might includes some fundamental problems.
For instance, environments are traced out in the reduced density operator, thus, it is not possible to characterize states of system including environments. Also, the method does not give any insights for correlation between system and environments.
On the other hand, Keldysh method is based on the existence of NESS and validity of the perturbation series, adiabatic switch-on for interaction.
However, those assumptions are not always valid.

There are other approaches to nonequilibrium systems, like steady state thermodynamics by Sasa and Tasaki~\cite{ST06}, Zubarev ensemble~\cite{Zubarev95}.
Though assumptions of those methods should be on debate~\cite{Hayakawa03}.

Contrary to the methods described above, the method of $C^*$ algebra does not require assumptions for the switch-on of interaction, extrapolation from equilibrium. Although it can treat only specific systems, one can study natural steady states derived from a time evolution of full systems.
$C^*$ algebra ${\cal A}$ is constructed as a set of observables with finite norms (the observables can be local elements in infinitely extended systems).

Time evolution on ${\cal A}$ is described by a linear map $\tau_t$:
${\cal A}\to {\cal A}$, and states~$\omega$ is described by a positive linear functional on ${\cal A}$:\ 
${\displaystyle 
\omega(A)\ge 0,\ \omega(\alpha A+\beta B)=\alpha\omega(A)+\omega(B)}$ .

Since norm of ${\cal A}$ is finite by definition and $\omega$ is positive, we have
\begin{eqnarray*}
|\omega(A)|\le ||A||<\infty\ \ .
\end{eqnarray*}
Therefore this method can describe behavior of infinite systems only with finite values.
For instance, hamiltonian of total systems and canonical ensembles are mathematically ill-defined.
Within the framework of $C*$ algebra, we do not explicitly use them, and the framework is mathematically well-defined.
\\
We would like to remark that this framework is contrary to conventional quantum mechanics; namely, in conventional quantum mechanics, Hilbert space for state is introduced first, and then, linear operators on the Hilbert space are defined.

The advantages of this method is that one can rigorously discuss the existence of NESS purely by dynamical time evolution, and it is not based on physically strong assumptions such as Markovianity and adiabatic switch-on
\footnote{
The method is usually applied with purely mathematical motivation. S. A. thinks that Professor Tasaki's motivation was to study physically interesting system by using the rigorous results of the $C^*$ algebra}.
In the following section, we are going to review the framework of $C^*$ algebra.

\subsection{The definition of $C^*$ algebra and some examples}
\begin{Def}
${\cal A}$ is a *~algebra if algebra ${\cal A}$ is together with 
an involution $*\:\ {\cal A}\rightarrow {\cal A}$:
\begin{itemize}
\item
$(A^*)^*=A,\ (AB)^*=B^* A^*,\ \forall A,B\in {\cal A}$

\item
$(\alpha A +\beta B)^*=\alpha^* A^* +\beta^* B^*,\ \forall A,B\in {\cal A},\ \forall \alpha,\beta\in {\rm \bf{C}}$ ,
\end{itemize}
where $\alpha^*$ denotes the complex conjugate of $\alpha$\footnote{Note that we assign a double meaning to $*$.
It is the usual complex conjugate operation for complex number and also the involution on the elements of the algebra.}.
\end{Def}

\begin{Def}
${\cal A}$ is a $C^*$ algebra if *~algebra ${\cal A}$ satisfies the following properties.
\begin{itemize}
\item
{\cal A} is a is a Banach 
algebra with respect to a norm $||\cdot||<\infty$.

\item
The following properties are satisfied for arbitrary $A,B\in {\cal A}$:
	\begin{itemize}
	\item[(i)]
	$||AB||\le ||A||\ ||B||$

	\item[(ii)]
	$||A^*||=||A||$

	\item[(iii)]
	$||A^* A||=||A||^2\ \ (C^*\ {\rm  property)}$
	\end{itemize}

\end{itemize}
\end{Def}

\begin{Ex}[CAR algebra of spinless fermion]
\label{CAR_ex}
Let $f\in L^2$, and $a_k$ the annihilation operator on its associated Fock space ${\cal F}$.
An algebra ${\cal A}$ generated by
\begin{eqnarray*}
a(f)\equiv \int dk\ f^*(k) a_k,\ f\in L^2
\end{eqnarray*}
$a(f)^*$ and identity element ${\bf 1}$
equipped with a norm
\begin{eqnarray*}
\Vert f \Vert_{L^2}\equiv
\sqrt{\int dk\ |f(k)|^2}\ ,
\end{eqnarray*}
is $C^*$ algebra.

\end{Ex}

\begin{Prop}
For the CAR algebra defined in the previous example, we have the following three properties.
\begin{itemize}
	\item[(i)]
	$a(c_1f+c_2g)=c_1^*a(f)+c_2^*a(g)
	,\ c_j \in{\bf C},\ f,g\in h 
	$\qquad (antilinearity)
	\item[(ii)]
	$[a(f),a(g)]_+=0,\ f,g\in h $
	\item[(iii)]
	$[a(f),a(g)^*]_+=\langle f,g \rangle_{L^2} {\bf 1}$, 
	where
	$\langle f,g\rangle_{L^2}=\int dx f(x)^* g(x)$.

\end{itemize}
\end{Prop}

Proof of (iii)
\begin{eqnarray*}
[a(f),a(g)^*]_+
&=&
\int dk\ dk'\ f(k)^* g(k') [a_{k},a_{k'}^*]_+
\\ &=&
\int dk\ f(k)^* g(k) {\bf 1}
\\ &=&
\langle f,g\rangle_{L^2} {\bf 1}
\end{eqnarray*}

\rk{1}
Elements of ${\cal A}$ can be approximated by
finite sum of identity element {\bf 1}, 
$a\left(f_1\right)^*\cdots a\left(f_n\right)^*$, 
and
$a\left(g_1\right)\cdots a\left(g_m\right)$ 
with arbitrary precision.
\\
\rk{2}
The following theorem is well-known~[Bratelli-Robinson~5.2.5] (Throughout this paper, Bratelli-Robinson means that statements are from Ref.~\cite{BR02}).\\
Let $h$ be a pre-Hilbert space with closure $\bar{h}$ and let 
${\cal A}_i\ (i=1,2)$ be two $C^*$ algebras generated by the identity {\bf 1} 
and elements $a_i(f),\ f\in h$, satisfying 
\begin{itemize}
	\item[(i)]
	$a_i(c_1f+c_2g)=c_1^*a_i(f)+c_2^*a_i(g)
	,\ c_j \in{\bf C}$
	\qquad (antilinearity)
	\item[(ii)]
	$[a_i(f),a_i(g)]_+=0$
	\item[(iii)]
	$[a_i(f),a_i(g)^*]_+=\langle f,g\rangle {\bf 1}\ ,$\\
\end{itemize}
for all $f,g\in h,\ i=1,2$.
It follows that there exists a unique *-isomorphism 
$\alpha:\ {\cal A}_1 \to  {\cal A}_2$ such that
\begin{eqnarray*}
\alpha (a_1(f))=a_2(f)
\end{eqnarray*}

for all $f\in h$. Thus, there exists a unique, 
up to *-isomorphism, $C^*$ algebra
${\cal A}={\cal A}(h)={\cal A}(\bar{h})$ generated by elements $f(f)$, 
satisfying the canonical anti-commutation relations over $\bar{h}$.

\begin{Ex}[Bratelli-Robinson~2.1.2]
Let ${\cal H}$ be a Hilbert space and ${\cal B}({\cal H})$ be the set of all bounded operators over ${\cal H}$.
Define sums and products of elements of ${\cal B}({\cal H})$ in the standard manner, and equip this set with the operator norm
\begin{eqnarray*}
\Vert A \Vert =\sup \{
\Vert A\phi \Vert;\ \phi\in {\cal H}, \Vert \phi \Vert=1\}\ .
\end{eqnarray*}
Then, the adjoint operation satisfies the properties of involution, and with respect to this involution and the operator norm,  ${\cal B}({\cal H})$ form a $C^*$ algebra.

\end{Ex}

\subsection{Spectral analysis on $C^*$ algebra}
In conventional quantum mechanics, observables are linear operators on the Hilbert space, and one measures their eigenvalues.
In $C^*$ algebra, corresponding vector space does not exist, and it is not possible to discuss eigenvalues.
In this subsection, we are going to discuss the spectrum of element $A\in {\cal A}$, which is analogous to a set of eigenvalues.

\begin{Def}
Let ${\cal A}$ be an algebra with identity ${\bf 1}$ (Hereafter, we shall call unital algebra). The resolvent set 
$r_{\cal A}(A)$ of an element $A\in {\cal A}$ is defined as the set of 
$\lambda\in {\bf C}$ such that $(\lambda {\bf 1}-A)$ is invertible and the 
spectrum $\sigma_{{\cal A}}$ of $A$ is defined as the complement of $r_{\cal A}(A)$ in {\bf C}.
The inverse $(\lambda \bf{1}-A)^{-1}$, where $\lambda\in r_{\cal A}(A)$ is called the
resolvent of $A$ at $\lambda$.
\end{Def}

One of the simplest approaches to analyze resolvents and spectra is expanding 
resolvents with respect to $\lambda$, and analytically continuing it.
For $\lambda\in {\bf C}$ and $|\lambda|>||A||$, resolvent can be expanded as
\begin{eqnarray}
(\lambda \bf{1}-A)^{-1}=
\lambda^{-1}\sum_{m\ge 0} \left(\frac{A}{\lambda}\right)^m\in {\cal A}\ ,\ 
({\rm completeness})
\label{expansion}
\end{eqnarray}
Therefore, 
$\{\lambda:\ \lambda\in{\bf C},\ |\lambda|> ||A||\}$ 
is a subset of $r_{\cal A}(A)$, and spectrum 
$\sigma_{{\cal A}}(A)\subset$ 
$\{\lambda:\ \lambda\in{\bf C},\ |\lambda|\le ||A||\}$ 
 is bounded.
\\Note\\
One can prove that $\sigma_{\cal A}(A)$ is not empty.
For spectrum of $A$, the spectral radius $\rho(A)$ of $A$ is defined by
\begin{eqnarray*}
\rho(A) = 
\sup \{ |\lambda|,\lambda\in \sigma_{\cal A} (A)\}
\ .
\end{eqnarray*}

\begin{Def}

\begin{itemize}
	\item[(i)]
	If an element of $C^*$ algebra satisfies $A^* A=A A^*$, then, 
	$A$ is defined to be a normal element.
	\item[(ii)]
	If an element of $C^*$ algebra satisfies $A^* =A$, then, 
	$A$ is defined to be a selfadjoint element.
	The set of all selfadjoint elements of ${\cal A}$ 
	is denoted by ${\cal A}_{s.a.}$.

	\item[(iii)]
	Suppose ${\cal A}$ has identity ${\bf 1}$.
	If an element of $C^*$ algebra satisfies $A^* A=AA^*={\bf 1}$, then, 
	$A$ is defined to be a unitary element.

\end{itemize}
\end{Def}

\begin{Prop} [Arai 3.8]
${\cal A}_{s.a.}$ is closed.
\end{Prop}
Throughout this paper, Arai means that statements are from Ref.~\cite{Arai08}.\\
\underline{proof}\\
Suppose $A_n\in {\cal A}_{s.a.}$, and $A_n\to A\in {\cal A}$ as $n\to\infty$.
\\ 
$\Vert A_n-A\Vert = \Vert A^*_n-A^*\Vert$. Combined with $A_n=A_n^*$,  we have $A_n\to A^*$. Therefore, $A=A^*$.

\QED{}
\\
\\
Let us show some properties of spectrum.
The following proposition is important.
\begin{Prop}[Bratteli Robinson~2.2.2, 2.2.5]\ \ 
\begin{itemize}
\item[(i)]
	If $A\in{\cal A}$ is unitary (i.e. $AA^*=A^*A={\bf 1}$), 
	then, $\sigma_{\cal A}(A)\subset  
 \{\lambda:\lambda\in{\bf C}, |\lambda|=1 \}$ .
\item[(ii)]
	If $A\in{\cal A}$ is selfadjoint (i.e. $A^*=A$), 
	then, \\
	$\spect{A}\subset [-||A||,||A||],\ \ \spect{A^2}\subset [0,||A||^2]$ .

\item[(iii)]
	$
	\rho(A)=\lim_{n\to\infty}
	||A^n||^{1/n}=\inf_n
	||A^n||^{1/n}\le ||A||
	$
\end{itemize}
\end{Prop}

We are going to prove (i) and (ii) with the aid of (iii).\\
\underline{proof of (i)}\\
$C^*$ property yields
 \begin{eqnarray*}
  \|A^n\|^2 &=& \|\left(A^n\right)^* A^n \| \qquad (C^*\ {\rm C^* property})
 \nonumber \\ &=&
 \|\left(A^{n-1}\right)^* A^* A A^{n-1} \| = \|\left(A^{n-1}\right)^* A^{n-1} \|  = \cdots = 1 \ .
 \end{eqnarray*}
Thus, 
 \begin{eqnarray*}
 \rho(A)=1\ ,
 \end{eqnarray*}
 and it follows
 \begin{eqnarray*}
 \sigma_{\cal A}(A)\subset \{\lambda:\lambda\in{\bf C}, |\lambda|\le 1\}\ .
 \end{eqnarray*} 
Similarly, $\rho\left(A^*\right)=1$, and $A^{-1}=A^*$ implies
 \begin{eqnarray*}
  \left(\lambda {\bf 1} - A\right) = \lambda A 
  \left(A^{*} - \frac{1}{\lambda}{\bf 1} \right)\ .
 \end{eqnarray*}
Therefore, we have
 \begin{eqnarray*}
  \exists \left(\lambda {\bf 1} - A\right)^{-1}
  \Leftrightarrow
  \exists \left(A^{*} - \frac{1}{\lambda}{\bf 1} \right)^{-1}
\ .
 \end{eqnarray*}
 It follows
 \begin{eqnarray*}
  \lambda\in \reso{A}
  \Leftrightarrow
  \frac{1}{\lambda}\in \reso{A^*}\ ,
 \end{eqnarray*}
 and
 \begin{eqnarray*}
  \lambda\in \spect{A}
  \Leftrightarrow
  \frac{1}{\lambda}\in \spect{A^*}\ .
 \end{eqnarray*}
Combining with $\rho(A)=\rho(A^*)=1$, we have 
 \begin{eqnarray*}
  \lambda\in \spect{A}\Rightarrow |\lambda|=1\ .
 \end{eqnarray*}
In a conventional quantum mechanics, the same results can be derived for the eigenvalues of unitary operators
on Hilbert space.
Property~(i) corresponds to the fact that the eigenvalues of unitary operators acting on a Hilbert space have modulus 1.
\\
\\
\underline{proof of (ii)}\\
\begin{eqnarray*}
\Bigg|\Bigg| \frac{(\mp A)^n}{(2i||A||)^{n+1}} \Bigg|\Bigg|
\le \frac{||A||^n}{(2||A||)^{n+1}}
\le \frac{1}{2^{n+1}||A||}
\end{eqnarray*}
implies that the Von Neumann series: 
\begin{eqnarray*}
\sum_{n=0}^\infty
\frac{(\mp A)^n}{(2i||A||)^{n+1}}
\end{eqnarray*}
is convergent. 
It follows that $(2i||A||{\bf 1}\pm A)^{-1}$ exists.
Thus, Cayley transform~$K$ of $A$ 
(this transform can be defined for unbounded operator and transform 
Hermiticity to unitarity):
\begin{eqnarray*}
K=(i\alpha{\bf 1}- A)^{-1}(i\alpha{\bf 1}+ A),\ \ \alpha\equiv 2||A||
\end{eqnarray*}
 is well-defined. 
By using, Hermiticity and the existence of  $(2i||A||{\bf 1}\pm A)^{-1}$, 
and the following equality:
\begin{eqnarray*}
(i\alpha {\bf 1}+ A)(-i \alpha {\bf 1}+ A) 
&=&
(i \alpha {\bf 1}- A)(-i\alpha {\bf 1}- A)\ ,
\end{eqnarray*}
we have 
\begin{eqnarray*}
K^* &=& %(-2i||A||{\bf 1}+ A) (-2i||A||{\bf 1}- A)^{-1}
(-i\alpha{\bf 1}+ A) (-i\alpha{\bf 1}- A)^{-1}
\\ &=&
(i\alpha{\bf 1}+ A)^{-1} (i\alpha{\bf 1}- A)\ .
\end{eqnarray*}
Thus, $KK^*={\bf 1}$ ($K$ is unitary).
One can also prove the existence of $(K+{\bf 1})^{-1}$, and 
inverse Cayley transformation reads
\begin{eqnarray*}
A=i\alpha (K-{\bf 1}) (K+{\bf 1})^{-1}\ .
\end{eqnarray*}
Thus, resolvent of $A$ is
\begin{eqnarray*}
\lambda {\bf 1}-A &=&
\lambda {\bf 1}-i\alpha(K-{\bf 1}) (K+{\bf 1})^{-1}
\\ &=&
\left\{
(\lambda+i\alpha) {\bf 1}+
(\lambda-i\alpha) K 
\right\}
(K+{\bf 1})^{-1}
\\ &=&
(-\lambda+i\alpha)
\left\{
\frac{\lambda+i\alpha}{-\lambda+i\alpha} {\bf 1} - K 
\right\}
(K+{\bf 1})^{-1}\ .
\end{eqnarray*}
Since $\rho(A)\le ||A||$, 
we only need to discuss $|\lambda|<||A||$. 
With the restriction, we have
\begin{eqnarray*}
|-\lambda+i\alpha|=\Big| -\lambda+2i||A||\Big|
\ge 2||A||-\lambda>0\ .
\end{eqnarray*}
It follows
\begin{eqnarray*}
\exists (\lambda {\bf 1}-A)^{-1}
\Leftrightarrow
\exists
\left\{
\frac{\lambda+i\alpha}{-\lambda+i\alpha} {\bf 1} - K 
\right\}^{-1}
\end{eqnarray*}
Therefore $\spect{A}$ is a subset of ${\bf R}$:
\begin{eqnarray*}
\lambda\in \spect{A}
&\Leftrightarrow&
\frac{\lambda+i\alpha}{-\lambda+i\alpha} \in \spect{K}\subset
\{\sigma:\sigma\in{\bf C}, |\sigma|=1 \}
\\
&\Leftrightarrow&
\Big|\lambda+i\alpha \Big|=\Big|-\lambda+i\alpha \Big|
\\
&\Leftrightarrow&
{\rm Im}\lambda+\alpha =-{\rm Im}\lambda+\alpha
\\
&\Leftrightarrow&
\lambda\in {\bf R}
\end{eqnarray*}
In addition, $\rho(A)=||A||$ gives $\spect{A}\subset [-||A||,||A||]$ .
Finally, let us study the resolvent of $A^2$.
\begin{itemize}
	\item $\lambda>0$\\
	$(\lambda{\bf 1}-A^2)=
	(\sqrt{\lambda}{\bf 1}-A)(\sqrt{\lambda}{\bf 1}+A)$ implies that
	$\lambda\in\sigma(A^2)$ is equivalent to 
	$\sqrt{\lambda}\in\sigma(A)$ or $-\sqrt{\lambda}\in\sigma(A)$.
	Thus, resolvent does not exist for 
	$\lambda\in [0,||A||^2]$.
%%%%
	\item $\lambda<0$\\
	$(\lambda{\bf 1}-A^2)=
	\left(i\sqrt{|\lambda|}{\bf 1}-A\right)
	\left(i\sqrt{|\lambda|}{\bf 1}+A\right)$.
	Combining with 
	$\spect{A}\subset [-||A||,||A||]$, 
	resolvent of $A^2$ exists for $\lambda<0$.

\end{itemize}
Thus, we have $\spect{A^2}\in [0,||A||^2]$.

\QED{}

We only give an explanation for (iii)
(see [Bratteli Robinson~2.2.2] for a complete proof).\\
The expansion of resolvent with respect to $\lambda^{-1}$
reads
\begin{eqnarray}
R_\lambda(A)={1\over\lambda}\left({\bf 1}-{A\over\lambda}\right)^{-1}
={1\over\lambda}\left\{{\bf 1}+\sum_{n=1}^\infty {1\over\lambda^n}A^n
\right\}\ .
\label{resolvent_series}
\end{eqnarray}
This series is absolute convergent when 
\begin{eqnarray*}
\sum_{n=0}^\infty {1\over|\lambda|^{n+1}} ||A^n||\ ,
\end{eqnarray*}
is convergent.
Thus, the series~(\ref{resolvent_series}) is convergent for
\begin{eqnarray*}
\frac{1}{a}=\lim_{n\to \infty}||A^n||^{1/n}<|\lambda|
\ .
\end{eqnarray*}
Hence,
\begin{eqnarray*}
\lim_{n\to\infty} ||A^n||^{1/n}\ge \rho(A).
\end{eqnarray*}

\begin{Prop}
	$\rho(A)=||A||$ for a normal element $A\in {\cal A}$.
\end{Prop}
\underline{proof}\\ 
\begin{eqnarray*}
 \rho(A) &=& \lim_{n\to\infty} ||A^n||^{1/n}
 \\ &=&
 \lim_{n\to\infty}||A^{2^n}||^{1/2^n}
\end{eqnarray*}
From $C^*$ property and normality, we have
\begin{eqnarray*}
 ||A^{2^n}||^2 &=& ||(A^{2^n})^* A^{2^n} ||\qquad (C^*\ {\rm property})
 \\ &=& 
 ||(A^*A)^{2^n}||\qquad ({\rm normality})
 \\ &=& 
 ||(A^*A)^{ 2^{n-1} }||^2
 \\ &=& 
 || A^*A ||^{2^n}
 \\ &=& 
 || A ||^{2^{n+1}}\ .
\end{eqnarray*}
Therefore, 
\begin{eqnarray*}
 ||A^{2^n}||^{1/2^n}=||A||\to||A||,\qquad {\rm as}\ n\to \infty\ .
\end{eqnarray*}
\QED{}

\begin{Def}
$A\in{\cal A}$ is defined to be positive if $A^*=A$ and 
$\spect{A}\subset {\bf R}_+\equiv
 \{\lambda:\lambda\in{\bf R}, \lambda\ge 0 \}
$. 
If $A$ is positive, we denote $A\ge 0$, and the set of all positive elements of ${\cal A}$ is denoted by ${\cal A}_+$.
\end{Def}

\begin{Prop}
\begin{itemize}
	\item[(i)] $\spect{\lambda {\bf 1}-A}=\lambda-\spect{A}$
	\item[(ii)] $\spect{\lambda A}=\lambda\spect{A}$
	\item[(iii)] Let $A\in A$ be an invertible element, then, 
	\begin{eqnarray*}
	 \spect{A^{-1}}=\{ \lambda|:\ 1/\lambda\in \spect{A}\}\ .
	\end{eqnarray*}
	\item[(iv)] $\spect{A^*} =\overline{\spect{A}}$
	\item[(v)] $\spect{AB}\cup\{0\}=\spect{BA}\cup\{0\}$
		
\end{itemize}
\end{Prop}

\underline{proof of (i)}\\ 
$\mu\in\spect{\lambda {\bf 1}-A}\Leftrightarrow$
$(\lambda-\mu){\bf 1}-A$ is not invertible.
\\
$\Leftrightarrow \lambda-\mu\in\spect{A}$
$\Leftrightarrow \mu\in\lambda-\spect{A}$
\\
\underline{proof of (ii)}\\ 
For $\lambda =0$, the relation obviously is satisfied because of $\spect{{\bf 0}}=\{0\}$. 
Let us suppose $\lambda \neq 0$, then, \\
$\mu\in\spect{\lambda A}\Leftrightarrow \lambda A- \mu$ is not invertible.
$\Leftrightarrow \left(A-\frac{\mu}{\lambda} {\bf 1}\right)$ is not invertible.
$\Leftrightarrow \mu\in\lambda\spect{A}$
\\
\underline{proof of (iii)}\\ 
Since $A$ is invertible, we have $0\in r_{\cal A}(A)$. Let us suppose 
$\lambda\neq 0$.
\\
Because of
$A^{-1}-\lambda^{-1} {\bf 1}=-\lambda^{-1}A^{-1}(A-\lambda{\bf 1})\ ,$
the existence of $(A^{-1}-\lambda^{-1} {\bf 1})^{-1}$ and the existence of $(A-\lambda{\bf 1})^{-1}$ is equivalent.
\\
\underline{proof of (iv)}\\ 
The claim follows from $(\lambda{\bf 1}-A)^*=(\bar{\lambda}{\bf 1}-A^*)$.
\\
\underline{proof of (v)}\\ 
Suppose $\lambda\in\reso{BA}$, then, 
$$(\lambda{\bf 1}-AB)({\bf 1}+A(\lambda {\bf 1}-BA)^{-1}B)=\lambda{\bf 1}\ .$$
Therefore, $\lambda {\bf 1}-AB$ is invertible for $\lambda\in\reso{BA}$ with a possible exception $\lambda=0$. 
Therefore, $\spect{BA}\cup \{0\}\supset\spect{AB}\cup \{0\}$. 
Similarly, we have $\spect{AB}\cup \{0\}\supset\spect{BA}\cup \{0\}$.

\QED{}

\begin{Prop}
\begin{itemize}
	\item[(i)]
	$\spect{A}\subset [0,||A||]$, $\forall A\in {\cal A}_+$

	\item[(ii)]
	$\forall \lambda<0$, $(A-\lambda {\bf 1})^{-1} \in {\cal A}_+$. 
	Moreover, 
	\begin{eqnarray*}
	 ||(A-\lambda {\bf 1})^{-1}|| \le \frac{1}{|\lambda|}\ .
	\end{eqnarray*}

\end{itemize}
\end{Prop}
\underline{proof of (ii)}\\ 
With the aid of
\begin{eqnarray*}
\spect{(A-\lambda {\bf 1})^{-1}}=
\left\{ \frac{1}{\mu-\lambda}:\ \mu\in\spect{A}\right\}\ ,
\end{eqnarray*}
and (i), we have
\begin{eqnarray*}
\spect{(A-\lambda {\bf 1})^{-1}}
\subset [1/(||A||+|\lambda|),1/|\lambda|]\ .
\end{eqnarray*}
\QED{}
\begin{Prop} [Arai 3.31]
${\cal A}_+$ is closed.
\label{prop8}
\end{Prop}

\underline{proof}\\
Suppose $A_n\in {\cal A}_+$, and $A_n\to A\in {\cal A}$ as $n\to\infty$.
Since ${\cal A}_{s.a.}$ is closed, we have $A\in{\cal A}_{s.a.}$.
\\ 
It is sufficient to show that $A-\lambda{\bf 1}$ is invertible for $\lambda<0$.
Let us assume that $\lambda$ is negative.
Then, positivity of $A_n$ yields $\lambda\in \reso{A_n}$, and thus, 
$A_n-\lambda {\bf 1}$ is invertible.
Therefore, 
\begin{eqnarray*}
A-\lambda{\bf 1} &=& (A-A_n)+(A_n-\lambda{\bf 1})
\\  &=& 
\{ (A-A_n)+(A_n-\lambda{\bf 1}) \}(A_n-\lambda{\bf 1})^{-1} (A_n-\lambda{\bf 1})\\  &=& 
\{ (A-A_n)(A_n-\lambda{\bf 1})^{-1} +{\bf 1} \}(A_n-\lambda{\bf 1})
\ .
\end{eqnarray*}
Since $A_n-\lambda {\bf 1}$ is invertible, $A-\lambda {\bf 1}$ is invertible, if and only if, $\{ (A-A_n)(A_n-\lambda{\bf 1})^{-1} +{\bf 1} \}$ is invertible.
Thus, it is sufficient to show 
$-1\in \reso{ (A-A_n)(A_n-\lambda{\bf 1})^{-1} }$.
For all positive number $\epsilon$, there exists 
a natural number $n_0$ such that 
\begin{eqnarray*}
\Vert A_n-A\Vert < \epsilon,\ \ \forall n>n_0\ .
\end{eqnarray*}
Combining the previous estimation with proposition, we have
\begin{eqnarray*}
\Vert (A-A_n)(A_n-\lambda{\bf 1})^{-1} \Vert
&\le& \Vert A-A_n \Vert\ \Vert(A_n-\lambda{\bf 1})^{-1} \Vert
\\ &\le&
\frac{\epsilon}{|\lambda|},\ \ \forall n\ge n_0
\ .
\end{eqnarray*}
By taking $\frac{\epsilon}{|\lambda|}<1$, we have $\rho\left((A-A_n)(A_n-\lambda{\bf 1})^{-1}\right)<1$ for $n>n_0$, and hence $-1\in \reso{ (A-A_n)(A_n-\lambda{\bf 1})^{-1} }$.

\QED{}

\begin{Prop}
\label{positivity_Prop}
Let $A$ be a $C^*$ algebra ${\cal A}$, then, the following five properties are satisfied:
\begin{itemize}
	\item[(i)]
	A selfadjoint element $A$ is positive if, and only if 
	$||{\bf 1}-A/||A||\ ||\le 1$.
	Moreover if a self-adjoint element $A$ satisfies
	$\Vert {\bf 1}-A\Vert\le 1$, then $A$ is positive, and $||A||<2$.

	\item[(ii)]
	A selfadjoint element $A$ is positive if, and only if $A=B^2$ for 
	some $B\in{\cal A}_{s.a}$. If $A$ is positive, 
	there exists a unique positive $B$ such that $A=B^2$.
	Moreover, if $A$ is commutable with $C$, i.e. $AC=CA,\ (C\in {\cal A})$, 
	then, $B$ is also commutable with $C$, i.e., $BC=CB$.
		
	\item[(iii)]
	Let $A_1\ A_2\in {\cal A}$ be positive elements,
	then, $c_1A_1$ and $c_1A_1+c_2A_2$ are positive for $c_1, c_2>0$.

	\item[(iv)]	
	If $A$ is self-adjoint, then, 
	$A_+ A_-=0$ and $A=A_+ - A_-$ for some positive elements $A_\pm$.

	\item[(v)]	
	$A$ is positive, if, and only if, 
	$A=B^*B$ for some $B\in{\cal A}$.
\end{itemize}
\end{Prop}

\underline{Note for (ii)}\\
From the uniqueness of $B$ for positive elements, the square root of $A\in{\cal A}_+$ is well-defined, and is usually denoted by $A^{1/2}$ or $\sqrt{A}$.
Since $A^2\in {\cal A}_+$ for selfadjoint element, we define modulus of the 
selfadjoint elements by $|A|=\sqrt{A^2}$.

\underline{Note for (iv)}\\
The decomposition $A=A_+-A_-$ is referred to as the orthogonal 
decomposition of $A$.
\\ \\
\noindent {\bf Lemma}
Let $A,B\in {\cal A}_+$. If $A+B={\bf 0}$, then, $A=B={\bf 0}$.
\\
\underline{proof}\\ 
$A=-B$ implies $\spect{A}\subset[-||B||,0]$. It follows
$$\spect{A}=\spect{B}=\{ {\bf 0} \}.$$ Thus, we have $A=B=0$.

\QED{}
\\
\underline{proof of (i)}\\
%Lemma follows\\
Suppose $A\in{\cal A}_+$, then, 
$\spect{{\bf 1}-A/||A||} \subset [0,1]$.
Thus, $\rho({\bf 1}-A/||A||)=||{\bf 1}-A/||A||\ ||\le 1$.\\
On the other hand, suppose $||{\bf 1}-A/||A||\ ||\le 1$.
Since $\spect{A}\subset [-||A||,||A||]$ for a selfadjoint element 
$A\in{\cal A}$, we have  
\begin{eqnarray*}
\spect{{\bf 1}-A/||A||}\subset[-1,1]\ .
\end{eqnarray*}
With the aid of Lemma, we have $\spect{A}\subset [0,2||A||\ ]$.
\\
\underline{proof of (ii)}\\ 
$\spect{B^2}\subset [0,||B||^2]$ implies that $B^2$ is a positive element for a positive element $B\in {\cal A}_+$.
Let us prove the existence of $B$ for positive elements $A$. 
Let us define $B$ as
\begin{eqnarray*}
&&B\equiv \sqrt{||A||}\left[
{\bf 1}+\sum_{n=1}^\infty c_n
\left(
{\bf 1}-\frac{A}{||A||}
\right)^n
\right]
\\
&&\sqrt{1-x}=1+\sum_{n=1}^\infty c_n x^n
\end{eqnarray*}
From (i), we have, $||{\bf 1}-A/||A||\ ||\le 1$, thus, 
the series in the definition of $B$ is convergent in ${\cal A}$.
Let $C={\bf 1}-A/||A||$, then, 
\begin{eqnarray*}
B^2 &=& ||A||
\left(
{\bf 1}+
\sum_{n=1}^\infty |c_n| C^n
\right)^2
\\ &=& 
||A|| ({\bf 1}-C)=A
\ .
\end{eqnarray*}
With the aid of (i), $||{\bf 1}-B/||B||\ ||\le 1$ is equivalent to the positivity of $B$.
Because of $C^*$property and the self-adjoint property of $B$, we have
\begin{eqnarray*}
||B||^2=||B^* B||=||B^2||=||A||.
\end{eqnarray*}
Combining with $c_n<0$ and $||C||\le 1$, we obtain a desired inequality:
\begin{eqnarray*}
\Bigg|\Bigg|{\bf 1}-\frac{B}{||B||} \Bigg|\Bigg|
&=& \Bigg|\Bigg|{\bf 1}-\frac{B}{ \sqrt{||A||} } \Bigg|\Bigg|
\\ &\le&
-\sum_{n=1}^\infty c_n\ ||C||^n
\\ &=&
1-\left(1+\sum_{n=1}^\infty c_n\ ||C||^n \right)
\\ &=&
1-\sqrt{1-||C||}\le 1
\end{eqnarray*}
Thus, $B$ is positive.
\\
From the definition of $B$, it is obvious that $B$ commutes with any elements which are commutable with $A$.
Next, we are going to prove the uniqueness of an element $B$.
Let us assume that the second positive element $B'$ 
exists.
Let $B=C^2, B'=C'^2$. 
One can easily prove that $A,\ B, B',\ C$ and $C'$ commute each other.
Thus, we have the following equality.
\begin{eqnarray*}
{\bf 0} &=& (A-A)(B-B')
\\ &=& 
(B^2-B'^2)(B-B')
\\ &=& 
(B-B')B(B-B')+(B-B')B'(B-B')
\\ &=& 
\{ (B-B')C \}^2 + \{ (B-B')C' \}^2\ .
\end{eqnarray*}
Since $(B-B')C$ is selfadjoint, $\{ (B-B')C \}^2$ and $\{ (B-B')C' \}^2$
 are positive. Applying a lemma, we have
\begin{eqnarray*}
\{ (B-B')C \}^2 = \{ (B-B')C' \}^2 = {\bf 0}\ .
\end{eqnarray*}
It follows
\begin{eqnarray*}
{\bf 0} &=& 
\{ (B-B')C \}^2 - \{ (B-B')C' \}^2
\\ &=& 
(B-B')^3\ ,
\end{eqnarray*}
and
\begin{eqnarray*}
(B-B')^4={\bf 0}\ .
\end{eqnarray*}
From $C^*$ property and Hermiticity, we have 
\begin{eqnarray*}
{\bf 0}=||(B-B')^4||=||(B-B')^2||^2=||B-B'||^4\ .
\end{eqnarray*}
Thus, $B$ is unique.
\\
\underline{proof of (iii)}\\ 
$c_1>0$ implies
$\sigma(c_1A_1)=\{\lambda : \lambda/c_1\in\sigma(A_1)\}\subset
\{\lambda : \lambda/c_1\in [0,\Vert A_1\Vert]\}=[0,c_1\Vert A_1\Vert]$.
Thus, $c_1A_1$ is a positive element.
Let $p\equiv c_1\Vert A_1\Vert/(c_1\Vert A_1\Vert+c_2\Vert A_2\Vert)<1$.
With the aid of 
$$
p{A_1\over \Vert A_1\Vert}+(1-p){A_2\over\Vert A_2\Vert}={c_1A_1+c_2A_2\over c_1\Vert A_1\Vert+c_2\Vert A_2\Vert}\ ,
$$
we have
\begin{eqnarray*}
\left\Vert {\bf 1}-\left\{p{A_1\over \Vert A_1\Vert}+(1-p){A_2\over\Vert A_2\Vert}\right\}
\right\Vert
&=&
\left\Vert {\bf 1}-
\frac{c_1 A_1+ c_2 A_2}{\Vert c_1 A_1+ c_2 A_2\Vert}
\right\Vert
\\ &\le& 
p\left\Vert {\bf 1}-{A_1\over \Vert A_1\Vert}\right\Vert
+(1-p)\left\Vert {\bf 1}-{A_2\over\Vert A_2\Vert}\right\Vert
\le 1\ .
\end{eqnarray*}
Therefore, $c_1A_1+c_2A_2$ is a positive element.
\\
\underline{proof of (iv)}\\ 
Let $A_\pm=\frac{|A|\pm A}{2}$, then, $A=A_+-A_-$ is obvious.
From the definition of $A$, one has $|A|^2=A^2$, and thus, $A|A|=|A|A$, and 
it follows
$$
4A_+A_-=(|A|+A)(|A|-A)=A^2+A|A|-|A|A-A^2=0\ .
$$
Next, we are going to prove a positivity of $A_\pm$.
Since proofs for $A_+$ and $A_-$ are the same, we only show
the positivity for $A_-$.
Because $A_\pm$ is selfadjoint element, one should prove 
$\Vert {\bf 1} - A_\pm/\Vert A_\pm \Vert \Vert\le 1$.
Since $nA^2\in {\cal A}_+$, ${\bf 1}+nA_\pm^2$ is invertible for $n\in{\bf N}$ 
(proposition ), 
$$
A_n=nA_-^2({\bf 1}+nA_-^2)^{-1}
$$
is well-defined. It follows 
$$
|A|A_n=A_-A_n\ , \quad
A_-A_n-A_-=-A_-({\bf 1}+nA_-^2)^{-1},
$$
and it implies
\begin{eqnarray*}
&&\Vert |A|A_n-A_-\Vert^2=\Vert A_-({\bf 1}+nA_-^2)^{-1}\Vert^2
=\Vert ({\bf 1}+nA_-^2)^{-1}A_-^2({\bf 1}+nA_-^2)^{-1}\Vert
\nonumber\\
&&\le \Vert ({\bf 1}+nA_-^2)^{-1}A_-^2\Vert\ 
\Vert ({\bf 1}+nA_-^2)^{-1}\Vert
={1\over n}
\Vert {\bf 1}-({\bf 1}+nA_-^2)^{-1}\Vert\ 
\Vert ({\bf 1}+nA_-^2)^{-1}\Vert,
\label{prop1}
\end{eqnarray*}
where we have used the $C^*$ property and
\begin{eqnarray*}
({\bf 1}+nA_\pm^2)^{-1}A_\pm^2=\frac{1}{n}
\left(
{\bf 1}-({\bf 1}+nA_\pm^2)^{-1}
\right)\ .
\end{eqnarray*}
$\spect{ {\bf 1}+nA_-^2 }\subset [1,\infty)$ yields that
$\spect{({\bf 1}+nA_-^2)^{-1}}$ and 
$\spect{ {\bf 1}-({\bf 1}+nA_-^2)^{-1}}$ are subsets of $[0,1]$, 
and thus, the right hand side of the equation~(\ref{prop1}) is less than or equal to $1/n$.
Therefore we have 
$$\lim_{n\to\infty} \ |A|A_n =A_-\ .$$
Moreover $|A|A_n$ reads
$$
|A|A_n=\Big(|A|^{1/4}|A_-|^{1/2}\big({\bf 1}/n+A_-^2\big)^{-1/2}|A_-|^{1/2}|A|^{1/4}\Big)^2\ ,
$$
hence, it is positive.
Since $A_+$ is closed (proposition 8), we conclude that $A_-\in {A_+}$.
\\
\underline{proof of (iv)}\\ 
Proposition implies that for any positive element $A$, 
there exists an elements $B$ such that $A=B^*B$.
Conversely, we need to prove the positivity of $B^* B$ for any $B\in{\cal A}$.
Since $B^*B\in {\cal A}_{s.a.}$ is unitary, there exists positive elements 
$C$ and $D$ such that
\begin{eqnarray*}
B^*B=C-D,\ CD={\bf 0}
\end{eqnarray*}
We need to prove$ D={\bf 0}$.
The positivity of 
$$
-(BD)^*(BD)=-DB^*BD=-D(C-D)D=D^3
$$
yields $\spect{(BD)^*(BD)}\subset (-\infty,0]$.
Next, we decompose $BD$ by selfadjoint elements $S,T$\footnote{
Let us consider the decomposition of $A\in{\cal A}$ into two selfadjoint operator.
Let us define $A_{{\rm re}}=(A+A^*)/2,\ A_{{\rm im}}=(A-A^*)/2i$, then, 
$A_{{\rm re}}$ and $A_{{\rm im}}$ are selfadjoint, and they satisfy
$$A = A_{{\rm re}} + i A_{{\rm im}}\ .$$
We note that $A_{{\rm re}}$ and $A_{{\rm im}}$
are referred to as real part and imaginary
part, respectively.
}.
$$BD=S+iT$$
Then, $(BD)(BD)^*$ reads
\begin{eqnarray*}
(BD)(BD)^*=(S+iT)(S-iT)=S^2+T^2+i(TS-ST)=-(BD)^*(BD)+2S^2+2T^2
\ .
\end{eqnarray*}
Since $-(BD)^*(BD)$, $2S^2$, and $2T^2$ are positive, we conclude that
$(BD)(BD)^*$ is positive.
Thus, we have 
$\spect{(BD)(BD)^*}\subset [0,||B||^2 ||D||^2]$.
With the aid of proposition4, we have $\spect{(BD)^*(BD)}\subset [0,||B||^2 ||D||^2]$.

Recall that we have already $-(BD)^*(BD)\in {\cal A}_+$, and hence we conclude 
$\spect{-(BD)^*(BD)}=\spect{D}=\{ 0\}$.
Since $D^3$ is normal, proposition 5 yields $||D^3||=0=||D||^3$, and hence 
$D=0$.

\QED{}

\section{Time evolution on $C^*$ algebra ${\cal A}$}
As we have stated in the introduction, one of the main motivations to introduce $C^*$ algebra was to avoid a mathematically ill-defined observables, such as 
hamiltonian of infinite systems. 
In addition, states also do not have asymptotic limit since $e^{-iHt}$ oscillate very frequently, and thus, 
it is crucial to establish a theory which describes the time evolution of local observables without using a hamiltonian.
Indeed, this divergence problems is crucial in mesoscopic systems, but still one can study interesting local observables such as density of local electrons, current etc.
In this section, we shall present the time evolution on $C^*$ algebra, which formally agree with the time evolution described with a hamiltonian.
Namely, the idea is to introduce a generator of time evolution $\tau_t(A)$ of local observable $A\in {\cal A}$ which formally matches conventional quantum mechanical time evolution $A(t)=e^{iHt} Ae^{-iHt}$ 
(R.H.S. is ill-defined and it only has a formal meaning).
We axiomatically impose the following properties on 
the time generator $\tau_t(\cdot):\ {\cal A}\to  {\cal A}$ 
(strong continuous *-isomorphism group).
\begin{itemize}
	\item[(i)] $\tau_t(\alpha A+\beta B)=\alpha \tau_t(A)+\beta \tau_t(B),
	\ \ 
	\forall \alpha,\beta\in{\bf C},\ \ \forall A,B\in {\cal A}$
	\qquad (linear)
	\\
	$\tau_t(AB)=\tau_t(A)\tau_t(B),\ \ \forall A,B\in{\cal A}$
	\\ $\tau_t(A^*)=\tau_t(A)^*$
	\item[(ii)] $\tau_s\left(\tau_t(A)\right)=\tau_{s+t}(A),
	\ \ \tau_0(A)=A$
	\qquad (group property)
	\item[(iii)]
	$\lim_{t\to 0}\Vert \tau_t(A)-A\Vert=0$,\qquad (strong continuity)
\end{itemize}
We remark that a map $\phi$ satisfying the condition (i) is referred to as a 
*-morphism.
Moreover, if *-morphism is one-to-one, then, it is called *-isomorphism.

\begin{Prop}
Let $\tau_t$ be a group of strong continuous *-isomorphism over ${\cal A}$.
Then, 
$\tau_t({\bf 1})={\bf 1}$ and 
$\Vert \tau_t(A)\Vert=\Vert A\Vert$.

\end{Prop}
{\bf Lemma}
$\spect{A}=\spect{\tau_t(A)}$ for $A\in{\cal A}_{s.a.}$.
\\
\underline{proof of lemma}\\ 
Let $A\in {\cal A}_{s.a.}$, and let $\lambda\in \reso{A}$, then, we have the following equality:
\begin{eqnarray*}
(\lambda{\bf 1}-\tau_t(A))\tau_t\big((\lambda{\bf 1}-A)^{-1}\big)
&=& \left\{ \tau_t(\lambda{\bf 1})-\tau_t(A) \right\}
\tau_t\big((\lambda{\bf 1}-A)^{-1}\big)
\\ &=& \tau_t(\lambda{\bf 1}-A)\tau_t\big((\lambda{\bf 1}-A)^{-1}\big)
\qquad {\rm (*-isomorphism)}
\\ &=& \tau_t\big((\lambda{\bf 1}-A)(\lambda{\bf 1}-A)^{-1}\big)
\qquad {\rm (*-isomorphism)}
\\ &=& \tau_t({\bf 1})={\bf 1}
\end{eqnarray*}
Thus, $\lambda{\bf 1}-\tau_t(A)$ is invertible, and it follows 
$\lambda\in r(\tau_t(A))$.
Conversely, let $\lambda\in \reso{\tau_t(A)}$, then, 
\begin{eqnarray*}
\tau_{-t}\big((\lambda{\bf 1}-\tau_t(A))^{-1}\big)(\lambda{\bf 1}-A)
&=& \tau_{-t}\big((\lambda{\bf 1}-\tau_t(A))^{-1}\big)\tau_{-t}\big(\tau_t(\lambda{\bf 1}-A)\big)
\qquad {\rm (group\ property)}
\\ &=&
\tau_{-t}\Big(\big(\lambda{\bf 1}-\tau_t(A)\big)^{-1}\big(\lambda{\bf 1}-\tau_t(A)\big)\Big)
\qquad {\rm (*-isomorphism)}
\\ &=&
\tau_{-t}({\bf 1})={\bf 1}
\ \ .
\end{eqnarray*}
Thus, we have $\lambda\in r(A)$.
We conclude $\reso{A}=\reso{\tau_t(A)}$, which follows $\spect{A}=
\spect{\tau_t(A)}$. 

\QED{}
\\
\underline{proof of proposition}\\ 
Applying $\tau_t(\tau_{-t}(A))=A$, we have
\begin{eqnarray*}
\tau_t({\bf 1}) &=& \tau_t({\bf 1})\tau_t(\tau_{-t}({\bf 1}))\qquad 
{\rm (group\ property)}
\\ &=& \tau_t({\bf 1}\tau_{-t}({\bf 1}))
\qquad {\rm (*-isomorphism)}
\\ &=&
\tau_t(\tau_{-t}({\bf 1}))={\bf 1}\ .\qquad 
{\rm (group\ property)}
\end{eqnarray*}

Suppose $A\in{\cal A}_{s.a.}$, then, following the previous lemma, we have 
$\spect{A}=\spect{\tau_t(A)}$.
\\
Applying proposition 5 for $A$, we have
$\Vert A\Vert=\rho(A)=\rho(\tau_t(A))=\Vert\tau_t(A)\Vert$. 
\\
Suppose $A$ is not selfadjoint, then, we have
$$\Vert A\Vert^2=\Vert A^*A\Vert=\Vert \tau_t(A^*A)\Vert=\Vert \tau_t(A)^*\tau_t(A)\Vert=\Vert\tau_t(A)\Vert^2\ .$$

\QED{}

 \begin{Prop}
\label{derivative_Prop}
Let ${\cal A}$ be a $C^*$ algebra.
For a group of strong continuous *-isomorphisms~$\tau_t$ over ${\cal A}$, 
there exists a dense subset ${\cal D}(\delta)$ of ${\cal A}$ and linear operator $\delta$ on ${\cal D}(\delta)$ such that
  \begin{equation}
   \lim_{t\to0} \left\|\frac{\tau_t\left(A\right)-A}{t} - \delta \left(A\right) \right\| = 0
,\ \ \forall A\in {\cal D}(\delta)\ .
  \label{derivative}
  \end{equation}
Moreover, 
\begin{itemize}
	\item[(i)] ${\bf 1}\in D(\delta)$ and $\delta({\bf 1})={\bf 0}$.

	\item[(ii)] If $A,\ B\in D(\delta)$, then, 
	$AB\in D(\delta)$ and 
	$\delta(AB)=A\delta(B)+\delta(A)B$.

	\item[(iii)] If $A\in D(\delta)$, then, 
	$A^*\in D(\delta)$ and $\delta(A)^*=\delta(A^*)$.
\end{itemize}

\end{Prop}
This element $\delta$ corresponds to a derivative ($\delta=\frac{d\tau_t*}{dt}|_{t=0}$), and is referred to as a generator of $\tau_t$ in the doming 
${\cal D}(\delta)$.
\\
{\bf Remark 1} \\
This proposition is proved without using differentiability.
Thus, if strong continuity is imposed on a self map, then, 
derivative exists for almost all elements in the $C^*$ algebra in the sense of this proposition.
\\
\underline{proof}\\ 
Let 
\begin{eqnarray*}
{\cal D}(\delta)=\left\{ A_\epsilon\equiv 
{1\over \epsilon}\int_0^\infty ds e^{-s/\epsilon}\tau_s(A):
\epsilon>0, A\in{\cal A}
\right\}\ .
\end{eqnarray*}
First let us prove that ${\cal D}(\delta)$ is dense in ${\cal A}$.
It is sufficient to show $||A_e-A||\to 0$ as $\epsilon\to 0$ for any $A\in {\cal A}$ (Any element in ${\cal A}$ can be approximated by $A_\epsilon\in {\cal D}(\delta)$ with arbitrary precision.).
With the aid of the strong continuity and the equality:
\begin{eqnarray*}
\frac{1}{\epsilon}\int^\infty_0 ds e^{-s/\epsilon}ds = 1
\ ,
\end{eqnarray*}
one can easily see $||A_\epsilon-A||\to 0$ as follows:
\begin{eqnarray}
\lim_{\epsilon\to 0}\Vert A_\epsilon -A\Vert 
&=& 
\lim_{\epsilon\to 0}
\left\Vert {1\over \epsilon}\int_0^\infty ds e^{-s/\epsilon}\tau_s(A)
-{1\over \epsilon}\int_0^\infty ds e^{-s/\epsilon}A\right\Vert
\nonumber \\
 &=& \lim_{\epsilon\to 0}
\left\Vert \int_0^\infty {ds\over\epsilon} e^{-s/\epsilon}\{\tau_s(A)-A\}
\right\Vert
\nonumber \\
 &=& \lim_{\epsilon\to 0}
\left\Vert \int_0^\infty ds e^{-s}\{\tau_{\epsilon s}(A)-A\}
\right\Vert
\nonumber \\
 &\le& \lim_{\epsilon\to 0}
 \int_0^\infty ds e^{-s}\Vert\tau_{\epsilon s}(A)-A\Vert
\nonumber \\
&=&0 
\label{rough}
\end{eqnarray}
Therefore we conclude $\overline{{\cal D}(\delta)}={\cal A}$, where 
$\overline{{\cal D}(\delta)}$ represents a closure of ${\cal D}(\delta)$ with respect to the norm we discussed\footnote{
Rigorously speaking, limit of (\ref{rough}) should be treated more carefully.
}
.
Next, we are going to prove equality $(\ref{derivative})$.
For any element $A_\epsilon\in {\cal D}(\delta)$, let us define $\delta(A_\epsilon)$ by 
\begin{eqnarray}
\delta(A_\epsilon)\equiv \frac{1}{\epsilon}(A_\epsilon-A)
\ .
\end{eqnarray}
Linearity of $\delta$ is obvious.
For arbitrary $A\in {\cal A}$, we have
\begin{eqnarray*}
\tau_t(A_\epsilon) &=& \int_0^\infty {ds\over \epsilon}e^{-s/\epsilon}\tau_{s+t}(A)
\nonumber\\ &=&
\int_t^\infty {ds\over \epsilon}e^{t-s\over\epsilon}\tau_s(A)
\nonumber\\ &=&
\left( \int_0^\infty-\int_0^t \right) {ds\over \epsilon}e^{t-s\over\epsilon}\tau_s(A)
\nonumber\\ &=&
e^{t/\epsilon}A_\epsilon-{t\over \epsilon}
\int_0^1 ds e^{(1-s)t\over\epsilon}\tau_{ts}(A)
\ .
\end{eqnarray*}
Therefore, we have the following inequality:
\begin{eqnarray*}
\left\Vert {\tau_t(A_\epsilon)-A_\epsilon\over t}
-\delta(A_\epsilon)
\right\Vert
&=&
\left\Vert {\tau_t(A_\epsilon)-A_\epsilon\over t}
-{A_\epsilon-A\over \epsilon}\right\Vert
\\
&=&
\left\Vert {e^{t/\epsilon}-1-t/\epsilon\over t}A_\epsilon
-{1\over \epsilon}
\int_0^1 ds e^{(1-s)t\over\epsilon}\tau_{ts}(A)
+{A\over \epsilon}\right\Vert
\nonumber\\
&\le& \left|{e^{t/\epsilon}-1-t/\epsilon\over t}\right|
\Vert A_\epsilon\Vert
+{1\over \epsilon}
\int_0^1 ds e^{(1-s)t\over\epsilon}\Vert\tau_{ts}(A)-A\Vert
\nonumber\\
&&\mskip 100mu +{1\over \epsilon}\left|1-
\int_0^1 ds e^{(1-s)t\over\epsilon}\right|\Vert A\Vert
\end{eqnarray*}
By taking a limit of $t\to 0$, we have the desired equality:
\begin{eqnarray*}
\lim_{t\to 0}
\left\Vert {\tau_t(A_\epsilon)-A_\epsilon\over t}
-\delta(A_\epsilon)
\right\Vert=0
\end{eqnarray*}
Let us prove the properties (i)-(iii).
Thanks to $\tau_t({\bf 1})={\bf 1}$, we obtain
\begin{eqnarray*}
\lim_{t\to 0}
\left\Vert {\tau_t({\bf 1})-{\bf 1}\over t}-{\bf 0}\right\Vert
=0
\ .
\end{eqnarray*}

Let $A,B\in {\cal D}(\delta)$, then,
\begin{eqnarray*}
&&\left\Vert {\tau_t(AB)-AB\over t}
-\{\delta(A)B+A\delta(B)\}\right\Vert
\nonumber\\
&&=\left\Vert {\tau_t(A)\tau_t(B)-A\tau_t(B)+A\tau_t(B)-AB\over t}
-\delta(A)\tau_t(B)+\delta(A)\tau_t(B)
-\{\delta(A)B+A\delta(B)\}\right\Vert
\nonumber\\
&&=
\left\Vert 
{\tau_t(A)-A\over t}\tau_t(B)-\delta(A)\tau_t(B)
+\delta(A)\tau_t(B)-\delta(A)B
+A{\tau_t(B)-B\over t}-A\delta(B)
\right\Vert
\nonumber\\
&&\le 
\left\Vert 
{\tau_t(A)-A\over t}-\delta(A)\right\Vert
\Vert\tau_t(B)\Vert
+\Vert\delta(A)\Vert \Vert\tau_t(B)-B\Vert
+\Vert A\Vert \left\Vert{\tau_t(B)-B\over t}-\delta(B)
\right\Vert
\nonumber\\
&&\to 0 \ \ (t\to 0)\ .
\end{eqnarray*}
Thus, (ii) is proved.  
For $A\in D(\delta)$, we have 
\begin{eqnarray*}
&&\left\Vert {\tau_t(A^*)-A^*\over t}
-\delta(A)^*\right\Vert
=
\left\Vert {\tau_t(A)^*-A^*\over t}
-\delta(A)^*\right\Vert
\nonumber\\
&&
=\left\Vert {\tau_t(A)-A\over t}
-\delta(A)\right\Vert
\to 0 \ \ (t\to 0),
\end{eqnarray*}
hence (iii) is proved.

\begin{Ex}[Spinless fermion in $d$ dimensional space]\ \ 
For this example, we make a rough argument which show that $D(\delta)$ is not equal to ${\cal A}$.
Let us think of a CAR algebra discussed in Example~\ref{CAR_ex}.
For $a(f)=\int f(k)^*a_k\ dk$, we define a following time evolution:
\begin{eqnarray*}
\tau_t\big(a(f)\big)=\int e^{-i\omega_k} f(k)^*a_k\ dk
\end{eqnarray*}
$\Vert a(f) \Vert=\sqrt{\int |f(k)|^2\ dk }$ implies
\begin{eqnarray*}
\Vert \tau_t\big(a(f)\big) \Vert=\sqrt{\int |f(k)|^2 dk }<\infty\ ,
\end{eqnarray*}
and thus, $\tau_t$ is well-defined for any $a(f)\in {\cal A}$.
On the other hand, formal calculation gives
\begin{eqnarray*}
\Vert \delta\big(a(f)\big) \Vert=\sqrt{\int \omega_k^2 |f(k)|^2\ dk }\ .
\end{eqnarray*}
R.H.S. can be divergent, and has meaning only for some $a(f)\in {\cal A}$.
It means that domain of derivative is not equal to ${\cal A}$.

\end{Ex}

Proposition~\ref{derivative_Prop} claims that a time generator $\delta$ exists for $\tau_t$. 
On the other hand, the existence of $\tau_t$ for a given time generator $\delta$ is much more difficult to prove, and individual problems are
usually discussed using the Hille Yoshida theory of semi-group.
However, there are some general results for perturbative systems.

\begin{Prop}
\label{generator_perturbation_Prop}
Let ${\cal A}$ be a $C^*$ algebra, and $\tau_t$ be a group of strong continuous *-isomorphisms over ${\cal A}$. 
Let us define $\Gamma_t\in {\cal A}$ as a solution of
  \begin{eqnarray}
	\Gamma_t={\bf 1}+i\int^t_0 ds\Gamma_s\tau_s(V)
	\label{e62}
  \end{eqnarray}
for $V\in {\cal A}_{s.a.}$.
Then, 
\begin{itemize}
	\item[(i)]
	$\Gamma_t$ is unitary.
	\item[(ii)]
	$\Gamma_{t+s}=\Gamma_t\tau_t(\Gamma_s)$
	\item[(iii)]
$$   \tau_t^V(A)\equiv \Gamma_t\tau_t(A)\Gamma_t^* \ \ ( ^\forall A\in{\cal A})
 $$	is a group of strong continuous *-isomorphisms.
	Thus, generator of $\tau_t^V$ exists for $A\in {\cal D}(\delta)$.
	Let $\delta^V$ be a generator of $\tau_t^V$ for $A\in {\cal D}(\delta)$.	Then, it has the following form:
	\begin{equation}
	\delta^V(A)=\delta(A)+i[V,A]
	\label{e61}
	  \end{equation}
\end{itemize}

\end{Prop}
{\bf Remark}\\
Here, let us formally discuss the relation between conventional quantum mechanics and this theorem.
Let us study perturbative systems $H=H_0+V$. 
Then, time evolution operator is denoted by $U=e^{-iHt}$ (or one can say that $U$ changes Schr\"odinger picture to Heisenberg picture).
Let $A_S$ be an arbitrary observables in Schr\"odinger picture, and 
let us define $U_0, A_H, A_I$ as 
$U_0=e^{-iH_0 t},\ A_H=U^\dag A U, A_I=U_0^\dag A U_0$.
Then, ensemble averages of observable $A_H(t)$ read
\begin{eqnarray*}
\ave{A_H(t)} &=& {\rm tr} \{ A_S U\rho_0 U^\dag\}
 \nonumber 
 \\ &=& 
{\rm tr} \{ U^\dag A_S U\rho_0\}
\\ &=& 
{\rm tr} \{ A_I(t) U_0^\dag U\rho_0 U^\dag U_0\}
\end{eqnarray*}
Usually, $U_I=U_0^\dag U$ is used as a time evolution operator in interaction picture. Here, we shall define $\Gamma_t =U^\dag U_0$, instead, then, 
$A_H(t)=\Gamma_t A_I(t) \Gamma^\dag$.
The equation of motion for$\Gamma_t$ reads 
\begin{eqnarray*}
i\frac{d \Gamma_t}{dt} &=& i(U^\dag U_0)'
\\ &=& i(e^{iHt} e^{-iH_0t})'
\\ &=& ie^{iHt}(iH-iH_0) e^{-iH_0 t}
\\ &=& -U^\dag V U_0
\\ &=& -U^\dag U_0 U_0^\dag V U_0
\\ &=& -\Gamma_t U_0^\dag V U_0
\end{eqnarray*}
We shall define $\tau_t(A)$ as $U_0^\dag A U_0$. 
Then, this equation corresponds to the time derivative of Eq.(\ref{e62}).
On the other hand, Heisenberg equation of motion for $A_H(t)$ is given by
\begin{eqnarray*}
\frac{d A_H(t)}{dt}\Bigg|_{t=0} &=& i[H,A_H(t)]\Big|_{t=0}
\\ &=& 
i[H_0,A_H(t)]\Big|_{t=0}+i[V,A_H(t)]\Big|_{t=0}
\ .
\end{eqnarray*}
The first term corresponds to $\delta(A)$ in equation (\ref{e61}).
Roughly speaking, we separate evolution with $H_0$ and $V$\footnote{This argument is only to have a intuition, and not correct statement.}, and 
$\tau(\cdot)$ is used to avoid free hamiltonian evolution~$H_0$.
\\
\underline{proof}\\ 

\begin{eqnarray}
\Gamma_t&=&{\bf 1}+i\int_0^t ds \tau_s(V)+i^2 \int_0^t ds_2 \int_0^{s_2} ds_1 \tau_{s_1}(V) \tau_{s_2}(V)
\nonumber\\
&&+
\cdots +i^n \int_0^t ds_n \int_0^{s_n} ds_{n-1} \cdots \int_0^{s_2} ds_1
\tau_{s_1}(V) \tau_{s_2}(V)\cdots \tau_{s_n}(V)
+\cdots
\label{DysonSeries1}
\end{eqnarray}
is a solution of the integral equation if the limit exists and it is continuous with respect to $t$ (This series corresponds to the Dyson series in the conventional quantum mechanics).
With the aids of an inequality: 
$\Vert \tau_{s_1}(V) \cdots \tau_{s_n}(V)\Vert\le
\Vert\tau_{s_1}(V)\Vert \cdots \Vert\tau_{s_n}(V)\Vert=\Vert V\Vert^n
$, we see that 
\begin{eqnarray*}
&&1+\left\Vert\int_0^t ds \tau_s(V)\right\Vert+
\cdots +\left\Vert\int_0^t ds_n \int_0^{s_n} ds_{n-1} \cdot\cdot \int_0^{s_2} ds_1
\tau_{s_1}(V) \tau_{s_2}(V)\cdot\cdot \tau_{s_n}(V)\right\Vert
+\cdots
\nonumber\\
&&\le
1+\int_0^t ds \Vert\tau_s(V)\Vert+
\cdots +\int_0^t ds_n \int_0^{s_n} ds_{n-1}\cdot\cdot \int_0^{s_2} ds_1
\Vert\tau_{s_1}(V) \tau_{s_2}(V)\cdot\cdot \tau_{s_n}(V)\Vert
+\cdots
\nonumber\\
&&\le
1+\int_0^t ds \Vert V\Vert+
\cdots +\int_0^t ds_n \int_0^{s_n} ds_{n-1}\cdot\cdot \int_0^{s_2} ds_1
\Vert V\Vert^n
+\cdots
\nonumber\\
&&=
1+t \ \Vert V\Vert +
\cdots +{t^n\over n!}
\Vert V\Vert^n
+\cdots=e^{\Vert V\Vert t}<+\infty
\end{eqnarray*}
is absolute convergent for $t>0$.
Similarly, one can also prove the convergence for $t<0$.
Thus, (\ref{DysonSeries1}) is norm-convergent, and it is a solution of the integral equation.
Since (\ref{DysonSeries1}) is norm-convergent and every term is differentiable, one concludes that this solution is differentiable, and hence
\begin{eqnarray*}
{d\over dt}\Gamma_t &=& i\Gamma_t\tau_t(V)\
\\
{d\over dt}\Gamma_t^* &=& -i\tau_t(V)\Gamma_t^*
 .
\end{eqnarray*}
Moreover, one can easily see that 
$(\Gamma_t \Gamma_t^*)$ is $t$ independent:
$$
{d\over dt}(\Gamma_t\Gamma_t^*)=
{d\Gamma_t\over dt}\Gamma_t^*+
\Gamma_t{d\Gamma_t^*\over dt}
=i\Gamma_t\tau_t(V)\Gamma_t^*-i\Gamma_t\tau_t(V)\Gamma_t^*={\bf 0}
\ ,
$$
and thus, we have $\Gamma_t\Gamma_t^*=\Gamma_0\Gamma_0^*={\bf 1}$.
Next, let us prove $\Gamma_t^* \Gamma_t={\bf 1}$.
$$
{d\over dt}(\Gamma_t^*\Gamma_t)=
-i[\tau_t(V),\Gamma_t^*\Gamma_t]
$$
implies
$$
\Gamma_t^*\Gamma_t-{\bf 1}=
-i\int_0^t dt[\tau_t(V),\Gamma_t^*\Gamma_t]
=-i\int_0^t dt[\tau_t(V),\Gamma_t^*\Gamma_t-{\bf 1}]\ .
$$
Let $R_1(t) \equiv \int_0^tds\Vert \Gamma_s^*\Gamma_s-{\bf 1}\Vert
$. With the aids of the previous equality, we have
the following inequality for $t>0$:
\begin{eqnarray*}
{dR_1(t)\over dt} &=& \Vert
\Gamma_t^*\Gamma_t-{\bf 1}\Vert\le
\int_0^t dt\Vert[\tau_t(V),\Gamma_t^*\Gamma_t-{\bf 1}]\Vert
\le
2\Vert V\Vert\int_0^t dt\Vert \Gamma_t^*\Gamma_t-{\bf 1}\Vert
=2\Vert V\Vert R_1(t)
\end{eqnarray*}
Multiplying the previous inequality by ${e^{-2\Vert V\Vert t}\over 2\Vert V\Vert}$, we obtain
\begin{eqnarray*}
&&0\le
{e^{-2\Vert V\Vert t}\over 2\Vert V\Vert}\Vert \Gamma_t^*\Gamma_t-{\bf 1}\Vert
={e^{-2\Vert V\Vert t}\over 2\Vert V\Vert}{dR_1(t)\over dt}
\le e^{-2\Vert V\Vert t}R_1(t)
\nonumber\\
&&
=\int_0^t ds {d\over ds}\{e^{-2\Vert V\Vert s}R_1(s)\}
=\int_0^t ds
e^{-2\Vert V\Vert s}
\Big\{
{dR_1(s)\over ds}-2\Vert V\Vert R_1(s)\Big\}\le 0
\ .
\end{eqnarray*}
We conclude that $\Gamma_t$ is unitary.

Next, we are going to prove the property (ii), 
i.e., $\Gamma_{s+t}=\Gamma_s\tau_s(\Gamma_t)$.

With the aids of $\Gamma_s^*\Gamma_{t+s}\Big|_{t=0}={\bf 1}$ and 
${\displaystyle 
{d\over dt}\Gamma_s^*\Gamma_{t+s}=i\Gamma_s^*\Gamma_{t+s}\tau_{t+s}(V)
}$, we have
\begin{eqnarray*}
\Gamma_s^*\Gamma_{t+s}\!
&=&
\!
{\bf 1}+i\int_0^t dt_1 \tau_{s+t_1}(V)+i^2 \int_0^t dt_2 \int_0^{t_2} dt_1 \tau_{s+t_1}(V) \tau_{s+t_2}(V)
\nonumber\\
&&
+
\cdots +i^n \int_0^t dt_n \int_0^{t_n} dt_{n-1} \cdots \int_0^{t_2} dt_1
\tau_{s+t_1}(V) \tau_{s+t_2}(V)\cdots \tau_{s+t_n}(V)
+\cdots
\nonumber\\
&=&
{\bf 1}+i\int_0^t dt_1 \tau_s\big(\tau_{t_1}(V)\big)+
i^2 \int_0^t dt_2 \int_0^{t_2} dt_1 \tau_s\Big(\tau_{t_1}(V) \tau_{t_2}(V)\Big)
\nonumber\\
&&
+
\cdots +i^n \int_0^t dt_n \int_0^{t_n} dt_{n-1} \cdots \int_0^{t_2} dt_1
\tau_s\Big(\tau_{t_1}(V) \tau_{t_2}(V)\cdots \tau_{t_n}(V)\Big)
+\cdots
\nonumber\\
&=&\tau_s\Big(
{\bf 1}+i\int_0^t dt_1 \tau_{t_1}(V)+
i^2 \int_0^t dt_2 \int_0^{t_2} dt_1 \tau_{t_1}(V) \tau_{t_2}(V)
\nonumber\\
&&
+
\cdots +i^n \int_0^t dt_n \int_0^{t_n} dt_{n-1} \cdots \int_0^{t_2} dt_1
\tau_{t_1}(V) \tau_{t_2}(V)\cdots \tau_{t_n}(V)
+\cdots\Big)
\nonumber\\
&=&\tau_s(\Gamma_t)\ .
\end{eqnarray*}
Unitarity of $\Gamma_t$ gives a desired equality.

Finally, we prove that $\tau_t^V(A)\equiv\Gamma_t\tau_t(A)\Gamma_t^*$
is a group of strong continuous *-isomorphisms.
It is obvious that $\tau_t^V$ is linear.
In addition, 
\begin{eqnarray}
&&\tau_t^V(AB)=\Gamma_t\tau_t(AB)\Gamma_t^*=\Gamma_t\tau_t(A)\tau_t(B)\Gamma_t^*
=\Gamma_t\tau_t(A)\Gamma_t^*\Gamma_t\tau_t(B)\Gamma_t^*=\tau_t^V(A)\tau_t^V(B)
\nonumber\\
&&\tau_t^V(A^*)=\Gamma_t\tau_t(A^*)\Gamma_t^*=\Gamma_t\tau_t(A)^*\Gamma_t^*
=\big(\Gamma_t\tau_t(A)\Gamma_t^*\big)^*=\tau_t^V(A)^*
\nonumber
\end{eqnarray}
follows that $\tau^V_t$ is a *-isomorphism. Moreover, it satisfies
\begin{eqnarray*}
&&\tau_t^V\big(\tau^V_s(A)\big)=\Gamma_t\tau_t\big(\Gamma_s\tau_s(A)\Gamma_s^*\big)\Gamma_t^*=
\Gamma_t\tau_t(\Gamma_s)\tau_t\big(\tau_s(A)\big)\tau_t(\Gamma_s^*)\Gamma_t^*
\nonumber\\
&&=\Gamma_t\tau_t(\Gamma_s)\tau_{t+s}(A)\tau_t(\Gamma_s)^*\Gamma_t^*
=\Gamma_{t+s}\tau_{t+s}(A)\Gamma_{t+s}^*=\tau^V_{t+s}(A)
\ .
\end{eqnarray*}
Thus, $\tau^V_t$ is a group. Strong continuity follows from
\begin{eqnarray*}
\Vert \tau_t^V(A)-A\Vert
 &=& 
\Vert \Gamma_t\tau_t(A)\Gamma_t^*-A\Vert
\\ &\le&
\Vert \Gamma_t(\tau_t(A)-A)\Gamma_t^*\Vert
+\Vert \Gamma_tA\Gamma_t^*-A\Gamma_t^*\Vert+\Vert A\Gamma_t^*-A\Vert
\nonumber\\
&\le& \Vert \tau_t(A)-A\Vert+\Vert A\Vert \Vert \Gamma_t^*-{\bf 1}\Vert
+\Vert A\Vert \Vert \Gamma_t-{\bf 1}\Vert
\nonumber\\
&\to& 0 \ \
,\qquad {\rm as}\ \ t\to 0\ .
\end{eqnarray*}
Recalling, $\Vert (\tau_t(A)-A)/t-\delta(A)\Vert\to 0$ and 
$\Vert (\Gamma_t-{\bf 1})/t-iV\Vert \to 0$ for the limit $t\to 0$, we have
\begin{eqnarray*}
&&{\tau^V_t(A)-A\over t}=
{\Gamma_t\tau_t(A)\Gamma_t^*-A\over t}=
\Gamma_t{\tau_t(A)-A\over t}\Gamma_t^*+
{\Gamma_t-{\bf 1}\over t}A\Gamma_t^*+
A{\Gamma_t^*-{\bf 1}\over t}
\nonumber\\
&&\to \delta(A)+iVA-iAV=\delta(A)+i[V,A],
\qquad
{\rm as}\ t\to 0\ ,
\end{eqnarray*}
for any $A\in {\cal D}(\delta)$ in a sense of norm convergence.
Therefore, 
generator $\delta^V$ of $\tau_t^V$
reads 
\begin{eqnarray*}
\delta^V(A)=\delta(A)+i[V,A]\ .
\end{eqnarray*}
\QED{}

Proposition~\ref{generator_perturbation_Prop} gives a time generator for autonomous systems.
Time evolution with non-autonomous perturbation can be described by the 
following proposition.

\begin{Prop}
Let ${\cal A}$ be a $C^*$ algebra, and $\tau_t$ be a group of strong continuous *-isomorphisms over ${\cal A}$.

Let us define $\Gamma_{t,s} \in {\cal A}$ as a solution of 
  \begin{equation}
   \Gamma_{t,s}={\bf 1}+i\int_0^t dt_1 \Gamma_{t_1,s} \tau_{t_1-s}\big(V(t_1)\big)
   \label{e76}
  \end{equation}
for $V\in {\cal A}_{s.a.}$.
Then, 
\begin{itemize}
	\item[(i)] $\Gamma_{t,s}$ is unitary.
	\item[(ii)] $\Gamma_{t,s}=\Gamma_{s_1,s}\tau_{s_1-s}(\Gamma_{t,s_1}),\ 
	\ \forall s_1,s,t\in {\bf R}$
	\item[(iii)] 
	$
	\tau_{t,s}^V(A)\equiv \Gamma_{t,s}\tau_{t-s}(A)\Gamma_{t,s}^* \ \ 
	$\\
	is a group of strong continuous *-isomorphisms which satisfies 
	\begin{equation*}
	{d\over dt}\tau_{t,s}^V(A)=\tau_{t,s}^V\Big(
	\delta(A)+i[V(t),A]\Big),\ \forall A\in {\cal D}(\delta)
	\ .
	\end{equation*}
\end{itemize}
\end{Prop}
\underline{proof}\\ 
\begin{equation}
\Gamma_{t,s}={\bf 1}+\sum_{n=1}^\infty
i^n \int_s^t dt_n \int_s^{t_n} dt_{n-1} \cdots \int_s^{t_2} dt_1
\tau_{t_1-s}\big(V(t_1)\big) \tau_{t_2-s}\big(V(t_2)\big)\cdots \tau_{t_n-s}\big(V(t_n)\big)
\label{DysonSeries2}
\end{equation}
is a solution of the integral equation if the limit exists, and it is continuous with respect to $t$.
$||\tau_t(A)||=||A||$ follows
\begin{eqnarray*}
\Vert \tau_{t_1-s}(V(t_1)) \cdots \tau_{t_n-s}(V(t_n))\Vert
&\le&
\Vert\tau_{t_1-s}(V(t_1))\Vert \cdots \Vert\tau_{t_n-s}(V(t_n))\Vert=K(t,s)^n
\\
K(t,s) &\equiv& \sup_{s\le \tau \le t}\Vert V(\tau)\Vert\ .
\end{eqnarray*}
With the aids of this inequality, we have
\begin{eqnarray*}
&&1+
\sum_{n=1}^\infty
\left\Vert
\int_s^t dt_n \int_s^{t_n} dt_{n-1} \cdots \int_s^{t_2} dt_1
\tau_{t_1-s}\big(V(t_1)\big) \tau_{t_2-s}\big(V(t_2)\big)\cdots \tau_{t_n-s}\big(V(t_n)\big)
\right\Vert
\nonumber\\
&&\le
1+
\sum_{n=1}^\infty K(t,s)^n
\int_s^t dt_n \int_s^{t_n} dt_{n-1} \cdots \int_s^{t_2} dt_1
\nonumber\\
&&=
1+
\sum_{n=1}^\infty{ K(t,s)^n (t-s)^n\over n!}
=e^{K(t,s) (t-s)}<+\infty
\end{eqnarray*}
for $t>s$. Thus, (\ref{DysonSeries2}) is absolute convergent.
In a similar way, one can prove that 
the series (\ref{DysonSeries2}) is norm-convergent for $t<s$, and thus
the series (\ref{DysonSeries2}) is a solution of the integral equation.

\underline{proof of (i)}\\
Since $V(t)$ is selfadjoint and $\tau_t(\cdot)$ is *-isomorphism, 
we have $\tau_t(V(t))^*=\tau_t(V(t))$. 
Thus, differentiating (\ref{e76}) with respect to $t$, we obtain
\begin{eqnarray*}
{d\over dt}\Gamma_{t,s} &=& i\Gamma_{t,s}\tau_{t-s}(V(t))\
\\
{d\over dt}\Gamma_{t,s}^* &=& -i\tau_{t-s}(V(t))\Gamma_{t,s}^*
\ .
\end{eqnarray*}
It follows that $(\Gamma_{t,s}\Gamma_{t,s}^*)$ is $t$ independent:
\begin{eqnarray*}
{d\over dt}(\Gamma_{t,s}\Gamma_{t,s}^*)
=i\Gamma_{t,s}\tau_{t-s}(V(t))\Gamma_{t,s}^*-i\Gamma_{t,s}\tau_{t-s}(V(t))\Gamma_{t,s}^*=0
\end{eqnarray*}
Therefore, we have 
$\Gamma_{t,s}\Gamma_{t,s}^*=\Gamma_{s,s}\Gamma_{s,s}^*={\bf 1}$.
Next let us prove $(\Gamma_{t,s}^*\Gamma_{t,s})={\bf 1}$.
Integrating 
\begin{eqnarray*}
{d\over dt}(\Gamma_{t,s}^*\Gamma_{t,s})=
-i[\tau_{t-s}(V(t)),\Gamma_{t,s}^*\Gamma_{t,s}]
\end{eqnarray*}
from $s$ to $t$, we obtain
\begin{eqnarray*}
\Gamma_{t,s}^*\Gamma_{t,s}-{\bf 1}=
-i\int_s^t dt[\tau_{t-s}(V(t)),\Gamma_{t,s}^*\Gamma_{t,s}]
=-i\int_s^t dt[\tau_{t-s}(V(t)),\Gamma_{t,s}^*\Gamma_{t,s}-{\bf 1}]
\ .
\end{eqnarray*}
Therefore, the following inequality is satisfied for
$t>s$:
\begin{eqnarray*}
{dR_1(t)\over dt}&=&\Vert
\Gamma_{t,s}^*\Gamma_{t,s}-{\bf 1}\Vert
\\
& \le &
\int_s^t dt'\Vert[\tau_{t'-s}(V(t')),\Gamma_{t',s}^*\Gamma_{t',s}-{\bf 1}]\Vert
\\
& \le&
2K(t,s)\int_s^t dt' \Vert \Gamma_{t',s}^*\Gamma_{t',s}-{\bf 1}\Vert
=2K(t,s) R_1(t)
\\
\Rightarrow&&
0 \le \Vert \Gamma_{t,s}^*\Gamma_{t,s}-{\bf 1}\Vert \le 2K(t,s) R_1(t)
\\
R_1(t) &\equiv& \int_s^tdt'\Vert \Gamma_{t',s}^*\Gamma_{t',s}-{\bf 1}\Vert
\end{eqnarray*}
From $R(s)={\bf 0}$ and the previous inequality, we have 
\begin{eqnarray*}
0&\le&
{e^{-2K(t,s) t}\over 2K(t,s)}\Vert \Gamma_{t,s}^*\Gamma_{t,s}-{\bf 1}\Vert
\\
&=&
\int_s^t dt' {d\over dt'}\{e^{-2K(t,s) t'}R_1(t')\}
\\
&=&
\int_s^t dt'
e^{-2K(t,s) t'}
\Big\{
{dR_1(t')\over dt'}-2K(t,s) R_1(t')\Big\}
\ .
\end{eqnarray*}
Combining $0 \le K(t',s)\le K(t,s)\ (t'\le t)$ and
${dR_1(t')\over dt'}-2K(t',s) R_1(t')\le 0$, we have
\begin{eqnarray*}
0&\le&
{e^{-2K(t,s) t}\over 2K(t,s)}\Vert \Gamma_{t,s}^*\Gamma_{t,s}-{\bf 1}\Vert
\\
&=&
\int_s^t dt'
e^{-2K(t,s) t'}
\Big\{
{dR_1(t')\over dt'}-2K(t,s) R_1(t')\Big\}
\nonumber
\\
&\le& 0\ .
\end{eqnarray*}
Therefore, $\Gamma_{t,s}^*\Gamma_{t,s} = {\bf 1}$, and thus $\Gamma_{t,s}$ is unitary.

\underline{proof of (ii)}\\
Let $y(t)_{s_1,s}\equiv \Gamma_{s_1,s}^*\Gamma_{t,s}$.
By integrating
${\displaystyle 
{d\over dt}\Gamma_{s_1,s}^*\Gamma_{t,s}=i\Gamma_{s_1,s}^*\Gamma_{t,s}\tau_{t-s}(V(t))
}$ from $s_1$ to $t$, we obtain
\begin{eqnarray}
y_{s_1,s}(t)={\bf 1}+i\int_{s_1}^t dt_1\ y_{s_1,s}(t_1) \tau_{t'-s}(V(t_1))
\ ,
\end{eqnarray}
where we have used  $\Gamma_{s_1,s}^*\Gamma_{t,s}\Big|_{t=s_1}={\bf 1}$.
The solution can be expressed by the following series (One prove the convergence of the following series with the same arguments we have done to prove the convergence of series (\ref{DysonSeries2}).):
\begin{eqnarray*}
y_{s_1,s}(t)
=
{\bf 1}+\sum_{n=1}^\infty
i^n \int_{s_1}^t dt_n \int_{s_1}^{t_n} dt_{n-1} \cdots \int_{s_1}^{t_2} dt_1
\tau_{t_1-s}(V(t_1)) \tau_{t_2-s}(V(t_2))\cdots \tau_{t_n-s}(V(t_n))
\end{eqnarray*}
It follows
\begin{eqnarray*}
&& {\!\!\!\!\!\!\!\!\!\!\!\!\!}
\Gamma_{s_1,s}^*\Gamma_{t,s}  \nonumber \\
&&{\!\!\!\!\!\!\!\!\!\!}
= {\bf 1}
+\sum_{n=1}^\infty
i^n \int_{s_1}^t dt_n \int_{s_1}^{t_n} dt_{n-1} \cdots \int_{s_1}^{t_2} dt_1
\tau_{s_1-s}\Big(
\tau_{t_1-s_1}(V(t_1)) \tau_{t_2-s_1}(V(t_2))\cdots \tau_{t_n-s_1}(V(t_n))\Big)
\nonumber\\
&&{\!\!\!\!\!\!\!\!\!\!}
=\tau_{s_1-s}\Big(
{\bf 1}
+\sum_{n=1}^\infty
i^n \int_{s_1}^t dt_n \int_{s_1}^{t_n} dt_{n-1} \cdots \int_{s_1}^{t_2} dt_1
\tau_{t_1-s_1}(V(t_1)) \tau_{t_2-s_1}(V(t_2))\cdots \tau_{t_n-s_1}(V(t_n))\Big)\Big)
\nonumber\\
&&{\!\!\!\!\!\!\!\!\!\!}
=\tau_{s_1-s}(\Gamma_{t,s_1})\ .
\end{eqnarray*}
Thus, we have $\Gamma_{t,s}=\Gamma_{s_1,s}\tau_{s_1-s}(\Gamma_{t,s_1})$.

\underline{proof of (iii)}\\
First, 
  \begin{equation*}
   \tau_{t,s}^V(A)\equiv \Gamma_{t,s}\tau_{t-s}(A)\Gamma_{t,s}^*, \ \  ^\forall A\in{\cal D}(\delta) 
  \end{equation*}
is obviously linear. In addition,
\begin{eqnarray}
&&\tau_{t,s}^V(AB)=\Gamma_{t,s}\tau_{t-s}(AB)\Gamma_{t,s}^*=\Gamma_{t,s}\tau_{t-s}(A)\tau_{t-s}(B)
\Gamma_{t,s}^*
\nonumber\\
&&\mskip 80mu
=\Gamma_{t,s}\tau_{t-s}(A)\Gamma_{t,s}^*\Gamma_{t,s}\tau_{t-s}(B)\Gamma_{t,s}^*=\tau_{t,s}^V(A)
\tau_{t,s}^V(B)
\nonumber\\
&&\tau_{t,s}^V(A^*)=\Gamma_{t,s}\tau_{t-s}(A^*)\Gamma_{t,s}^*=\Gamma_{t,s}\tau_{t-s}(A)^*
\Gamma_{t,s}^*
=\big(\Gamma_{t,s}\tau_{t-s}(A)\Gamma_{t,s}^*\big)^*=\tau_{t,s}^V(A)^*
\nonumber
\end{eqnarray}
implies that $\tau^V_{t,s}$ is a *-isomorphism. 
Strong continuity follows from
\begin{eqnarray*}
\Vert \tau_{t,s}^V(A)-A\Vert
 &=& 
\Vert \Gamma_{t,s} \tau_{t-s} (A)\Gamma_{t,s}^*-A\Vert
\\ &\le&
\Vert \Gamma_{t,s} (\tau_{t-s}(A)-A)\Gamma_{t,s}^*\Vert
+\Vert (\Gamma_{t,s}-{\bf 1}) A\Gamma_{t,s}^*\Vert
+\Vert A(\Gamma_{t,s}^*-{\bf 1)} \Vert
\nonumber\\
&\to& 0 \ \
,\qquad {\rm as}\ \ t\to s\ .
\end{eqnarray*}
Next, let us prove the group property.
Since $\tau_t$ is a *-isomorphism, we have
\begin{eqnarray*}
\tau^V_{t_1,s}\big(\tau^V_{t,t_1}(A)\big) &=& \Gamma_{t_1,s}\tau_{t_1-s}\big(\Gamma_{t,t_1}
\tau_{t-t_1}(A)\Gamma_{t,t_1}^*\big)\Gamma_{t_1,s}^*
\nonumber\\
&=&
\Gamma_{t_1,s}\tau_{t_1-s}(\Gamma_{t,t_1})\tau_{t_1-s}\big(\tau_{t-t_1}(A)\big)
\tau_{t_1-s}(\Gamma_{t,t_1}^*)\Gamma_{t_1,s}^*\qquad ({\rm *-isomorphism})
\nonumber\\
&=&
\Gamma_{t_1,s}\tau_{t_1-s}(\Gamma_{t,t_1}) \tau_{t-s}(A)
\tau_{t_1-s}(\Gamma_{t,t_1}^*)\Gamma_{t_1,s}^*\qquad ({\rm *-isomorphism})
\ .
\end{eqnarray*}
Moreover, the equality $\Gamma_{t,s}=\Gamma_{t_1,s}\tau_{t_1-s}(\Gamma_{t,t_1})$ yields
$\Gamma_{t,s}^*=\tau_{t_1-s}(\Gamma_{t,t_1}^*)\Gamma_{t_1,s}^* $ because  $\tau_t$ is a *-isomorphism.
Hence, we have
\begin{eqnarray*}
\tau^V_{t_1,s}\big(\tau^V_{t,t_1}(A)\big) &=& 
\Gamma_{t,s} \tau_{t-s}(A)
\Gamma_{t,s}^*= \tau^V_{t,s}(A)
\ .
\end{eqnarray*}
Therefore, we conclude that $ \tau^V_{t,s}(\cdot)$ satisfies the group property.
Finally, let us discuss the form of the generator for $A\in {\cal D}(\delta)$.
For $^\forall A\in D(\delta)$, we have
\begin{eqnarray*}
{\tau_{t+h,s}^V(A)-\tau_{t,s}^V(A)\over h} &=&
\tau^V_{t,s}\left(
{\tau_{t+h,t}^V(A)-A\over h}\right)\qquad ({\rm group\ property})
\\ &=&\tau^V_{t,s}\left(
{\Gamma_{t+h,t}\tau_h(A)\Gamma_{t+h,t}^*-A\over h}\right)
\nonumber\\
&=& \tau^V_{t,s}\left(\Gamma_{t+h,t}
{\tau_h(A)-A\over h}\Gamma_{t+h,t}^*\right)
+\tau^V_{t,s}\left(\Gamma_{t+h,t}A
{\Gamma_{t+h,t}^*-{\bf 1}\over h}\right)
\nonumber
\\
&&\mskip 235 mu+\tau^V_{t,s}\left(
{\Gamma_{t+h,t}-{\bf 1}\over h}A\right)
\nonumber\\
&\to&
\tau^V_{t,s}\left(\delta(A)\right)
+\tau^V_{t,s}\left(-iAV(t)\right)
+\tau^V_{t,s}\left(
iV(t)A\right)
,\qquad {\rm as}\ \ h\to 0\ ,
\end{eqnarray*}
where we have used $\Gamma_{t,t}={\bf 1}$, 
$\Vert (\tau_h(A)-A)/h-\delta(A)\Vert\to 0$,\ 
$\Vert (\Gamma_{t+h,t}-{\bf 1})/h-iV(t)\Vert \to 0\ (h\to 0)$.
In conclusion, the generator is expressed by
\begin{equation}
{d\over dt}\tau_{t,s}^V(A)=\tau_{t,s}^V\Big(
\delta(A)+i[V(t),A]\Big)\ .
  \end{equation}

\section{State on $C^*$ algebra}
Conventional quantum mechanics starts from Hilbert space, and states of quantum systems are represented by unit vectors in Hilbert space ${\cal H}$, and observables are selfadjoint operators on ${\cal H}$. 
Then, expectation values with state $\Psi$ are given by an inner product:
\begin{eqnarray}
\ave{A}_\Psi \equiv \langle \Psi, A\Psi \rangle
\label{expechilbert}
\end{eqnarray}
From the expression~(\ref{expechilbert}), states can be interpreted as a  positive linear functional~$\omega_\Psi:\ {\cal A}_S\to {\bf C}$, where ${\cal A}_S$ represents a *-algebra generated by bounded operators on the Hilbert space ${\cal H}$.
This functional gives the correspondence between
expectation values $\omega_\Psi(A)$ and observables $A\in {\cal A}_S$ 
at state $\Psi$.
Stating from this interpretation, we can generalize a framework of quantum mechanics (the method of $C^*$ algebra).
In the method of $C^*$ algebra, we no longer start from Hilbert space (it is not even necessary to have Hilbert space in the theory), and start from $C^*$ algebra~${\cal A}$, where selfadjoint elements in ${\cal A}$ are defined to be physical observables (it means that a set of observables are a sub-algebra of ${\cal A}$).
Then, we redefine state as a positive linear functional with a
normalization condition \footnote{The same definition can be used for *algebra, and one can study wider class of systems. For instance, it is more natural to start from *algebra if one want t
o study unbounded observables.} 
(see definition~\ref{state_Def}).
This state gives a correspondence between observables and their expectation values at state $\omega$.
As we shall explain, Hilbert space structure will appear {\it later} as a representation of $C^*$ algebra ${\cal A}$.
Roughly speaking, we consider that we know states of quantum systems when we have all the correspondence between expectation values and local observables.
For infinitely extended systems, this definition works better in a mathematical sense.
\\
\begin{Def}
\label{state_Def}
Let ${\cal A}$ be a * algebra, and let ${\cal A^*}$ be a set of linear functionals from ${\cal A}$ to ${\bf C}$. Then, $\omega\in {\cal A}^*$ is said to be a positive linear functional if $\omega(A^* A)\ge 0,$ for all $A\in {\cal A}$.
Moreover, positive linear functional is called faithful if 
$A={\bf 0}$ for all $A$ satisfying $\omega(A^*A)=0$. States are positive linear functionals that satisfies the normalization:
$\omega({\bf 1})=1$. 
\end{Def}

We note that for any positive elements $A$ in $C^*$ algebra, there exists $B\in {\cal A}$ such that $A=B^*B$, and thus, positivity of $\omega$ is equivalent to imposing $\omega(A)\ge 0$ for any positive elements $A\in{\cal A}_+$.
\\
\\
\ Although we introduced state from the correspondence with (\ref{expechilbert}), it is easy to see the connection between state on $C^*$ algebra and density operator in a conventional quantum mechanics.
In particular, there is one-to-one correspondence between density operators and states on $C^*$ algebra in finite systems.
More general results are given by example \ref{traceclass}.
Before stating the proposition, 
let us first present density operator in conventional quantum mechanics, and then its connection to states in $C^*$ algebra (Ex.~\ref{density ope}).

\begin{Ex}
\label{traceclass}
Let ${\cal B(H)}$ be a set of bounded linear operators on Hilbert space ${\cal H}$. Then, if $T\in {\cal B(H)}$ satisfies 
\begin{eqnarray*}
||T||_1\equiv {\rm Tr}|T|=\sum_{n=1}^\infty
\langle e_n, |T|e_n \rangle<\infty
\end{eqnarray*}
for some CONS $\{e_n\}_{n=1}^\infty$, then, 
$T$ is called a trace class operator. 
A set of trace class operators over ${\cal H}$ will be denoted by ${\cal C}_1{\cal (H)}$.
%\end{Def}

%\begin{Def}
\label{density ope}
Let $T\in {\cal C}_1{\cal (H)}$ be defined as positive trace class operator if 
$T\neq {\bf 0}$ and $T\ge 0$.
Moreover, if positive trace class operator $\rho$ is normalized (i.e. ${\rm Tr}\rho=1$), then, it is called a density operator.
%\end{Def}

%\begin{Prop}
Let ${\cal A}$ be a 
unital $C^*$ subalgebra of ${\cal B(H)}$.
For arbitrary density operator $\rho\in{\cal C}_1{\cal (H)}$, we define a map 
$\omega^\rho:\ {\cal A}\to{\bf C}$ as 
\begin{eqnarray}
\omega^\rho(A)={\rm Tr}(\rho A)\ .
\end{eqnarray}
Then, $\omega^\rho(\cdot)$ is a state on ${\cal A}$.
%\end{Prop}
\end{Ex}
We note that $\omega^\rho$ is referred to as the normal state associated to a density operator $\rho$.
\\
\underline{proof}\\
Linearity of $\omega^\rho(\cdot)$ comes from the linearity of trace.
$\omega^\rho({\bf 1})$ follows from ${\rm Tr}\ \rho=1$.
Since ${\rm Tr}(\rho A^* A)={\rm Tr}(A\rho A^* )$ for $^\forall A\in {\cal A}$, we have $\omega^\rho(A^* A)\ge 0$.

\QED{}

Positivity and $C^*$ algebra is a very strong condition, and for instance, one can show the continuity from positivity (proposition~\ref{positivecont}), and some 
more properties (proposition~\ref{Cauchy}).
\begin{Prop}[Bratteli Robinson~2.3.11]\ \ 

\label{positivecont}
Let $\omega$ be a positive linear functional over a $C^*$ algebra ${\cal A}$.
Then, $\omega$ is continuous.

\end{Prop}
see [Bratteli Robinson] for the proof.

\begin{Prop}
\label{Cauchy}
Let ${\cal A}$ be a $C^*$ algebra and $\omega(\cdot)$ be a state over ${\cal A}$.
Then, the following properties are satisfied:
\begin{itemize}
	\item[(i)] $\omega(A)^*=\omega(A^*)$

	\item[(ii)] $|\omega(A^*B)|\le \sqrt{\omega(A^*A)\omega(B^*B)}$\ 
	(Cauchy-Schwarz inequality)

	\item[(iii)] $\omega(A^*B^*BA)\le \Vert B\Vert^2 \omega(A^*A)$

	\item[(iv)] $|\omega(A)|\le \Vert A\Vert$
\end{itemize}

\end{Prop}
\underline{proof of (i)}\\
Positivity of $\omega$ follows
$$
\omega\left((A+\lambda{\bf 1})^*(A+\lambda{\bf 1})\right)=\omega(A^*A)+\lambda^*\omega(A)+\lambda\omega(A^*)+|\lambda|^2
\in {\bf R}\ ,
$$
for arbitrary complex number $\lambda$.
Thus, we have
$$
{\rm Im}\left(\lambda^*\omega(A)+\lambda\omega(A^*)\right)=0\ .
$$
By substituting $\lambda=1$ and $\lambda=i$, we obtain
\begin{eqnarray*}
{\rm Im}\omega(A^*)=-{\rm Im}\omega(A)
,\qquad
{\rm Re}\omega(A^*)={\rm Re}\omega(A)\ .
\end{eqnarray*}
We conclude $\omega(A^*)=\omega(A)^*$.
\\
\\
\underline{proof of (ii)}\\
Positivity of $\omega$ follows
\begin{eqnarray}
0\le
\omega\left((A+\lambda B)^*(A+\lambda B)\right)=\omega(A^*A)+\lambda^*\omega(B^*A)+\lambda\omega(A^*B)+|\lambda|^2
\omega(B^*B)
\ ,
\label{CS1}
\end{eqnarray}
for arbitrary complex number $\lambda$.
Let us take $\lambda=t\omega(A^*B)^*, t\in {\bf R}$, then, (i) yields
$$\omega(B^*A)=\omega(A^*B)^*\ .$$
Thus, inequality~(\ref{CS1}) reads
\begin{eqnarray}
0\le
\omega(A^*A)+2t|\omega(A^*B)|^2+t^2|\omega(A^*B)|^2
\omega(B^*B)\ .
\label{CS2}
\end{eqnarray}
Since inequality (\ref{CS2}) is satisfied for arbitrary $t\in {\bf R}$, 
discriminant of the R.H.S. of (\ref{CS2}) should be less than or equal to 0:
\begin{eqnarray}
|\omega(A^*B)|^4-\omega(A^*A)|\omega(A^*B)|^2\omega(B^*B)
\le 0
\label{CS3}
\end{eqnarray}
If $\omega(A^*B)=0$, then, Cauchy-Schwarz inequality is satisfied, else 
inequality (\ref{CS3}) gives Cauchy-Schwarz inequality.
\\
\\
\underline{proof of (iii)}\\
\begin{eqnarray*}
\left\Vert {\bf 1}-\left({\bf 1}-{B^*B\over\Vert B\Vert^2}\right)\right\Vert
\le {\Vert B^*B\Vert\over\Vert B\Vert^2}=1
\end{eqnarray*}
implies ${\bf 1}-B^*B/\Vert B\Vert^2\in{\cal A}_+$.
Therefore, there exists $B'$ such that
\begin{eqnarray*}
{\bf 1}-\frac{B^*B}{\Vert B\Vert^2}=B^{\prime *}B'
\ .
\end{eqnarray*}
Positivity of $\omega$ gives
\begin{eqnarray*}
0 &\le& \omega\Big(A^* B'^* B'A\Big)
\\ &=&
\omega(A^*A)-{\omega(A^*B^*BA)\over\Vert B\Vert^2}
\ .
\end{eqnarray*}
Thus, we have $\omega(A^*B^*BA)\le \Vert B\Vert^2\omega(A^*A)$.
\\
\\
\underline{proof of (iv)}\\
\begin{eqnarray*}
|\omega(A)| &=& |\omega(A^*)|\qquad (i)
\\ &\le& 
\sqrt{\omega(A^*A)}\qquad ({\rm substitute}\ B={\bf 1}\ {\rm into}\ (ii))
\\ &\le& 
||A||\qquad ({\rm substitute}\ A={\bf 1},\ B=A\ {\rm into}\ (iii))
\end{eqnarray*}
\QED{}

\subsection{GNS representation\label{GNS_sec}}
We have discussed states and time evolution on $C^*$ algebra without using the Hilbert space structure.
In conventional quantum mechanics, discussion starts from setting a Hilbert space, and states are unit vector in the Hilbert space. 
Therefore, we do not need to set up different Hilbert space for different states.

In $C^*$ algebra, states are positive linear functional over $C^*$ algebra ${\cal A}$, and Hilbert space appears as a representation of states.
Namely, Hilbert space is introduced for each state, and there is a strong connection between state $\omega$ and the introduced Hilbert space. 
Elements in ${\cal A}$ are connected to a linear operator on the Hilbert space.

In this subsection, we briefly discuss the representation theory, which gives the connection between Hilbert space and positive linear functional.
\begin{Def}
A representation of a $C^*$ algebra ${\cal A}$ is defined to be a pair 
$({\cal H},{\cal D},\pi)$, where ${\cal H}$ is a complex Hilbert space, 
${\cal D}$ is a dense subspace of ${\cal H}$, and $\pi$ is a *-morphism of ${\cal A}$ into ${\cal B(D)}$. 
A representation $({\cal H},{\cal D},\pi)$ is defined to be faithful if $\pi$ is a *-isomorphism between ${\cal A}$ and $\pi({\cal A})$, i.e. ker~$\pi=\{{\bf 0}\}$.
Moreover, a representation is defined to be cyclic, if there exists a 
vector $\Omega\in {\cal D}\setminus \{ {\bf 0}  \}$ such that
$\pi({\cal A}) \Omega$ is dense in ${\cal H}$.
$\Omega\in {\cal D}$ is referred to be as a cyclic vector.

\end{Def}
Clearly, faithful representation is one of the most important classes of representations, and it is  well-known that each representation $({\cal H},{\cal D},\pi)$ of a $C^*$ algebra defines a faithful representation of the quotient algebra ${\cal A}/{\rm ker}\ \pi$(see discussion of [Bratteli Robinson] before the definition 2.3.2).
Moreover, the next proposition gives criteria for faithfulness.

\begin{Prop} [Bratteli Robinson~2.3.3]\ 

Let (${\cal H},{\cal D},\pi)$ be a representation of $C^*$ algebra ${\cal A}$.
The representation is faithful, if and only if, it satisfies each of the following equivalent conditions:
\begin{itemize}
	\item[(i)] ker~$\pi={\bf 0}$.
	\item[(ii)] $||\pi(A)||=||A||,\ ^\forall A\in {\cal A}$.
	\item[(iii)] $\pi(A)>0$ for all $A>0$.
\end{itemize}

\end{Prop}

Once the representation~$({\cal H},\pi)$ is given, we can define 
a linear functional $\omega_{\Psi}(\cdot):\ {\cal A}\to {\bf C}$ by
\begin{eqnarray*}
\omega_{\Psi}=\langle \Psi, A\Psi\rangle
\ ,
\end{eqnarray*}
for some $\Psi\in {\cal H}$.
Then, one can easily prove that $\omega_{\Psi}$ is positive. 
Moreover, $\omega_{\Psi}$ is a state over ${\cal A}$, if and only if, $\Psi$ is a unit vector in ${\cal H}$.
For a unit vector $\Psi$, $\omega_\Psi$ is said to be a vector state of representation $({\cal H},{\cal D},\pi)$.

In short, this argument gives states from representations; however the existence of the representation itself is not so obvious.
In the rest of this subsection, we demonstrate one of the most common methods to construct a representations from states (GNS representation), and then, show an example of the representation.
Namely, positive linear functional with normalization ensure the existence of representations~(GNS representation).
\begin{Prop} [GNS representation]
Let ${\cal A}$ be a unital $C^*$ algebra, and $\omega$ be a state over ${\cal A}$.
Then, there exist a Hilbert space~${\cal H}_\omega$, its element
$\Omega_\omega\in {\cal H}_\omega$, and a *-morphism $\pi_\omega$ of 
${\cal A}$ into ${\cal B}({\cal H}_\omega)$ such that
\begin{eqnarray}
&&\omega(A)=\langle \Omega_\omega,\pi_\omega(A)\Omega_\omega\rangle  \ \ \ (^\forall A\in{\cal A})
\\
&&{\cal H}_\omega=
\overline{ \{\pi_\omega(A)\Omega_\omega \ : \ A\in{\cal A}\} }
\label{cyclic}
\end{eqnarray}
where bar in the 
R.H.S. of Eq.~(\ref{cyclic}) represents a closure with respect to a norm in Hilbert space.
In addition, $\Omega_\omega$ is a cyclic vector of ${\cal H}_\omega$, i.e., 
$\{\pi_\omega(A)\Omega_\omega| A\in {\cal A} \}\subset {\cal H}_\omega$ is dense in 
${\cal H}_\omega$.
A pair $({\cal H}_\omega,\pi_\omega,\Omega_\omega)$ is referred to as a 
GNS representation of $({\cal A},\omega)$.
Moreover, GNS representation is unique up to unitary transformation.

\end{Prop}
{\bf Remark 1}\\
Suppose that (${\cal H}'_\omega,\pi_\omega,\Omega_\omega)$ is a representation of 
$({\cal A},\omega)$. Then, there exists a unitary transformation 
$U:\ {\cal H}'_\omega\to {\cal H}_\omega$, such that
\begin{eqnarray*}
\left\{
\begin{array}{c}
U^{*} \pi'_\omega(A)U = \pi_\omega (A)
\\
\ \ \ \ \ U^{*}\Omega_\omega' = \Omega_\omega
\end{array}
\right.\ .
\end{eqnarray*}
It follows 
\begin{eqnarray*}
\langle \Omega_\omega, \pi_\omega(A)\Omega_\omega\rangle 
=\langle U^{*}\Omega_\omega' , U^{*}\pi'_\omega(A)\Omega_\omega'\rangle 
=\langle \Omega_\omega', \pi'_\omega(A)\Omega_\omega'\rangle \ .
\end{eqnarray*}
Therefore, inner products are preserved for different hamiltonian, which is transformed by unitary transformation.
\\{\bf Remark 2} (Inequivalent representation)\\
Let $\omega$ and $\bar{\omega}$ be different states over $C^*$ algebra ${\cal A}$.
Let $({\cal H}_\omega,\pi_\omega,\Omega_\omega)$ and 
$({\cal H}_{\bar{\omega}}, \pi_{\bar{\omega}}, \bar{\Omega}_\omega)$ be their representations, respectively.
In general, there does not exist a unitary transformation~$U$ satisfying 
\begin{eqnarray*}
\left\{
\begin{array}{c}
U^{*} \pi_{\bar{\omega}}(A)U = \pi_{\omega} (A)
\\
\ \ \ \ \ U^{*} \bar{\Omega}_\omega = \Omega_\omega
\end{array}
\right.\ .
\end{eqnarray*}
In some sense, different states over a single $C^*$ algebra are not connected.
\\
\underline{proof}\\
Let us start from constructing a Hilbert space.
$\omega(A)^*=\omega(A^*)$ yields $\omega(A^*B)=\omega(B^*A)^*$.
Thus, a map $\langle\cdot , \cdot \rangle:\ {\cal A}\times {\cal A}\to {\bf C}$
\begin{eqnarray*}
\langle A , B \rangle\equiv \omega(A^*B),\ A,B\in {\cal A}
\end{eqnarray*}
is sesquilinear and satisfy Hermitian symmetry~$\langle A , B \rangle$.
Moreover, it satisfies $\langle A , A \rangle\ge 0$ for any elements
$A\in{\cal A}$.
Thus, the map~$\langle\cdot , \cdot \rangle$ satisfies all the properties of inner product except that $\langle A, A \rangle=0$ does not imply $x = {\bf 0}$.
To construct an inner product, let us form a quotient 
algebra ${\cal D}_\omega={\cal A}\setminus {\cal N}$, where ${\cal N}$ represents 
${\cal N}\equiv \{ A\in {\cal A}|\ \langle A,A\rangle=0\}$.
The elements of ${\cal D}_\omega$ are classes $[A]=\{A+n|\ n\in{\cal N}\}$.
One can easily show ${\cal A}_\pi$ forms a vector space with respect to 
the following operation:
\begin{eqnarray*}
[A]+[B]\equiv [A+B],\qquad
\alpha[A]\equiv [\alpha A],\ \ A,B\in {\cal A},\ \alpha\in {\bf C}
\end{eqnarray*}
Let us introduce a map 
$\langle\cdot , \cdot\rangle_\omega:\ {\cal D}_\omega\times  {\cal D}_\omega\to {\bf C}$ as 
\begin{eqnarray*}
\langle [A], [B]\rangle_\omega \equiv \langle A, B\rangle
,\ [A],[B]\in {\cal D}_\omega\ .
\end{eqnarray*}
This map does not depend on the choice of representatives and therefore is well-defined:
\begin{eqnarray*}
|\langle A_1, B_1\rangle - \langle A_2, B_2\rangle|
&\le& |\langle A_1-A_2, B_1\rangle| + |\langle A_2, B_1-B_2\rangle|
\\ &\le& \sqrt{\langle A_1-A_2,A_1-A_2\rangle}\sqrt{\langle B_1,B_1\rangle}
\\ && +\sqrt{\langle A_2,A_2\rangle}\sqrt{\langle B_1-B_2,B_1-B_2\rangle}
\qquad ({\rm Cauchy-Schwarz\ inequality})
\\&=& 0,\qquad ^\forall A_1,\ A_2\in[A],\ ^\forall B_1,\ B_2\in[B]
\ .
\end{eqnarray*}
Obviously, $([A],[A])_\omega=0$ implies $[A]=[{\bf 0}]$; therefore
${\cal D}_\omega$ is a pre-Hilbert space (inner product space) with a inner product $\langle \cdot , \cdot \rangle_\omega$.
Then we have a Hilbert space ${\cal H}_\omega$ as the Cauchy completion of ${\cal D}_\omega$ 
with respect to the norm $|[A]|\equiv\sqrt{([A],[A])}$.

Next, we are going to discuss *-morphism~$\pi_\omega(\cdot)$.
$\pi_\omega(A)\in {\cal B(H}_\omega)$ is defined to be $\pi_\omega(A)[B]\equiv [AB]$ for $A\in{\cal A}$, then $\pi_\omega(\cdot)$ can be extended to bounded operator over ${\cal H}_\omega$ as we shall explain below.

Let $\psi\in{\cal H}_\omega$, then, there exists a sequence 
$\{B_n\}_{n=1}^\infty\subset {\cal D}_\omega$
%{\cal A}$ 
such that
\begin{eqnarray*}
\lim_{n\to\infty}\Big| [B_n]-\psi \Big|=0\ \ .
\end{eqnarray*}
For that sequence, we have
\begin{eqnarray}
&&\Big| \pi_\omega(A) [B_n]-\pi_\omega(A) [B_m] \Big|
=\Big|[A(B_n-B_m)]\Big|
=\sqrt{\omega\big((B_n-B_m)^*A^*A(B_n-B_m)\big)}
\nonumber\\
&&\le \Vert A\Vert \sqrt{\omega\big((B_n-B_m)^*(B_n-B_m)\big)}
=\Vert A\Vert \Big| [B_n]-[B_m] \Big|\to 0,\qquad
{\rm as}\ n,m\to\infty\ .
\nonumber
\end{eqnarray}
Therefore, $\{[B_n]\}_{n=1}^\infty$ is a Cauchy sequence in 
${\cal H}_\omega$, and thus completeness follows that the limit exists.
We define the limit $\lim_{n\to\infty}\pi_\omega(A)[B_n]$ as $\pi_\omega(A)\psi$.
This map~$\psi\to\pi_\omega(A)\psi$ is obviously linear, 
and $\pi_\omega(A)$ is an element of ${\cal B}({\cal H}_\omega)$
due to $|\pi_\omega(A)\psi|\le\Vert A\Vert~|\psi|$:
$$
\Big| \pi_\omega(A) [B_n] \Big|=\sqrt{\omega(B_n^*A^*AB_n)}\le
\Vert A\Vert\sqrt{\omega(B_n^*B_n)}=\Vert A\Vert \Big|[B_n]\Big|
$$
We need to prove that the map 
$\pi_\omega:\ {\cal A}\ni A\to\pi_\omega(A)\in{\cal B}({\cal H}_\omega)$ is *-morphism, i.e., the map is linear, $\pi_\omega(A)^*=\pi_\omega(A^*)$, and
$\pi_\omega(A)\pi_\omega(B)=\pi_\omega(AB)$.
Linearity comes from\footnote{Strictly speaking, it should be proved for 
$C\notin {\cal A}$ as well, but the discussion is similar to the extension of 
$\pi_\omega$ we have discussed.}
$$\pi_\omega(c_1A+c_2B)[C]=[c_1AC+c_2BC]
=c_1 [AC]+c_2 [BC]=\{c_1\pi_\omega(A)+c_2\pi_\omega(B)\} [C]\ .$$
$\pi_\omega(A)^*=\pi_\omega(A^*)$ comes from
\begin{eqnarray*}
\langle [B], \pi_\omega(A)^* [C]\rangle  &=& \langle \pi_\omega(A) [B],[C]\rangle =\langle [AB],[C]\rangle 
=\omega(B^*A^*C)
\\
 &=& \langle [B],[A^*C]\rangle =\langle [B],\pi_\omega(A^*) [C]\rangle \ .
\end{eqnarray*}

Finally $\pi_\omega(A)\pi_\omega(B)=\pi_\omega(AB)$ comes from 
\begin{eqnarray*}
\pi_\omega(A_1)\pi_\omega(A_2) [B]=[A_1A_2B]=\pi_\omega(A_1A_2) [B]\ .
\end{eqnarray*}
Thus, $\pi_\omega$ is a *-morphism of ${\cal A}$ into ${\cal B(H}_\omega)$.

Let us prove that $\Omega_\omega=[{\bf 1}]$  is a cyclic vector, and satisfies
\begin{eqnarray*}
\omega(A)=
\langle \Omega_\omega,\pi_\omega(A)\Omega_\omega\rangle , \ \ \ ^\forall A\in{\cal A}\ .
\end{eqnarray*}
$\{\pi_\omega(A)\Omega_\omega : \ A\in{\cal A}\}
=\{[A] : \ A\in{\cal A}\}={\cal D}_\omega$ yields that $\Omega_\omega$ 
is a cyclic vector.
In addition, we have
\begin{eqnarray*}
\langle \Omega_\omega,\pi_\omega(A)\Omega_\omega\rangle =
\langle [{\bf 1}] , [A]\rangle =\omega(A)\ .
\end{eqnarray*}
Moreover, $\omega({\bf 1})=1$ implies that $\Vert \Omega_\omega\Vert=1$, and thus, $\Omega_\omega$ is a unit vector.
Finally, we are going to prove that 
another representation~$({\cal H}_\omega',\pi_\omega',\Omega_\omega')$
is transformed to 
$({\cal H}_\omega,\pi_\omega,\Omega_\omega)$ by a unitary transformation.
Let us define a map~$U:\ {\cal H}_\omega\to{\cal H}_\omega'$ by
\begin{eqnarray}
U \pi_\omega(A)\Omega_\omega = \pi_\omega'(A)\Omega_\omega'
\ \ ^\forall A\in{\cal A}\ .
\label{UDef}
\end{eqnarray}
Since $\{\pi_\omega(A)\Omega_\omega:A\in{\cal A}\}$ is dense in 
${\cal H}_\omega$, we can extend the domain of $U$ into ${\cal H}_\omega$.
Then, we have
\begin{eqnarray*}
&&\Big\langle U^* \pi_\omega'(A)\Omega_\omega' ,\pi_\omega(B)\Omega_\omega\Big\rangle =
\Big\langle \pi_\omega'(A)\Omega_\omega',U \pi_\omega(B)\Omega_\omega \Big\rangle 
\nonumber\\
&&=
\Big\langle \pi_\omega'(A)\Omega_\omega',\pi_\omega'(B)\Omega_\omega'\Big\rangle =
\Big\langle \Omega_\omega',\pi_\omega'(A)^*\pi_\omega'(B)\Omega_\omega'\Big\rangle 
=
\Big\langle \Omega_\omega',\pi_\omega'(A^*B)\Omega_\omega'\Big\rangle 
\nonumber\\
&&=\omega(A^*B)
=\Big\langle \Omega_\omega,\pi_\omega(A^*B)\Omega_\omega\Big\rangle =
\Big\langle \pi_\omega(A)\Omega_\omega,\pi_\omega(B)\Omega_\omega\Big\rangle 
\ .
\end{eqnarray*}
It follows $U^* \pi_\omega'(A)\Omega_\omega'=\pi_\omega(A)\Omega_\omega$.
Combining with a definition of $U$, we have 
\begin{eqnarray*}
UU^*\big(\pi_\omega'(A)\Omega_\omega'\big) &=& U\big(\pi_\omega(A)\Omega_\omega\big)
=\pi'_\omega(A)\Omega_\omega'
\\
U^*U\big(\pi_\omega(A)\Omega_\omega\big) &=& U^*\big(\pi_\omega'(A)\Omega_\omega'
\big)=\pi_\omega(A)\Omega_\omega\ .
\end{eqnarray*}
Thus, we conclude that $U$ is unitary. Moreover,
$$
U^*\pi_\omega'(A)U\pi_\omega(B)\Omega_\omega=U^*\pi_\omega'(A)\pi_\omega'(B)\Omega_\omega'=
U^*\pi_\omega'(AB)\Omega_\omega'=\pi_\omega(AB)\Omega_\omega=\pi_\omega(A)\pi_\omega(B)\Omega_\omega
$$
follows $U^*\pi_\omega'(A)U=\pi_\omega(A)$.
Thanks to this equality, the definition of $U$~(\ref{UDef}) reads 
$\Omega_\omega=U^* \Omega_\omega'$.

\QED{}

As an example of GNS representation, let us demonstrate a construction of representation of Hilbert-Schmidt space.
\begin{Def}
Let ${\cal H}$ be a Hilbert space.
A bounded operator $T\in {\cal H}$ on ${\cal H}$ is defined to be 
a Hilbert-Schmidt operator if 
$T^*T\in {\cal C}_1({\cal H})$.
A set of Hilbert-Schmidt operators over ${\cal H}$ will be denoted by ${\cal C}_2{\cal (H)}$.
A norm $||T||_2=\sqrt{{\rm Tr}\ (T^*T)}$ is said to be a Hilbert-Schmidt norm.

\end{Def}

\begin{Prop}[Arai 1.40,\ 1.41]
Let ${\cal H}$ be a Hilbert space.
Let us define a map~$\langle \cdot, \cdot\rangle_2:\ {\cal C}_2{\cal (H)}\times {\cal C}_2{\cal (H)}\to {\bf C}$ by
\begin{eqnarray*}
\langle T , S \rangle_2 \equiv {\rm Tr}(T^* S)\ .
\end{eqnarray*}
Then, $\langle \cdot,\cdot\rangle_2$ is an inner product over ${\cal C}_2{\cal (H)}$.
Moreover, let us define ${\cal L}_2({\cal H})$ as an pre-Hilbert 
of ${\cal C}_2{\cal (H)}$ equipped with the inner product $\langle \cdot,\cdot\rangle_2$. Then, ${\cal L}_2({\cal H})$ is a Hilbert space.

\end{Prop}
See Arai for the proof.

We are going to construct a GNS representation of 
$C^*$-sub-algebra~${\cal A}$ of ${\cal B(H)}$, where  ${\cal H}$ is Hilbert space.
Suppose $T$ is an element of ${\cal L}_2({\cal H})$, then, $AT\in {\cal L}_2({\cal H})$ for arbitrary elements $A$ in ${\cal A}$ 
(see [Arai~1.36] for example).
Only to avoid confusion, we denote $[T]$ for $T\in {\cal C}_2{\cal (H)}$ when we use it as a element of ${\cal L}_2({\cal H})$\footnote{$T$ was used in the lecture of Prof. Tasaki}. 
Let us define a map $l(A)$ by
\begin{eqnarray*}
l(A) [T]\equiv [AT],\ T\in {\cal L}_2 ({\cal H})
\end{eqnarray*}
One can easily show 
\begin{eqnarray*}
|| l(A) [T]||_2\le \Vert A\Vert\ \Vert [T]\Vert_2\ ,
\end{eqnarray*}
therefore, $l(A)\in {\cal B}( {\cal L}_2 ({\cal H}) )$.
Moreover, $l(\cdot):\ {\cal A}\to {\cal B}\big({\cal L}_2({\cal H})\big)$ is easily proved to be a morphism.
Combining with
\begin{eqnarray*}
\langle [T], l(A)[S]\rangle_2={\rm Tr}\ (T^*AS)=
\langle l(A^*) [T], [S]\rangle_2\ ,
\end{eqnarray*}
we conclude that $l(\cdot):\ {\cal A}\to {\cal B}( {\cal L}_2 ({\cal H}) )$ 
is a *-morphism.
Thus, $({\cal L}_2 ({\cal H}) ),l)$ is a representation of ${\cal A}$.
Let $\rho$ be a density operator over ${\cal H}$, and 
$\omega^\rho$ be its associated normal state:
\begin{eqnarray*}
\omega^\rho( A ) \equiv {\rm Tr}\ (\rho A),\ \ A\in {\cal A} 
\end{eqnarray*}
It follows
\begin{eqnarray} 
\omega^\rho( A ) = {\rm Tr}\ (\rho^{1/2} A\rho^{1/2})=
\langle \big[\rho^{1/2}\big], l(A) \big[\rho^{1/2}\big] \rangle_2
\ .
\label{HS-GNS}
\end{eqnarray}
Namely, expectation value of $A$ with a state $\omega^\rho$ is expressed 
by expectation value of $l(A)$ with respect to $\big[\rho^{1/2}\big]$ in the Hilbert space~${\cal L}_2 ({\cal H})$.
Let ${\cal H}_\rho$ be a closure of ${\cal D}_\rho=l({\cal A})\big[\rho^{1/2}\big]$, then, ${\cal H}_\rho$ is a closed subset of ${\cal L}_2 ({\cal H})$, and is a Hilbert space.
One can prove that 
$\Big({\cal H}_\rho,l,\big[ \rho^{1/2}\big]\Big)$ is a GNS representation of $({\cal A},\omega^\rho)$ (see [Arai \S 4.4]).

\subsection{Equilibrium state}
In the previous subsection, we have discussed the representation.
There, first Hilbert space~${\cal H}_\omega$ was constructed from $C^*$ algebra~${\cal A}$ and a state~$\omega(\cdot)$ over ${\cal A}$, then, an element $A$ in $C^*$ algebra was mapped into a bounded operator on the Hilbert space.
Expectation value of elements $A\in{\cal A}_{s.a.}$ is expressed by
\begin{eqnarray}
\omega(A)=\langle \Omega_\omega, \pi_\omega(A)\Omega_\omega\rangle
\ .
\label{cf quantum}
\end{eqnarray}
In this expression, $\Omega_\omega$ is a unit vector in the Hilbert space~${\cal H}_\omega$, and (\ref{cf quantum}) corresponds to an average in conventional quantum mechanics.

In this subsection, we are going to discuss the characterization of equilibrium
states in $C^*$ algebra.
First we give a GNS representation of equilibrium state for finite system, then, we discuss KMS (Kubo-Martin-Schwinger) condition for finite systems.
The condition will be generalized as a condition to characterize equilibrium for generic systems (including infinite systems).

\subsubsection{Equilibrium state and GNS representation}
In this subsection, we discuss a GNS representation of finite system.
Let ${\cal H}$ be a Hilbert space, and ${\cal A}$ be a $C^*$sub-algebra of ${\cal B(H)}$.
Then, a normal state associated to a density operator
\begin{eqnarray*}
\rho_\beta=\frac{e^{-\beta H}}{ {\rm Tr}\ e^{-\beta H}}
\end{eqnarray*}
is defined to be a Gibbs state.
Let ${\cal H}_\rho$ be a closure of 
${\cal D}_\rho=l({\cal A})\big[ \rho^{1/2}\big]$.
Let us define $\Omega_\beta$ and $l_\beta:\ {\cal A}\to{\cal B(H}_\beta)$ by $\Omega_\beta\equiv \big[ \rho_\beta^{1/2}\big]$, and $l_\beta(A)=l(A)|{\cal H}_\beta$.

Following the discussion in the previous subsection, $({\cal H}_{\beta},l_\beta,\Omega_\beta)$ is the GNS representation of $({\cal A},\omega^{\beta})$.
Then, expectation value at Gibbs states~$\langle\cdot\rangle_\beta$ reads
\begin{eqnarray*}
\langle A \rangle_\beta=\langle \Omega_\beta, 
l_\beta(A)\Omega_\beta\rangle_2\ .
\end{eqnarray*}

\subsubsection{correlation function for finite systems\label{sec_corre} }
In this subsection, we are going to discuss equilibrium states for 
finite systems.
One of the most important properties which characterize Gibbs states are time-correlation function:
\begin{eqnarray*}
C_{A,B}(t,s)\equiv 
\langle 
\alpha^H_t(A) \alpha^H_s(B) 
\rangle_\beta,\qquad A,B\in {\cal A},\ t,s\in{\bf R}
\ ,
\end{eqnarray*}
where $\alpha^H_t(A)$ is defined by $\alpha^H_t(A)=e^{itH} A e^{-itH}$.
We note that $H$ is well-defined since we only discuss finite systems in this subsection.
In this subsection, we will show the restriction on time-correlation function
Gibbs state (KMS condition).
This condition will be generalized later in \S~\ref{KMS_sec}, and, conversely, KMS condition is shown to derive Gibbs state.

Thanks to $[H,\rho_\beta]={\bf 0}$, we have 
\begin{eqnarray*}
C_{A,B}(t,s) &=& 
{\rm tr}\left( \rho_\beta e^{iHt} A e^{-iHt} e^{iHs} B e^{-iHs}\right)
\\ &=& 
{\rm tr}\left( \rho_\beta A e^{iH(s-t)} B e^{-iH(s-t)}\right)
=\langle A\alpha^H_{s-t}(B) \rangle_\beta
\\ &=& 
{\rm tr}\left( \rho_\beta e^{iH(t-s)} A e^{-iH(t-s)} B \right)
=\langle \alpha^H_{t-s}(A) B\rangle_\beta
\ .
\end{eqnarray*}
Therefore, the following two correlation functions 
are of fundamental interest:
\begin{eqnarray*}
F_{A,B}(t) &\equiv& \langle A\alpha^H_{t}(B) \rangle_\beta
=G_{A,B}(-t)
\\
G_{A,B}(t) &\equiv& \langle \alpha^H_{t}(A) B \rangle_\beta
=F_{A,B}(-t)
\end{eqnarray*}
Let us first extend $F_{A,B}(t)$ to complex plane.
Suppose $y\in [0,\beta]$. We define 
$\widetilde{F}_{A,B}(t+iy)$ by
\begin{eqnarray*}
\widetilde{F}_{A,B}(t+iy) &\equiv& \frac{1}{Z_\beta}
{\rm Tr}\ \left(e^{-\beta H}A e^{i(t+iy)H} B e^{-i(t+iy)H} \right)
\\
Z_\beta &\equiv& {\rm Tr}\ \left( e^{-\beta H} \right)
\ .
\end{eqnarray*}

\begin{Prop} [Arai 7.14, 7.15]\ 

Let $I_\beta$ be a strip:
\begin{eqnarray*}
I_\beta\equiv \{ z=t+iy|t\in{\bf R}, 0<y<\beta\}\ .
\end{eqnarray*}
Then, $\widetilde{F}_{A,B}(z)$ is an analytical continuation of
$F_{A,B}(z)$ to $I_\beta$.
$\widetilde{F}_{A,B}(t)$ is holomorphic in $I_\beta$, 
and is bounded and continuous in $\bar{I}_\beta$.
Moreover, the following inequality is satisfied:
\begin{eqnarray*}
|\widetilde{F}_{A,B}(z)|\le C_\beta ||A||\ ||B||,\ \ z\in \bar{I}_\beta
\ ,
\end{eqnarray*}
where $C_\beta$ is a constant independent of $A$ and $B$.

\end{Prop}
Since $\widetilde{F}_{A,B}(z)$ is an analytical continuation of $F_{A,B}(t)$, 
we shall write $F_{A,B}(z)$.
We remark that $G_{A,B}(t)$ can be analytically continued to $I_{-\beta}$ 
and it satisfy $G_{A,B}(z)=F_{A,B}(-z)\ (z\in \bar{I}_\beta)$ 
due to the symmetry $F_{A,B}(t)=G_{A,B}(-t)$.

\begin{Prop}
\begin{eqnarray}
F_{A,B}(t) &=& G_{B,A}(t-i\beta)
\label{KMS_fin1}
\\
F_{A,B}(t+i\beta) &=& G_{B,A}(t)
\label{KMS_fin2}
\end{eqnarray}

\end{Prop}

\underline{proof}\footnote{One need to make more rigorous arguments. See [Arai~7.18] for example}
\\
\begin{eqnarray*}
F_{A,B}(t) &=& 
\frac{1}{Z_\beta}{\rm Tr}\ \left( e^{-\beta H} A e^{itH} B e^{-itH} \right)
\\ &=& 
\frac{1}{Z_\beta}{\rm Tr}\ \left( e^{itH} B e^{-itH} e^{-\beta H} A \right)
\end{eqnarray*}
On the other hand, $G_{B,A}(t-i\beta)$ reads
\begin{eqnarray*}
G_{B,A}(t-i\beta) &=& 
 \frac{1}{Z_\beta}{\rm Tr}\ 
\left( e^{-\beta H}e^{i(t-i\beta)H} A e^{-i(t-i\beta)H}  B \right)
\\ &=& 
\frac{1}{Z_\beta}{\rm Tr}\ 
\left( e^{itH} A e^{-itH}e^{-\beta H}  B \right)
\ .
\end{eqnarray*}
Thus, we have (\ref{KMS_fin1}).
$G_{A,B}(z)=F_{A,B}(-z)\ (z\in \bar{I}_\beta)$  and 
 (\ref{KMS_fin1}) follow
\begin{eqnarray*}
F_{A,B}(t+i\beta) = G_{A,B}(-t-i\beta) =F_{B,A}(-t)=G_{B,A}(t)\ .
\end{eqnarray*}
\QED{}

\subsubsection{KMS condition\label{KMS_sec}}
In \S~\ref{sec_corre}, we have discussed a restriction on time-correlation function at equilibrium state.
Since density operator is ill-defined in infinite systems, we cannot characterize equilibrium state by Gibbs state.
Instead, we employ the equality (\ref{KMS_fin2}) to characterize equilibrium states (Later we will show that this condition gives Gibbs state for finite systems).
Roughly speaking, we impose 
\begin{eqnarray*}
\omega\left( A \sigma_{t+i\beta}(B) \right) \Big|_{t=0}
=
\omega\left( \sigma_t(B) A \right)
\Big|_{t=0}\ ,
\end{eqnarray*}
for equilibrium states.

To give more rigorous definition, let us start from giving a definition of 
analytic elements in $C^*$ algebra~${\cal A}$.

\begin{Def} 

Let $\sigma_t$ be a strong continuous group of *-isomorphisms over 
$C^*$ algebra~${\cal A}$.
Let $S_\lambda$ be a strip 
\begin{eqnarray*}
S_\lambda=\{z: \ |{\rm Im}z|<\lambda\}
\ .
\end{eqnarray*}
$A\in {\cal A}$ is called analytic for $\sigma_t$, 
if there exists a function $f:S_\lambda\to {\cal A}$ which satisfies the following condition:

\begin{itemize}
\item[(i)] $f(t)=\sigma_t(A),\ t\in{\bf R}$.
\item[(ii)] The following limit exists for $z\in I_\lambda$ in norm (strong analyticity):
$$
\lim_{h\to 0}{f(z+h)-f(z)\over h} 
$$
\end{itemize}
If $A\in {\cal A}$ is analytic for $\sigma_t$, we denote $f(z)=\sigma_z(A)$.

\end{Def}

Similar to the construction of generator, one can construct 
a dense set in ${\cal A}$ which consist of entire analytic elements for any $\sigma_t$. Namely, for almost all $A\in {\cal A}$, $\sigma_t(A):\ t\to{\cal A}$ can be analytically continued to a entire complex plane: 
proposition~\ref{dense analytical set}).

\begin{Prop}
\label{dense analytical set}
Let $\sigma_t$ be a strong continuous group of *-isomorphisms over 
$C^*$ algebra~${\cal A}$.
There exists entire analytic elements for $\sigma_t$.
Thus, we can define a set~${\cal A}_\sigma$ of analytic elements for $\sigma_t$.
${\cal A}_\sigma$ is dense *-subalgebra in ${\cal A}$, and is 
$\sigma_t$-invariant.

\end{Prop} 
\underline{proof}\\
(Existence) \\
Let $\epsilon$ be a positive number and $A_\epsilon$ be an element of ${\cal A}$ defined by
\begin{eqnarray*}
A_\epsilon\equiv \int_{-\infty}^\infty 
{ds\over \sqrt{\pi}\epsilon}e^{-s^2/\epsilon^2}\sigma_s(A)
\ .
\end{eqnarray*}
We are going to prove that $A_\epsilon$ is analytic for $\sigma_t$, and a set of $A_\epsilon$ is dense in ${\cal A}$.
Let us start from analyticity.
It is sufficient to prove a function $f(z):\ ${\bf C}$ \to {\cal A}$:
\begin{eqnarray}
f(z) \equiv \int_{-\infty}^\infty 
{ds\over \sqrt{\pi}\epsilon}e^{-(s-z)^2/\epsilon^2}\sigma_s(A)
\label{analytic_f}
\end{eqnarray}
is strong analytic in a entire complex plane, and matches with $\sigma_t(A)$ for $t\in {\bf R}$.

Thanks to $|e^{-(s-z)^2/\epsilon^2}|=\exp[-(s^2-2s{\rm Re}z
+{\rm Re}z^2)/\epsilon^2]$, 
R.H.S. of (\ref{analytic_f}) is convergent, and is an element of ${\cal A}$.
Similarly, a function $g(z):\ ${\bf C}$ \to {\cal A}$:
\begin{eqnarray*}
g(z) \equiv \int_{-\infty}^\infty 
{ds\over \sqrt{\pi}\epsilon}{2(s-z)\over \epsilon^2}
e^{-(s-z)^2/\epsilon^2}\sigma_s(A)
\end{eqnarray*}
is an element of ${\cal A}$.
With the aids of an inequality:
\begin{eqnarray*}
|e^\alpha-1-\alpha|\le
\sum_{n=0}^\infty|\alpha|^{n+2}/(n+2)!
\le
\sum_{n=0}^\infty|\alpha|^{n+2}/n!=|\alpha|^2e^{|\alpha|}
\ ,
\end{eqnarray*}
we have
\begin{eqnarray*}
&&\left\Vert {f(z+h)-f(z)\over h}-g(z)\right\Vert
\\
=&&\left\Vert
\int_{-\infty}^\infty 
{ds\over \sqrt{\pi}\epsilon}
\Big\{
e^{{2h(s-z)-h^2\over\epsilon^2}}-1
-{2h(s-z)\over \epsilon^2}
\Big\}
{e^{-{(s-z)^2\over\epsilon^2}}\over
h}
\sigma_s(A)
\right\Vert
\nonumber\\
\le&&
\int_{-\infty}^\infty 
{ds\over \sqrt{\pi}\epsilon}
\left|
e^{{2h(s-z)-h^2\over\epsilon^2}}-1
-{2h(s-z)-h^2\over \epsilon^2}
-{h^2\over \epsilon^2}
\right|
\left|
{e^{-{(s-z)^2\over\epsilon^2}}\over
h}\right|
\Vert
\sigma_s(A)
\Vert
\nonumber\\
\le&&
|h|
\Vert A\Vert
\int_{-\infty}^\infty 
{ds\over \sqrt{\pi}\epsilon}
\Big\{
{|2(s-z)-h|^2\over\epsilon^4}
\exp\Big(\left|{2h(s-z)-h^2\over\epsilon^2}\right|\Big)
+{1\over \epsilon^2}
\Big\}
\left|
e^{-{(s-z)^2\over\epsilon^2}}\right|
\\
\to&& 0, \qquad {\rm as}\ h\to 0\ ,
\end{eqnarray*}
for $z\in {\bf C}$. Thus, $f(z)$ is strong analytic.

Moreover, the following equality:
\begin{eqnarray*}
f(t)=\int_{-\infty}^\infty 
{ds\over \sqrt{\pi}\epsilon}e^{-(s-t)^2/\epsilon^2}\sigma_s(A)
=\int_{-\infty}^\infty 
{ds\over \sqrt{\pi}\epsilon}e^{-s^2/\epsilon^2}\sigma_{t+s}(A)
=\sigma_t(A_\epsilon)
\end{eqnarray*}
yields $f(z)=\sigma_z(A_\epsilon)$ ($A_\epsilon$ is an element of ${\cal A}_\sigma$). Thus, ${\cal A}_\sigma$ is not empty.
\\
\\
\ Let us prove that ${\cal A}_\sigma$ is dense in ${\cal A}$ \\
For every $A\in {\cal A}$, we define a sequence $\{B_n\}_{n=1}^\infty$: 
\begin{eqnarray*}
B_n\equiv A_{1/n}=
n
\int_{-\infty}^\infty 
{ds\over \sqrt{\pi}}e^{-n^2s^2}\sigma_s(A)
\end{eqnarray*}

By definition, $B_n$ is an element of ${\cal A}_\sigma$, and the equality:
\begin{eqnarray*}
\left\Vert B_n-A\right\Vert
 &=& \left\Vert 
n\int_{-\infty}^\infty 
{ds\over \sqrt{\pi}}e^{-n^2s^2}\{\sigma_s(A)-A\}
\right\Vert
\\ &=& 
\left\Vert 
\int_{-\infty}^\infty 
{ds\over \sqrt{\pi}}e^{-s^2}\{\sigma_{s/n}(A)-A\}
\right\Vert
\nonumber\\
&\le&
\int_{-\infty}^\infty 
{ds\over \sqrt{\pi}}e^{-s^2}\Vert\sigma_{s/n}(A)-A\Vert
\to 0, \qquad {\rm as}\ n\to\infty
\nonumber
\end{eqnarray*}
follows $B_n\to A$ in norm, thus, ${\cal A}_\sigma$ is dense in ${\cal A}$.

Next, we prove that ${\cal A}_\sigma$ is a *-subalgebra of ${\cal A}$\\
Let $A,B$ be elements of ${\cal A}_\sigma$, i.e., there exists strong continuous functions $f,h:\ {\bf C}\to {\cal A}$ s. t.
\begin{eqnarray*}
f(t)=\sigma_t(A), \qquad
h(t)=\sigma_t(B)\ ,
\end{eqnarray*}
for any $t\in {\bf R}$.
Let $c_1$ and $c_2$ be complex numbers.
Obviously, $c_1f(z)+c_2h(z),\ f(z)h(z)$, and $f(z^*)^*$ are strong analytic, and satisfy
\begin{eqnarray*}
&&\left. c_1f(z)+c_2h(z)\right|_{z=t}=c_1f(t)+c_2h(t)=
c_1\sigma_t(A)+c_2\sigma_t(B)=\sigma_t(c_1A+c_2B)
\nonumber\\
&&\left. f(z)h(z)\right|_{z=t}=f(t)h(t)=\sigma_t(A)\sigma_t(B)=\sigma_t(AB)
\nonumber\\
&&\left. f(z^*)^*\right|_{z=t}=f(t)^*=\sigma_t(A)^*=\sigma_t(A^*)
\ .
\end{eqnarray*}
Therefore $c_1A+c_2B,\ AB$, and $A^*$ are elements of ${\cal A}_\sigma$, and thus, ${\cal A}_\sigma$ is a *-sub-algebra of ${\cal A}$.
\\
\\
\ Finally, we prove that ${\cal A}_\sigma$ is $\sigma_t$-invariant.\\
Let $A$ be elements of ${\cal A}_\sigma$, i.e., there exists strong continuous functions $f,g:\ {\bf C}\to {\cal A}$ such that
\begin{eqnarray*}
&&f(t)=\sigma_t(A),\ t\in {\bf R}
\\
&&\lim_{h\to 0}\left\Vert{f(z+h)-f(z)\over h}-g(z)\right\Vert=0,\ z\in{\bf C}
\ .
\end{eqnarray*}
The following equality
\begin{eqnarray*}
\lim_{h\to 0}\left\Vert{\sigma_s(f(z+h))-\sigma_s(f(z))\over h}-\sigma_s(g(z))\right\Vert
&=&
\lim_{h\to 0}\left\Vert \sigma_s\left({f(z+h)-f(z)\over h}-g(z)\right)\right\Vert
\nonumber\\
&=&\lim_{h\to 0}\left\Vert {f(z+h)-f(z)\over h}-g(z)\right\Vert
\nonumber\\
&=&0
\nonumber
\end{eqnarray*}
yields that $f_s(z)\equiv \sigma_s(f(z))$ is strong analytic.

On the other hand,
\begin{eqnarray*}
f_s(z)\Big|_{z=t} = \sigma_s(\sigma_t(A))=\sigma_t(\sigma_s(A))
\end{eqnarray*}
implies $\sigma_s(A)$ is an element of ${\cal A}_\sigma$.
Thus, we conclude that ${\cal A}_\sigma$ is $\sigma$-invariant.

\QED{}

Using the idea of analytical elements, we can define the KMS condition.

\begin{Def}

Let $\sigma_t$ be a strong continuous group of *-isomorphisms over $C^*$ algebra~${\cal A}$, and ${\cal A}_\sigma$ be a *-sub-algebra of entire analytic elements for $\sigma_t$.
A state $\omega$ over ${\cal A}$ is said to be a $\sigma$-KMS state at $\beta$ (or $(\sigma,\beta)$-KMS state), if there exists a dense (in norm) and $\sigma$-invariant *-sub-algebra ${\cal A}_\sigma^\omega$ of ${\cal A}_\sigma$ such that
 \begin{equation*}
  \omega\left(A\sigma_{i\beta}\left(B\right)\right) = \omega\left(BA\right)
 \end{equation*}
for any $A,B \in {\cal A}_\sigma^\omega$.
 \end{Def}

\begin{Prop}
Let be the $C^*$ algebra of $n\times n$ matrices, and let 
$\sigma_s(A)\equiv e^{iHs}Ae^{-iHs},\ H\in{\cal A}_{s.a.}$.
Then, $(\sigma,\beta)$-KMS state is 
the normal state associated to a density operator $\rho=e^{-\beta H}/Z_\beta$.
\end{Prop}
\underline{proof}\\
Since there is one-to-one correspondence between density operators and states on $C^*$ algebra in finite systems, state $\omega$ can be expressed with density operator $\rho$:
\begin{eqnarray*}
\omega(A)={\rm Tr}\ (\rho A)
\end{eqnarray*}

$\sigma_{i\beta}\left(B\right)=e^{-\beta H}Be^{\beta H}$ 
and KMS condition follow
\begin{eqnarray*}
{\rm Tr}\Big(\rho Ae^{-\beta H}Be^{\beta H}\Big)=
{\rm Tr}(\rho BA)\ .
\end{eqnarray*}
Selfadjoint property for $H$ implies that $H$ has real eigenvalues 
$h_j\ (j=1\cdots n)$ with associated eigenvectors $\ket{j}$.
Take $A=\ket{i}\bra{j}$ and $B=\ket{k}\bra{l}$.
Then, KMS condition reads
\begin{eqnarray}
\bra{l} \rho \ket{i} e^{-\beta(h_j-h_l)}\delta_{jk}=
{\rm Tr}\Big(\rho Ae^{-\beta H}Be^{\beta H}\Big)=
{\rm Tr}(\rho BA)=
\bra{j} \rho \ket{k} \delta_{il}\ .
\label{KMS_finMat}
\end{eqnarray}
By substituting $i=l$ and $k=j$ into (\ref{KMS_finMat}), we have
\begin{eqnarray*}
\bra{l} \rho \ket{l} e^{-\beta h_j}
e^{\beta h_l}
=
\bra{j} \rho \ket{j} 
\ .
\end{eqnarray*}
Therefore, $\bra{j} \rho \ket{j} e^{\beta h_j}$ does not depend on $j$.
Let us define the constant as $1/C_{\beta}$.

By taking $k=j$, (\ref{KMS_finMat}) reads
\begin{eqnarray*}
\bra{l} \rho \ket{i} e^{-\beta(h_j-h_l)}
=\bra{j} \rho \ket{j} \delta_{il}=\delta_{il}{e^{-\beta h_j}\over C_\beta}
\ ,
\end{eqnarray*}
where we have used $\bra{l} \rho \ket{l} e^{\beta h_l}=1/C_{\beta}$.
Thus, we obtain
\begin{eqnarray*}
\bra{l} \rho \ket{i} =\frac{\delta_{il}\ e^{-\beta h_l}}{C_\beta}\ .
\end{eqnarray*}
Therefore, density operator reads
\begin{eqnarray*}
\rho &=& 
\sum_{li} \ket{l} \bra{l} \rho \ket{i} \bra{i}
\\ &=& 
\sum_i \frac{e^{-\beta h_i}} {C_\beta}\ket{i}\bra{i}
\\ &=& 
\sum_i \frac{e^{-\beta H}} {C_\beta}\ket{i}\bra{i}
\\ &=& 
\sum_i \frac{e^{-\beta H}} {C_\beta}
\ .
\end{eqnarray*}
Normalization condition ${\rm Tr}\ \rho=1$ gives $C_\beta=Z_\beta$.

We conclude that $\rho$ is 
the normal state associated to $\rho=e^{-\beta H}/Z_\beta$.

\QED{}

Next, we discuss grand canonical ensemble of spinless Fermion.
\begin{Prop}
\label{Grandcanonical}
Let ${\cal A}_{\rm CAR}$ be a unital CAR algebra 
generated by $a(f)\equiv\int f(k)^* a_k dk,\ f\in L^2$, $a(f)^*$ and unit element ${\bf 1}$.
Let $\Theta$ be a *-isomorphism over ${\cal A}_{\rm CAR}$, and 
$\sigma_t$ be a group of strong continuous *-isomorphisms defined by
\begin{eqnarray*}
\Theta\big(a(f)\big) &\equiv& -a(f)
\\
\sigma_t\big(a(f)\big) &\equiv& \int e^{-i (\omega_k-\mu)t}f(k)^* a_k dk
\ .
\end{eqnarray*}

Let $\omega$ be a $\Theta$-invariant, i.e.,  $\omega(\Theta(A))=\omega(A)$, 
$(\sigma,\beta)$-KMS state.

Then, $\omega$ satisfies Wick's theorem:
\begin{eqnarray}
&&\omega\big(a(f_1)^*\cdots a(f_n)^*a(g_m)\cdots a(g_1)\big)
\nonumber
\\
&&=
\delta_{n,m}\sum_{\sigma} (-1)~{|\sigma|}
\omega\Big( a (f_1)^* a\big(g_{|\sigma|(1)}\big) \Big)
\cdots
\omega\Big( a (f_n)^* a\big(g_{|\sigma|(n)}\big) \Big)
\ ,
\label{Wickoriginal}
\end{eqnarray}
where $\sigma=(\sigma(1)\cdots \sigma(n))$ represents a permutation of $1,\ldots,n$, and 
$|\sigma|$ is 0 for even permutation and 1 for odd permutation.

Moreover, the following equality is satisfied:
\begin{eqnarray}
\omega\big(a(f_1)^*\cdots a(f_n)^*a(g_m)\cdots a(g_1)\big)=
\begin{cases}
\det\Big\{\omega\big(a(f_i)^*a(g_j)\big)
\Big\}_{1\le
i,j\le n} &(n=m)\\
0 &(n\not= m)
\end{cases}
\ ,
\label{Wick}
\end{eqnarray}
where $\omega\big(a(f_i)^*a(g_j)\big)$ is given by
\begin{eqnarray}
\omega\big(a(f_i)^*a(g_j)\big)=\int dk {f_i(k)g_j(k)^*\over e^{\beta(\omega_k-\mu)}+1}
\ .
\label{2-point fermi}
\end{eqnarray}
In general, state satisfying (\ref{Wick}) is referred to as a 
quasi-free state.

  \end{Prop}
Formally speaking, 
we consider a time evolution 
\footnote{
$\sim$ can be replaced by equal for finite systems.
For infinite systems, we use $\sim$ to represent formal equalities.} $\sigma_t\sim e^{i(H-\mu N)t} A e^{-i(H-\mu N)t}$, 
then, $(\sigma,\beta)$-KMS states correspond to
a state described by a density operator $\rho\propto e^{-\beta(H-\mu N)}$ (Grand canonical state).
\\
\\
\underline{proof}\\
\underline{proof of (\ref{2-point fermi})}\\
KMS condition for $a(f)^*$ and $a(G)$ gives
\begin{eqnarray}
\omega 
\Big( a(f)^* \sigma_{i\beta}\big( a(G)\big) \Big)
=
\omega 
\Big(  a(G) a(f)^* \Big)
\label{KMS_grand}
\end{eqnarray}
Let $\widetilde{G}(k)\equiv e^{\beta(\omega_k-\mu)}G(k)$, then, 
$\sigma_{i\beta}\big( a(G)\big)$ reads
\begin{eqnarray}
\sigma_{i\beta}\big( a(G)\big)
=
 \int e^{\beta(\omega_k-\mu)}G(k)^* a_k dk
=
a(\widetilde{G})\ .
\label{KMS_grandL}
\end{eqnarray}
Thanks to the CAR relation, R.H.S. of (\ref{KMS_grand}) reads
\begin{eqnarray}
\omega \Big(  a(G) a(f)^* \Big)
=-\omega \Big(  a(f)^* a(G) \Big)+\langle G,f\rangle_{L^2}\ .
\label{KMS_grandR}
\end{eqnarray}
Substituting (\ref{KMS_grandL}) and (\ref{KMS_grandR}) into 
KMS condition~(\ref{KMS_grand}), 
one obtains
\begin{eqnarray*}
\omega \Big(  a(f)^* a(G+\widetilde{G}) \Big)
=\langle G,f\rangle_{L^2}\ .
\end{eqnarray*}
Let $g(k)=G(k)+\widetilde{G}(k)$, then, we have
\begin{eqnarray*}
G(k)=\frac{g(k)}{ 1+e^{\beta(\omega_k -\ \mu)} }\ ,
\end{eqnarray*}
and thus, KMS condition reads
\begin{eqnarray*}
\omega \Big(  a(f)^* a(g) \Big) &=& 
\langle G,f\rangle_{L^2}
\\ &=& 
\int dk\ \frac{g(k)^* f(k)}{ 1+e^{\beta(\omega_k -\ \mu)} }
\ .
\end{eqnarray*}
Thus, (\ref{2-point fermi}) is proved.
\\
\underline{proof of (\ref{Wick})}\\
Suppose $n+m$ is odd number.
Then, (\ref{Wick}) immediately follow from $\Theta$-invariance of $\omega$:
\begin{eqnarray*}
\omega\big(a(f_1)^*\cdots a(f_n)^*a(g_m)\cdots a(g_1)
\big) &=& \omega\Big(\Theta\big(a(f_1)^*\cdots a(f_n)^*a(g_m)\cdots a(g_1)
\big)\Big)
\nonumber\\ &=&
-\omega\big(a(f_1)^*\cdots a(f_n)^*a(g_m)\cdots a(g_1)\big)
\end{eqnarray*}
Hereafter, we prove (\ref{Wick}) in case $n+m=2l$.
First let us prove (\ref{Wick}) for $n=0$.
From KMS condition and CAR relation follow
\begin{eqnarray}
\omega\big(a(g_{2l})\cdots a(g_2)
a(G)\big) &=& 
\omega\big((a(\widetilde{G}))
a(g_{2l})\cdots a(g_2)
\big)
\nonumber\\ &=& 
-
\omega\big(a(g_{2l})\cdots a(g_2)a({\widetilde G})\big)
\ .
\label{KMS_grand0}
\end{eqnarray}
By taking $g_1(k)\equiv G(k)+{\widetilde G}(k)$, we have
$\omega\big(a(g_{2l})\cdots a(g_1)\big)=0$.
It follows
\begin{eqnarray*}
\omega\big(a(g_1)^*\cdots a(g_{2l})^*\big)=0\ .
\end{eqnarray*}
Thus, (\ref{Wick}) is proved for $(n=0,m)$ and $(n,m=0)$.

The last possibility is the case of $n\neq 0, m\neq 0$ and $n+m=2l$.
To prove (\ref{Wick}) for this case, 
we will use the following equality:
\begin{eqnarray}
&&\omega\big(a(f_1)^*\cdots a(f_n)^*a(g_m)\cdots a(g_1)
\big)\nonumber\\
&&=\sum_{j=1}^n(-1)^{j-1}\omega(a(f_j)^*a(g_1))\
\omega\big(
\underbrace{a(f_1)^*\cdots a(f_n)^*}_{{\rm except}\ a(f_j)^*}
a(g_m)\cdots a(g_2)
\big)\ ,
\label{reduction}
\end{eqnarray}
which can be easily obtained using
the similar argument of the proof of (\ref{2-point fermi}).
\footnote{Use CAR relation for $\omega\big(a(G)a(f_1)^*\cdots a(f_n)^*
 a(g_m)\cdots a(g_2)\big)=
\omega\big(a(f_1)^*\cdots a(f_n)^*
 a(g_m)\cdots a(g_2)a(\widetilde{G})\big)$.}.

Since (\ref{reduction}) takes out $\omega(a(f_j)^*a(g_1))$ from 
$\omega\big(a(f_1)^*\cdots a(f_n)^*a(g_m)\cdots a(g_1)
\big)$, we have 
\begin{eqnarray*}
\omega\big(a(f_1)^*\cdots a(f_n)^*a(g_m)\cdots a(g_1)
\big)=0
\end{eqnarray*}
for $n<m$. Similarly, one can prove (\ref{Wick}) for $n>m$.

Finally let us prove (\ref{Wick}) for $n=m$ by induction.
\\
$n=m=2$
\begin{eqnarray}
&&\omega\big(a(f_1)^*a(f_2)^*a(g_2)a(g_1)
\big)\nonumber\\
&&=
\omega\big(a(f_1)^*a(g_1)\big)\ \omega\big(a(f_2)^*a(g_2)\big)
-\omega\big(a(f_2)^*a(g_1)\big)\ \omega\big(a(f_1)^*a(g_2)\big)
\nonumber\\
&&=
{\rm
det}\left(\begin{matrix}
\omega\big(a(f_1)^*a(g_1)\big) & \omega\big(a(f_1)^*a(g_2)\big)
\cr
\omega\big( a(f_2)^*a(g_1) \big) & \omega\big(a(f_2)^*a(g_2)\big)
\end{matrix}\right)
\nonumber
\end{eqnarray}
Suppose (\ref{Wick}) is satisfied for $n=m\le k-1$.

Since 
\begin{eqnarray}
(-1)^{j+1}
\omega\big(
\underbrace{a(f_1)^*\cdots a(f_k)^*}_{{\rm except}\ a(f_j)^*}
a(g_k)\cdots a(g_2)
\big)
\end{eqnarray}
is a $(j,1)$ cofactor $\Delta_{j1}$ of a matrix 
$\{\omega(a(f_i)^*a(g_j))\}_{1\le i,j\le k}$, we have
\begin{eqnarray}
&&\omega\big(a(f_1)^*\cdots a(f_k)^*a(g_k)\cdots a(g_1)
\big)\nonumber\\
&&=\sum_{j=1}^k(-1)^{j-1}\omega(a(f_j)^*a(g_1))\
\omega\big(
\underbrace{a(f_1)^*\cdots a(f_k)^*}_{a(f_j)^* except}
a(g_k)\cdots a(g_2)
\big)
\nonumber\\
&&=\sum_{j=1}^k\omega(a(f_j)^*a(g_1))\
\Delta_{j1}
={\rm det}\{\omega(a(f_i)^*a(g_j))\}_{1\le i,j\le k}
\ .
\end{eqnarray}
Thus, we conclude that (\ref{Wick}) is satisfied.

Obviously, the Wick theorem (\ref{Wickoriginal}) is satisfied due to 
(\ref{Wick}) and (\ref{reduction}).

\QED{}

To describe equilibrium states of single system, we have considered $(\sigma,\beta)$-KMS state with a time evolution $\sigma_t(A)\sim e^{i(H-\mu N)t} A e^{-i(H-\mu N)t}$.
This statement needs to be modified to describe systems contain independent subsystems $[H_j,H_k]={\bf 0}\ (j\neq 0)$ with different temperatures.
For this purpose, it is convenient to include temperature in time evolution: 
$\widetilde{\sigma}_t(A) \sim 
e^{i\sum_j \beta_j (H_j-\mu N_j)t} A e^{-i\sum_j \beta_j (H_j-\mu N_j)t}$.
Then, $(\widetilde{\sigma},1)$-KMS state\footnote{Since temperature is formally included in the time-evolution, $\beta$ in the KMS condition does not represent temperature}
is equivalent to 
$(\sigma,\beta)$-KMS state for single systems. 
For instance, the KMS condition 
for $a(f)^*$ and $a(g)$ in proposition~\ref{Grandcanonical} reads
\begin{eqnarray}
\omega 
\Big( a(f)^* \sigma_{i\beta}\big( a(g)\big) \Big)
=
\omega 
\Big(  a(g) a(f)^* \Big)\ .
\label{KMS1}
\end{eqnarray}
On the other hand, $(\widetilde{\sigma},1)$ KMS condition reads
\begin{eqnarray}
\omega 
\Big( a(f)^* \widetilde{\sigma}_{i}\big( a(g)\big) \Big)
=
\omega 
\Big(  a(g) a(f)^* \Big)\ .
\label{KMS2}
\end{eqnarray}
Take 
\begin{eqnarray*}
\sigma_t\big(a(f)\big) &\equiv& \int e^{-i (\omega_k-\mu)t}f(k)^* a_k dk
\\
\widetilde{\sigma}_t\big(a(f)\big) &\equiv& \int e^{-i \beta (\omega_k-\mu)t}f(k)^* a_k dk
\ ,
\end{eqnarray*}
then, (\ref{KMS1}) and (\ref{KMS2}) are obviously equivalent.
Thanks to this idea, we have the next proposition for systems containing subsystems in different equilibria.
\begin{Prop}
Let $a_{k1}$ and $a_{k2}$ be annihilation operators satisfying 
\begin{eqnarray*}
\{a_{k\lambda},a_{k'\lambda'}^*\}=\delta_{\lambda\lambda'}\delta(k-k')
,\qquad
\{a_{k\lambda},a_{k'\lambda'}\}=0
\end{eqnarray*}
Let ${\cal A}_{\rm CAR}$ be a unital CAR algebra 
generated by 
${\displaystyle a(f)\equiv \sum_{\lambda}\int f_{\lambda}(k)^* a_{k\lambda} dk,\ f_{\lambda}\in L^2}$ $(\lambda=1,2)$, $a(f)^*$ and unit element ${\bf 1}$.

Let $\Theta$ and $\widetilde{\sigma}_t$ be *-isomorphism and a strong continuous group of *-isomorphisms defined by
\begin{eqnarray*}
\Theta\big(a(f)\big) &=& -a(f)
\\
\widetilde{\sigma}_t\big(a(f)\big) &=& 
\sum_\lambda\int e^{-i \beta(\omega_{k\lambda}-\mu_\lambda)t}f_\lambda(k)^* 
a_{k\lambda} dk
\ .
\end{eqnarray*}

Take $\omega$ as a $\Theta$-invariant ($\widetilde{\sigma}_t,1$)-KMS state.
Then, $\omega$ satisfy
\begin{eqnarray*}
\omega\big(a(f_1)^*\cdots a(f_n)^*a(g_m)\cdots a(g_1)\big)=
\begin{cases}
\det\Big\{\omega\big(a(f_i)^*a(g_j)\big)
\Big\}_{1\le
i,j\le n} &(n=m)\\
0 &(n\not= m)
\end{cases}\ ,
\end{eqnarray*}
where $\omega\big(a(f_i)^*a(g_j)\big)$ is given by
\begin{eqnarray*}
\omega\big(a(f_i)^*a(g_j)\big)=\sum_\lambda\int dk {f_{i\lambda}(k)g_{j\lambda}(k)^*\over e^{\beta_\lambda(
\omega_{k\lambda}-\mu_\lambda)}+1}
\ .
\end{eqnarray*}

  \end{Prop}
We skip the poof of this proposition since it is almost the same as the proof of proposition~\ref{Grandcanonical}.
Roughly speaking, 
we consider a time evolution 
$\widetilde{\sigma}_t\sim e^{i\beta (H-\mu N)t} A e^{-i\beta(H-\mu N)t}$, 
then, $(\widetilde{\sigma},1)$-KMS states correspond to
a state described by a density operator 
$\rho\propto e^{-\sum_\lambda\beta_\lambda(H_\lambda-\mu_\lambda N_\lambda)}$.

\section{Mixing, Return to equilibrium, and Asymptotic abelian}
 \begin{Def}
  Let $\tau_t$ a *-automorphism over $C^*$ algebra ${\cal A}$.
 $\tau_t$ is said to be asymptotic abelian, if the following condition is satisfied:
  \begin{equation*}
   \lim_{|t|\to\infty} \|\left[A,\tau_t\left(B\right)\right]_\pm\| = 0, \,\,
    \forall A,B \in {\cal A}\ ,
  \end{equation*}
, where we take $+$ if $A,B$ contains odd number of fermions, and 
$-$ for the other cases.
 \end{Def}
The condition for asymptotic abelian looks quite strong, 
but it is not so strong. 
For instance, let us study a CAR algebra and free-time evolution $\sigma_t$ defined 
in proposition~\ref{Grandcanonical}.
Take $A=a(f)^*a(f)$, and $B=a(g)^* a(g)$, and, let  $\tau_t(a(g))=g_t$, then, 
\begin{eqnarray*}
[A,\tau_t(B)]_- &=& [a(f)^* a(f), a(g_t)^* a(g_t)]_-
\\ &=&
a(f)^* [a(f), a(g_t)^* a(g_t)]_-
+[a(f)^*, a(g_t)^* a(g_t)]_- a(f)
\\ &=&
a(f)^* [a(f), a(g_t)^*]_+ a(g_t)
-a(g_t)^*[a(f)^*, a(g_t)]_+ a(f)
\\ &=&
\langle g_t,f\rangle_{L^2}^* a(f)^* a(g_t)
-\langle g_t,f\rangle_{L^2} a(g_t)^* a(f)
\end{eqnarray*}
Thus, we have
\begin{eqnarray*}
\Vert [A,\tau_t(B)]_- \Vert =
2\left| \langle g_t,f \rangle_{L^2} \right|\ \Vert a(f)\Vert\ \Vert a(g_t)\Vert 
\end{eqnarray*}
Thanks to the Riemann Lebesgue theorem, we have
\begin{eqnarray*}
\langle g_t,f \rangle_{L^2} = \int dk f(k)g(k)^* e^{i(\omega_k-\mu) t}
\to 0,\ {\rm as}\ |t|\to\infty
\ .
\end{eqnarray*}
It follows $\Vert [A,\tau_t(B)]_- \Vert\to 0$.

\begin{Def} 

Let ${\cal H}$ be a Hilbert space, and 
${\cal M}$ be a sub-algebra of ${\cal B}({\cal H})$ (${\cal M}$ is not necessary to be a *-sub-algebra).
\begin{equation}
{\cal M}'=\{ a\in{\cal B}({\cal H}):\
[a,b]\equiv ab-ba=0 \ \ ^\forall b\in{\cal M}\}
\end{equation}
is said to be a commutant of ${\cal M}$, and 
$({\cal M}')'={\cal M}''$ is said to be a bicommutant of ${\cal M}$.

Since arbitrary element in ${\cal M}$ commute with ${\cal M}'$, 
we have ${\cal M}''\subset {\cal M}$.
A Von Neumann algebra on ${\cal H}$is a *-subalgebra ${\cal M}$ of 
${\cal B}({\cal H})$ such that
${\cal M}''={\cal M}$.

\end{Def}

\begin{Prop}
\label{Neumann class}
For any subset ${\cal M}$ of ${\cal B}({\cal H})$, 
\begin{eqnarray*}
{\cal M}' &=& {\cal M}^{(3)}={\cal M}^{(5)}=\cdots
\\
{\cal M}'' &=& {\cal M}^{(4)}={\cal M}^{(6)}=\cdots\ ,
\end{eqnarray*}
where ${\cal M}^{(n)}$ is defined by 
${\cal M}^{(n+1)}=\big( {\cal M}^{(n)} \big)',\ {\cal M}^{(2)}={\cal M}''$.

\end{Prop}
\underline{proof}\\
If $B\in{\cal B}({\cal H})$ commutes with any $m\in{\cal M}''$, then, 
$B$ commutes with any elements in a subset of ${\cal M}''$.
Thus, ${\cal M}\subset {\cal M}''$ implies ${\cal M}^{(3)}\subset{\cal M}'$. 
On the other hand, replacing ${\cal M}$ by ${\cal M}'$ in ${\cal M}\subset {\cal M}''$, we have ${\cal M}'\subset {\cal M}^{(3)}$.

Thus, we have 
\begin{eqnarray}
{\cal M}'= {\cal M}^{(3)}\ .
\label{commutant13}
\end{eqnarray}
Moreover, by replacing ${\cal M}'$ by ${\cal M}^{(n)}$ in (\ref{commutant13}), 
we have the proposition.

\QED{}

From proposition\ref{Neumann class}, we have the following example of 
a Von Neumann algebra.
\begin{Ex}
Let $\omega$ be a state over $C^*$ algebra ${\cal A}$ and 
$({\cal H}_\omega,\pi_\omega,\Omega)$ be GNS representation of 
$({\cal A},\omega)$.
Then,
\begin{eqnarray*}
{\cal M}\equiv \{\pi_\omega(A): \ ^\forall A\in {\cal A}\}''
\equiv \pi_\omega({\cal A})''
\end{eqnarray*}
is a Von Neumann algebra.

\end{Ex}

\begin{Def}
The center ${\cal Z}({\cal M})$ of a Von Neumann algebra ${\cal M}$ 
on ${\cal H}$ is defined by
\begin{eqnarray*}
{\cal Z}({\cal M})={\cal M} \cap {\cal M}'
\end{eqnarray*}

\end{Def}

\begin{Def}
\label{factor_Def}
Let $\omega$ be a state over $C^*$ algebra ${\cal A}$ and 
$({\cal H}_\omega,\pi_\omega,\Omega)$ be GNS representation of 
$({\cal A},\omega)$.
Let ${\cal M}$ its associated von Neumann algebra
${\cal M}\equiv \pi_\omega(\mathfrak{A})''$.
A state $\omega$ is called factor\footnote{A von Neumann algebra is called a factor if it has a trivial center.}, 
if ${\cal M}$ has a trivial center, i.e., 
\begin{eqnarray*}
{\cal Z}({\cal M})
= {\bf C}\ {\bf 1}\equiv
\{c{\bf 1}: \ ^\forall c\in{\bf C}\}
\ .
\end{eqnarray*}

\end{Def}
{\bf Remark 1}\\
Let us restrict on finite systems.
Roughly speaking, a factor state $\omega$ means that its associate density 
operator has an inverse.
Let us roughly explain it for special case.

Let us think of a Hilbert Schmidt space and its associated GNS representation.
For finite systems, there exists a density operator associated to the state.
As we discussed in \S~\ref{GNS_sec}, we attach a Hilbert space structure with the Hilbert Schmidt norm.
Let us define $l(A)$ and $l'(A)$ as
\begin{eqnarray*}
l(A) [T]&\equiv& [AT],\ T\in {\cal L}_2({\cal H})
\\
l'(A) [T]&\equiv& [TA],\ T\in {\cal L}_2({\cal H})
\ .
\end{eqnarray*}
Then, expectation value reads
\begin{eqnarray*}
\omega(A^*B)={\rm Tr}\ (\rho A^* B)
={\rm Tr}\ \big( (A\rho^{1/2})^* (B\rho^{1/2}) \big)=
\langle l(A)\Omega , l(B)\Omega \rangle_2 
\ , 
\end{eqnarray*}
where $\Omega\equiv [\rho^{1/2}]$.
For any elements $[C]\in {\cal H}_\omega$, we have
\begin{eqnarray*}
l'(A) l(B)[C]=[BCA]=l(B) l'(A) [C]
\ .
\end{eqnarray*}
Thus, we have $[l'(A) , l(B)]$.
It follows\footnote{More rigorous arguments is necessary.}
\begin{eqnarray*}
l({\cal A})'= \{\l'(A):\ A\in {\cal A} \}\ .
\end{eqnarray*}
Namely algebra of right products is commutant of the algebra of left products.
Let $X\in l({\cal A})' \cap l({\cal A})''$, i.e., 
$[X,l(A)]=[X,l'(B)]=0,\ ^\forall A,B$; it means
\begin{eqnarray}
&& X\big([A\rho^{1/2}]\big)=X(l[A]\Omega) = l[A] X(\Omega)
\label{factor1}
\\
&& X\big([\rho^{1/2}A]\big)=X(l'(A)\Omega) = l'(A) X(\Omega)
\label{factor2}
\end{eqnarray}
Suppose there exists $Y\in {\cal A}$ such that
\begin{eqnarray*}
X(\rho^{1/2})=\big[ Y\rho^{1/2} \big]
\ .
\end{eqnarray*}
Then, (\ref{factor1}) reads
\begin{eqnarray*}
\big[ YA\rho^{1/2} \big]=\big[ AY\rho^{1/2} \big]
\ .
\end{eqnarray*}
If there exists inverse of $\rho$, then, it follows
\begin{eqnarray*}
\big[ YA\big]=\big[ AY\big]
\ .
\end{eqnarray*}
Since $Y$ commutes with any $A$, it is restricted to 
$Y=\alpha {\bf 1}\ (\alpha\in{\bf C})$.
Namely, if $\rho$ is invertible, then, $\omega$ is a factor.

{\bf Remark 2}\\
If $\omega$ is a unique KMS state\footnote{
Physically speaking, if no phase transition takes place, the KMS state is unique, else, several KMS states exist as a result of spontaneous symmetry breaking.}
, then, $\omega$ is a factor.
Let us explain it with rough arguments.
First let us write KMS condition:
$\omega\big( A\sigma_{i\beta}(B) \big)=\omega(BA)
$ in terms of the GNS representation:
\begin{eqnarray*}
\Big\langle  \Omega,\pi_\omega (A) \bar{\sigma}_{i\beta}(\pi_\omega(B)) \Omega \Big\rangle 
=
\Big\langle  \Omega, \pi_\omega(B)\pi_\omega(A) \Omega \Big\rangle 
\ ,
\end{eqnarray*}
where $\bar{\sigma}$ represents a extension of $\sigma$ into ${\cal B}({\cal H}_\omega)$\footnote{It is possible to construct this extension, but we skip it in this paper.}.
Let $T\in l({\cal A})' \cap l({\cal A})''$.
Suppose $T\Omega$ is a KMS state, then, the uniqueness follows $T\Omega=c\Omega\ (c\in{\bf C})$.
Thus, we have
\begin{eqnarray*}
T\pi_\omega(A)\Omega=\pi_\omega (A) T\Omega=c\pi_\omega(A)\Omega
\ .
\end{eqnarray*}
Since $\pi_\omega(A)\Omega$ is arbitrary, we have $T=c{\bf 1}$.
Therefore, it is sufficient to prove that $T\Omega$ is a KMS state.
It follows from the following equality:
\begin{eqnarray*}
\Big\langle  T\Omega,\pi_\omega (A) \bar{\sigma}_{i\beta}(\pi_\omega(B)) T\Omega \Big\rangle 
&=&
\Big\langle  \Omega,T^* \pi_\omega (A) T\bar{\sigma}_{i\beta}(\pi_\omega(B)) \Omega \Big\rangle  \qquad (T\in \pi_\omega({\cal A})' )
\\
&=&
\Big\langle  \Omega,\pi_\omega(B) T^* \pi_\omega (A) T  \Omega \Big\rangle  
\qquad ({\rm KMS\ condition})
\\
&=&
\Big\langle  T\Omega,\pi_\omega(B) \pi_\omega (A) T  \Omega \Big\rangle   \qquad 
(T^*\in \pi_\omega({\cal A})' )
\end{eqnarray*}
Therefore, $T\Omega$ is $\sigma$-KMS.

\QED{}

\begin{Prop}[Bratteli Robinson~4.3.24 (Strong Mixing)]\ 

\label{Mixing}
  Let $\tau_t$ be an asymptotic abelian over $C^*$ algebra ${\cal A}$.
If a state $\omega$ is factor, then,
\begin{eqnarray*}
\lim_{|t|\to\infty}
\left\{
\omega\big( A\tau_t(B) \big)-\omega(A)\omega\big(\tau_t(B)\big)
\right\}=0
\qquad ({\rm Cluster\ Property})
\end{eqnarray*}

\end{Prop}
\underline{Explanation}\\
Let $({\cal H}_\omega,\pi_\omega,\Omega)$ be a GNS representation of 
$({\cal A},\omega)$, then, we need to prove
\begin{eqnarray*}
\Big\langle \Omega,
\pi_\omega(A)
\left\{
\pi_\omega\big( \tau_t(B) \big)-\omega\big(\tau_t(B)\big){\bf 1}
\right\}\Omega
\Big\rangle \to 0,
\qquad {\rm as}\ |t|\to \infty
\ .
\end{eqnarray*}
For this purpose, it is sufficient to prove
\begin{eqnarray*}
\Big\langle \psi,
\left\{
\pi_\omega\big( \tau_t(B) \big)-\omega\big(\tau_t(B)\big){\bf 1}
\right\}\phi
\Big\rangle \to 0,\qquad {\rm as}\ |t|\to \infty
\end{eqnarray*}
for any elements $\phi,\ \psi$ in ${\cal H}_\omega$.

Let $X_t\equiv \pi_\omega\big( \tau_t(B) \big)-
\omega\big(\tau_t(B)\big){\bf 1}$. Then, 
\begin{eqnarray*}
\Vert X_t \Vert\le \Vert \pi_\omega\big(\tau_t(B)\big)\Vert + |\omega(\tau_t(B))|\le 2\Vert B \Vert \le \infty
\end{eqnarray*}
yields that there exists a sequence $\{t_j \}_{j=1}^{\infty}$ 
and an element $C$ in ${\cal B}({\cal H})$ such that
\footnote{
In infinite systems, it is not possible to say the existence in a sense of norm, and $X_{t_j}\to C$ only in a weak sense. This property is called Tychonov theorem.}
\begin{eqnarray}
&&\lim_{j\to \infty} \big(\psi, (X_{t_j}-C)\phi\big)=0
\nonumber
\\
&&\lim_{j\to\infty}t_j = \infty
\ .
\label{divergent_sub}
\end{eqnarray}
It follows
\begin{eqnarray*}
\Big|\Big\langle  \psi, [C,\pi_\omega(A)]\phi \Big\rangle \Big| 
&=&
\Big|\lim_{j\to \infty}
\Big\langle \psi,
\big[\pi_\omega\big( \tau_{t_j}(B) \big),\pi_\omega (A)\big]\phi
\Big\rangle \Big|
\\ &=&
\lim_{j\to \infty} K\Vert [\tau_{t_j}(B),A]\Vert
\end{eqnarray*}
for some positive $K\in{\bf R}$.
Asymptotic abelian property follows $[C,\pi_\omega(A)]={\bf 0}$, thus, 
$C\in \pi_\omega ({\cal A})'$.
On the other hand, $X_{t_j}\in \pi_\omega({\cal A})+c{\bf 1}$ implies 
$C\in \pi_\omega ({\cal A})''$.
Therefore, $C$ is an element of ${\cal Z}({\cal A})$.
Since $\omega$ is a factor, we have $C=\alpha{\bf 1}$ for some $\alpha\in{\bf C}$.
It follows
\begin{eqnarray*}
\alpha=\langle \Omega,C\Omega\rangle 
=\lim_{j\to\infty}
\Bigg[
\Big\langle \Omega,\pi_\omega\big(\tau_{t_j}(B)\big)\Omega\Big\rangle -
\omega\big(\tau_{t_j}(B)\big)
\Big)\Bigg]=0
\end{eqnarray*}
Thus, we have $C={\bf 0}$, and it follows
\begin{eqnarray*}
\lim_{j\to\infty}(\psi,X_{t_j}\phi)=0
\end{eqnarray*}
for any ${t_j}_{j=1}^\infty$ which satisfies (\ref{divergent_sub}); therefore, we have
\begin{eqnarray*}
\lim_{t\to\infty}(\psi,X_{t}\phi)=0\ .
\end{eqnarray*}
\QED{}

With the aid of this proposition, we can discuss a condition with which perturbed system returns to equilibrium.
\begin{Prop}[Bratteli Robinson~5.4.10]\ 

 Let $\tau_t$ be an asymptotic abelian over $C^*$ algebra ${\cal A}$, and $\omega$ be a unique $(\tau,\beta)$-KMS state. Let $V$ be a local perturbation.
Then, we have
\begin{eqnarray*}
\lim_{|t|\to\infty}\omega\big( A\tau_t(B) \big)=\omega(A)\omega(B)
\label{Mixinng2}
\end{eqnarray*}

\end{Prop}
\underline{Explanation}\\
Because of a remark~2 after Def.~\ref{factor_Def}, we say that $\omega$ is a factor.
Thus, the assumption of the proposition~\ref{Mixing} is satisfied, and it follows
\begin{eqnarray*}
\lim_{|t|\to\infty}
\left\{
\omega\big( A\tau_t(B) \big)-\omega(A)\omega\big(\tau_t(B)\big)
\right\}=0
\ .
\end{eqnarray*}
We need to prove $\omega\big( \tau_t(B) \big)=\omega(B)$ for (\ref{Mixinng2}).
Here, let us just make a formal argument:
\begin{eqnarray*}
\omega\big(\tau_t(B)\big) &\sim& \frac{1}{Z_\beta}
{\rm tr}\ \left(e^{-\beta H}e^{iHt} B e^{-iHt}\right)
\\ &\sim&
\frac{1}{Z_\beta}
{\rm tr}\ \left(e^{-\beta H} B \right)
\\ &\sim&
\omega(B)
\end{eqnarray*}

 \begin{Prop}
 $\omega$ is said to be a grand canonical state if it is
 $\sigma$-KMS state for $\sigma_x=\tau_x\alpha_{-\mu x}$, where
$\alpha_s \sim e^{iNs} A e^{-iNs}$.
Let $\omega$ be a grand canonical state.
  If interaction $V$ is analytic for $\sigma_x$,  
 $\alpha\left(V\right)=V$ (We shall call it Gauge invariance), and 
 perturbative time evolution $\tau_t^V$ is asymptotic abelian.
Then, 
  \begin{equation*}
   \lim_{t\to\pm\infty} \omega\left(\tau_t^V \left(A\right)\right) = \omega_V \left(A\right)
  \end{equation*}
 is satisfied, where $\omega_V$ is a KMS state for $\sigma_x^V$, 
and $\tau_t^V, \sigma_x^V$ is defined by
  \begin{eqnarray*}
   \frac{d}{dt} \tau_t^V\left(A\right) &=& \tau_t^V \left\{\delta_0 \left(A\right) i\left[V,A\right] \right\},
    \,\, \tau_{t=0}^V \left(A\right) =A\\
   \frac{d}{dx} \tau_x^V\left(A\right) &=& \sigma_x^V
    \left\{\delta_0 \left(A\right) -\mu\kappa\left(A\right) + i\left[V,A\right] \right\},
    \,\, \sigma_{t=0}^V \left(A\right) =A\ ,
  \end{eqnarray*}
  where $\kappa$ is a generator of the gauge transformation.

  \end{Prop}
Physically, this proposition means that if reservoirs are locally perturbed, then, total system reaches a new equilibrium $\omega_V$.

\underline{explanation}

Formally speaking,
the average is expressed by
  \begin{eqnarray*}
   \omega &\sim& \frac{1}{\Theta} e^{-\beta\left(H-\mu N\right)}\\
   \omega_V &\sim& \frac{1}{\Theta_V} e^{-\beta\left(H+V-\mu N\right)}
 \ .
  \end{eqnarray*}
Let us formally define $\Gamma_\beta$ by $\Gamma_\beta \sim e^{-\beta \left(H-\mu N\right)} e^{-\beta \left(H+V-\mu N\right)}$\footnote{
We note that $\Gamma_\beta$ can be defined without $H$.}.
Then, we have
  \begin{eqnarray*}
   \Theta &\sim & tr e^{-\beta\left(H-\mu N\right)} = tr \Gamma_\beta e^{-\beta\left(H+V-\mu N\right)}\\
   \omega &\sim& \frac{1}{\Theta} e^{-\beta\left(H-\mu N\right)}
    = \frac{\Gamma_\beta e^{-\beta\left(H+V-\mu N\right)}}
    {tr \Gamma_\beta e^{-\beta\left(H+V-\mu N\right)}}\\
   &\sim& \frac{1}{\Theta} \frac{\Gamma_\beta}{\langle \Gamma_\beta\rangle _V}
    e^{-\beta\left(H+V-\mu N\right)}
\ .
  \end{eqnarray*}
Hence, 
  \begin{eqnarray*}
   \omega\left(A\right) = \frac{\omega_V\left(A\Gamma_\beta\right)}{\omega_V\left(\Gamma_\beta\right)}
\ .
  \end{eqnarray*}
Combining with $\tau_t^V\left(C\right)=e^{i\left(H+V\right)t} C e^{-i\left(H+V\right)t}$, we conclude
  \begin{equation*}
   \omega\left(\tau_t^V \left(A\right)\right)
    = \frac{\omega_V\left(\tau_t^V\left(A\right)\Gamma_\beta\right)}{\omega_V\left(\Gamma_\beta\right)}
    \xrightarrow{t \to \infty}
    \frac{\omega_V\left(\tau_t^V\left(A\right)\right) \omega_V\left(\Gamma_\beta\right)}{\omega_V\left(\Gamma_\beta\right)}
    = \omega_V\left(A\right)
\ .
  \end{equation*}

% $\omega$ is said to be a grand canonical state if it is
% $\sigma$-KMS state for $\sigma_x=\tau_x\alpha_{-\mu x}$, where
%$\alpha_s \sim e^{iNs} A e^{-iNs}$.

% \begin{Prop}
%Let $\omega$ be a grand canonical state.
%  If interaction $V$ is analytic for $\sigma_x$,  
% $\alpha\left(V\right)=V$ (We shall call it Gauge invariance), and 
% perturbative time evolution $\tau_t^V$ is asymptotic abelian.
%Then, 
%  \begin{equation}
%   \lim_{t\to\pm\infty} \omega\left(\tau_t^V \left(A\right)\right) = \omega_V \left(A\right)
%  \end{equation}
% is satisfied, where $\omega_V$ is a KMS state for $\sigma_x^V$, 
%and $\tau_t^V, \sigma_x^V$ is defined by
%  \begin{eqnarray}
%   \frac{d}{dt} \tau_t^V\left(A\right) &=& \tau_t^V \left\{\delta_0 \left(A\right) i\left[V,A\right] \right\},
%    \,\, \tau_{t=0}^V \left(A\right) =A\\
%   \frac{d}{dx} \tau_x^V\left(A\right) &=& \sigma_x^V
%    \left\{\delta_0 \left(A\right) -\mu\kappa\left(A\right) + i\left[V,A\right] \right\},
%    \,\, \sigma_{t=0}^V \left(A\right) =A\ ,
%  \end{eqnarray}
%  where $\kappa$ is a generator of the gauge transformation.

%  \end{Prop}

\subsection{Nonequilibrium steady state}
As we have mentioned, the method of $C^*$ algebra was first applied to equilibrium systems and thermodynamics.
For instance, 
the first and second law of thermodynamics\cite{Tasaki06,PW05}, 
the relative-entropy reformulated by Ojima\cite{Oji89} was shown to be non-negative at steady state\cite{Rue01,JP01,Oji89,OHI88}, where its existence is proved under $L^1$-asymptotic abelian property\cite{Rue00,JP02a,JP02b}.

In this subsection, we will concern nonequilibrium state, and show some useful results without giving rigorous proofs.

The system is coupled to some reservoirs, and reservoirs are in different equilibria. Here, we are interested in a finite systems coupled to some reservoirs.
Those reservoirs are assumed to be in equilibrium at initial state, but we do assume this property only at the initial state.
To be more precise, initially state are written in terms of a product state 
$\rho_{tot}=(\rho_0)_{system} \otimes \big(\rho_{1}\otimes \rho_{2}\otimes \cdots\otimes \rho_{N})_{reservoirs}$, and each $\rho_{i} (i\ge 1)$ is 
in equilibrium at initial time.
We assume that at $t=0-$ reservoirs are in different equilibria, and we contact finite system to reservoirs at $t=0+$.
We are interested in a finite system at $t=+\infty$.

Let us start from defining a Field algebra~$\mathfrak{F}$ (One can find detailed discussion in \cite{TM03}).
This Field algebra is defined to be a $C^*$ algebra having 
the following *-automorphisms:
 \begin{description}
  \item[1.] Time-evolution~$\tau_t$
%	     \begin{equation}
%	      \tau_t \sim e^{iHt} \cdot e^{-iHt}
%	     \end{equation}
  \item[2.] Gauge transformation~$\alpha_{\vec{\phi}}$ which satisfies
%	     \begin{equation}
%	      \alpha_{\vec{\phi}} \sim \exp\left[i\sum \phi_\lambda N_\lambda\right]
%	       \cdot \exp\left[-i\sum \phi_\lambda N_\lambda\right]
%	     \end{equation}
%	     $B$O(B
	     $\alpha_{\vec{\phi_1}} \alpha_{\vec{\phi_2}}
	     = \alpha_{\vec{\phi_1}+\vec{\phi_2}}$
	     ,\ $\vec{\phi} \in \vec{R}^L$.
  \item[3.]
	     Involutive transformation $\Theta=\alpha_{\vec{\phi_0}}$.
	     %Fermi$BN3;R$N6v4q(B $\Theta$
	     %\begin{equation}
	     % \Theta^2 = \vec{1}
	     %\end{equation}
  \item[4.] Time reversal operator:\ $\iota$
	     \begin{equation*}
	      \iota \tau_t \iota = \tau_{-t}, \,\, \iota^2 = \vec{1}
	     \end{equation*}
 \end{description}
 We note that transformation 1, 2, and 3 commute each other.

Intuitively, time-evolution $\tau_t$ reads
 \begin{equation*}
  \tau_t\left(A\right) \sim e^{iHt} A e^{-iHt}\ ,
 \end{equation*}
gauge transformation 
$\alpha_{\vec{\phi}}$ reads 
 \begin{equation*}
  \alpha_{\vec{\phi}}\left(A\right)
   = \exp\left[i\sum \phi_\lambda N_\lambda\right] A \exp\left[-i\sum \phi_\lambda N_\lambda\right]\ ,
 \end{equation*}
and $\Theta$ represents a parity of fermion.

Next, we split total system into system and reservoirs.
 \begin{description}
  \item[1.] Observables:
	     \begin{equation*}
	      \mathfrak{F} = \left(\mathfrak{F}_0\right)_{system} \otimes
	       \left(\mathfrak{F}_1 \otimes \cdots \otimes \mathfrak{F}_N\right)_{reservoirs}
	     \end{equation*}
  \item[2.]
	Time-evolution:
	     \begin{equation*}
	      \tau_t^V = \tilde{\tau}_t^{\left(1\right)} \otimes
	       \cdots \otimes \tilde{\tau}_t^{\left(N\right)}\ ,
	     \end{equation*}
	construct from a generator:
	     \begin{equation*}
	      \delta^V\left(A\right)=\delta\left(A\right)-i\left[V,A\right]\ ,
	     \end{equation*}
	where $\tilde{\tau}_t^{\left(j\right)}$ acts on $\mathfrak{F}_j$, and
	  $\tilde{\tau}_t^{\left(j\right)}$ satisfies
    \begin{eqnarray*}
	      \tilde{\tau}_t^{\left(j\right)} \left(A\right) &=& A, \,\,\,\left(\forall A\in\mathfrak{F}_k,\,\
	       j \ne k\right)\\
	      \tilde{\tau}_t^{\left(j\right)}\tilde{\tau}_s^{\left(k\right)} &=& \tilde{\tau}_s^{\left(k\right)}\tilde{\tau}_t^{\left(j\right)} \,\,
	       \left(t,s \in \vec{R}, \,\, j \ne k\right)\ .
	     \end{eqnarray*}
  \item[3.]
	Gauge transformation:
	     \begin{equation*}
	      \alpha_{\vec{\phi}}=\tilde{\alpha}_{\vec{\phi}}^{\left(0\right)}
	       \otimes \tilde{\alpha}_{\vec{\phi}}^{\left(1\right)} \otimes \cdots
	       \otimes \tilde{\alpha}_{\vec{\phi}}^{\left(N\right)}\ .
	     \end{equation*}
	Similarly, it satisfies
	     \begin{eqnarray*}
	      \tilde{\alpha}_{\vec{\phi}}^{\left(j\right)} \left(A\right) &=& A, \,\,\,\left(\forall A\in\mathfrak{F}_k,\,\
	       j \ne k\right)\\
	      \tilde{\alpha}_{\vec{\phi}_1}^{\left(j\right)}
	       \tilde{\alpha}_{\vec{\phi}_2}^{\left(k\right)} &=&
	       \tilde{\alpha}_{\vec{\phi}_2}^{\left(k\right)}
	       \tilde{\alpha}_{\vec{\phi}_1}^{\left(j\right)} \,\,
	       \left(\vec{\phi}_1,\vec{\phi}_2 \in \vec{R}, \,\, j \ne k\right)
		\ .
	     \end{eqnarray*}
 \end{description}
Moreover,
 \begin{equation*}
  \tilde{\tau}_t^{\left(j\right)}\tilde{\alpha}_{\vec{\phi}}^{\left(k\right)}
   = \tilde{\alpha}_{-\vec{\phi}}^{\left(j\right)} \tilde{\tau}_t^{\left(k\right)}, \,\,
   \left(\forall j,k=1,\cdots,N,\,t\in\vec{R},\vec{\phi}\in\vec{R}^L\right)
 \end{equation*}
holds, and time-reversal operator $\iota$ satisfies
 \begin{equation*}
  \iota\tilde{\tau}_t^{\left(j\right)}\iota = \tilde{\tau}_{-t}^{\left(j\right)},\,\,\,\,
   \iota\tilde{\alpha}_t^{\left(j\right)}\iota = \tilde{\alpha}_{-t}^{\left(j\right)}\ .
 \end{equation*}

Since $\tau_t^V$ was separated as stated below, we have $D\left(\delta^V\right)=D\left(\delta\right)$ (see \cite{TM03,TT06c} for detailed discussion).
It follows
 \begin{eqnarray*}
  \delta\left(A\right) &=& \delta^V \left(A\right) + i\left[V,A\right] ,\,\, \forall A\in D\left(\delta\right)\\
  \delta^V\left(A\right) &=& \sum^N_{j=1} {\delta}_j \left(A\right) ,\,\, \forall A\in D\left(\delta\right)\ ,
 \end{eqnarray*}
where ${\delta}_j$ is a generator of $\tilde{\tau}_t^{\left(j\right)}$.

The following classes of subalgebra of $\mathfrak{F}$ are important:
 \begin{description}
  \item[1.] Subalgebra of observables:\ 
     $\mathfrak{A} \in \mathfrak{F}$, a set of gauge invariant elements.
  \item[2.] Parity subalgebra:\ 
	     \begin{equation*}
	      \mathfrak{F}_\pm = \left\{A \in \mathfrak{F}
				  | \Theta\left(A\right) = \pm A \right\}
	     \end{equation*}
	Observables with odd/even annihilation and creation operators.

  \item[3.]
	Subalgebra $\mathfrak{F}_L$ is norm dense and it satisfies
	     \begin{eqnarray*}
	      && \int_{-\infty}^{+\infty} dt \|\left[A,\tau_t\left(B\right)\right]\| < +\infty
	       \,\, \left(A\in\mathfrak{F}_L, \, B \in\mathfrak{F}_L \cap \mathfrak{F}_+ \right),\\
	       && \int_{-\infty}^{+\infty} dt \|\left[A,\tau_t\left(B\right)\right]_+\| < +\infty
	       \,\, \left(A, B \in\mathfrak{F}_L \cap \mathfrak{F}_- \right).
	     \end{eqnarray*}
	  If $\mathfrak{F}_L$ exists, then, time-evolution $\tau_t$ is said to satisfy $L^1\left(\mathfrak{F}_L\right)$ asymptotic abelian.
For instance, the interaction between system and reservoir $V$ is an element of $\mathfrak{F}_L$. This property implies rapid decay of correlations and is satisfied for free fermions in ${\bf R}^d (d\ge 1)$ (Bratteli Robinson~5.4.9).

 \end{description}

Each reservoir is infinitely extended, and is in a equilibrium with different thermodynamic valuables such as temperatures and chemical potentials.
To describe such state, we characterize each reservoir by KMS condition.

 Let us define $\sigma_x^\omega$ by
  \begin{equation}
  \sigma_{x}^\omega \left(B\right) \equiv \prod^N_{j=1} \tilde{\tau}^{\left(j\right)}_{-\beta_jx}
   \tilde{\alpha}^{\left(j\right)}_{-\beta_j\vec{\mu}_jx} \left(e^{iD_sx} Be^{-iD_sx} \right)\ ,
 \end{equation}
 where $\beta_j^{-1}$ is a inverse temperature of $j$th reservoir, 
 $\vec{\mu}_j=\left(\mu_j^{\left(1\right)},\cdots,\mu_j^{\left(L\right)}\right)$ is a chemical potential of $j$th reservoir, and 
 $\exp \left(D_s\right)$ represents a initial state of a finite system.

We are interested in a case which $\omega$ is a 
 $(\sigma_x^\omega,-1)$ KMS state, i.e., 
 \begin{equation*}
  \omega\left(A \sigma_{-i}^\omega \left(B\right)\right) = \omega\left(AB\right) \end{equation*}

Then, generator  ${\delta}_\omega$ is given by
 \begin{equation*}
  {\delta}_\omega\left(A\right) =
   -\sum \beta_j \left(\beta_j
   \left({\delta}_j\left(A\right)-\mu_\lambda^{\left(j\right)}{g_\lambda^{\left(j\right)}}\left(A\right)\right)\right)+ i\left[D_S,A\right]\ ,
 \end{equation*}
where ${g_\lambda^{\left(j\right)}}$ is a generator for ${\alpha}^{\left(j\right)}_{s\vec{e}_\lambda}$ ($\vec{e}_{\lambda}\in{\bf R}^L$ is the unit vector whose $\lambda$th element is 1, and $s$ is a real number.).
We note that the KMS state for $\sigma^\omega_x$ corresponds to the Maclennan-Zubarev ensembles\footnote{The problem of divergence in Maclennan-Zubarev ensemble\cite{Zubarev95,Maclennan63} was reformulated by Tasaki\cite{TM03}. 
The validity of the ensemble was proved for spinless electron
model of a single-level quantum dot\cite{TT06c}. } 
in an appropriate sense\cite{TM03,TT06c}.

As we shall describe in Proposition \ref{partitionNESS}
, we are interested in the effects of the way of splitting systems.
For this purpose, we introduce a 
locally modified state $\omega'$.
Namely, we consider a different partition:
 \begin{equation*}
  \mathfrak{F} = \mathfrak{F}_0'\otimes
	       \mathfrak{F}_1' \otimes \cdots \otimes \mathfrak{F}_N'
\ ,
 \end{equation*}
and define $\omega'$ as a ($\sigma_x^{\omega'}$,-1)~KMS state for this partition. 
In the same way, we introduce $\sigma_{x}^{\omega'}$ as
 \begin{equation*}
  \sigma_{x}^{\omega'} \left(B\right) = \prod^N_{j=1} \tilde{\tau}'^{\left(j\right)}_{-\beta_j x}
   \tilde{\alpha}'^{\left(j\right)}_{-\beta_j\vec{\mu}_jx} \left(e^{iD_s'x} Be^{-iD_s'x} \right)\ .
 \end{equation*}

This state $\omega'$ is defined to be a locally modified state, if 
${\delta}_{\omega'}$ and ${\delta}_\omega$ satisfy
 \begin{equation*}
  {\delta}_{\omega'}\left(A\right) - {\delta}_\omega\left(A\right) =
   i\left[W,A\right], \,\,\, \forall A \in D\left({\delta}_\omega\right),\ 
 \end{equation*}
for some selfadjoint element $W\in \mathfrak{F}$.

 \begin{Prop}[ Existence of steady states (Ruelle \cite{Rue00})\ ]\ 

 Suppose that the time evolution $\tau_t$ satisfies the $L^1\left(\mathfrak{F}_L\right)$
property, then, 
  \begin{equation}
   \lim_{t\to\pm\infty} \omega \circ \tau_t\left(A\right) \equiv \omega_\pm\left(A\right),\,\,
    \left(\forall A \in \mathfrak{F}_L\right)
\label{natural steady state}
  \end{equation}
 exists for each initial condition $\omega$, and $\omega_\pm$ is $\tau_t$-invariant.%\cite{ruelle00}
 \end{Prop}
 
 \begin{Prop}[ Tasaki and Matsui theorem~2 in \cite{TM03} ]
\label{partitionNESS}
 Suppose that the time evolution $\tau_t$ satisfies the $L^1\left(\mathfrak{F}_L\right)$
property, and there exists a unique KMS state for $\sigma_x^\omega$.
Then, for locally perturbed $\omega'$, the previous equality holds:
  \begin{equation*}
   \lim_{t\to\pm\infty} \omega' \circ \tau_t\left(A\right)
    =\lim_{t\to\pm\infty} \omega \circ \tau_t\left(A\right)
    =\omega_\pm\left(A\right),\,\, \left(\forall A \in \mathfrak{F}_L\right)
  \end{equation*}

 \end{Prop}
Interestingly, proposition~\ref{partitionNESS} concludes that the steady state 
$\omega_\pm$ is determined by thermodynamic properties such as 
temperature and chemical potential, and it does not depend on
the partition, initial condition in the finite system.

 \begin{Prop}[ Tasaki and Matsui theorem~3 in \cite{TM03} ]
 Suppose that the time evolution $\tau_t$ satisfies the $L^1\left(\mathfrak{F}_L\right)$
property, and there exists a unique KMS state for $\sigma_x^\omega$.
Then, the following equality holds:
  \begin{equation*}
   \omega_\pm=\iota^* \omega_\mp
\ ,
  \end{equation*}
where $\iota^*$ is defined by
  $\iota^* \omega\left(A\right) = \omega\left(\iota\left(A^*\right)\right)$.

 \end{Prop}

 \begin{Prop}[ Stability of steady states (Tasaki and Matsui theorem~4 in \cite{TM03})\ ]\ 

 Suppose that the time evolution $\tau_t$ satisfies the $L^1\left(\mathfrak{F}_L\right)$
property, and there exists a unique KMS state for $\sigma_x^\omega$.
If  M{\o}ller operator\cite{BR02,Rue00,TM03,TT06c}
$\gamma_\pm \left(\equiv \lim_{t\to \pm \infty} {\tau_t^{V}}^{-1} \tau_t\right)$ is invertible, then, a steady state $\omega_+$
is stable against local perturbations in the following sense:
  \begin{equation}
   \lim_{t\to\pm \infty}\frac{\omega_+ \left(B^* \tau_t\left(A\right)B\right)}{\omega_+\left(B^*B\right)}
    = \omega_+\left(A\right), \,\, \left(\forall A,B \in \mathfrak{F}\right).
  \end{equation}
The same arguments hold for $\omega_-$.
 \end{Prop}

\section{Landauer formula and the existence of unique steady states for bilinear hamiltonian\label{sec_Landauer} }
In this section, we will study particle current and condition with which unique steady state exists.
We use rigorous results of $C^*$ algebra but try to  demonstrate an application in more physical manner. 
We also describe a hamiltonian having a infinite norm; however as we explained, it is possible to define time evolution without using the hamiltonian (One can use the rigorous discussion of single dot couple to two reservoirs \cite{TT06c}).
The hamiltonian is used only to formally give a time evolution over $C^*$ algebra.
In this section, We will restrict our analysis on a finite bilinear hamiltonian coupled to infinite reservoirs:
\begin{eqnarray}
H &=& H_S+H_B+V
\nonumber
\\
H_S &=& \sum_\lambda \epsilon_\lambda f_\lambda^\dag f_\lambda
\nonumber
\\
H_B &=& \int\ dk 
\left(\omega_{k} a_{k}^\dag a_{k}
+
\mu_{k} b_{k}^\dag b_{k}
\right)
\nonumber
\\
V &=& \sum_{\lambda}\int dk\  
\left(
u^{L}_{k} w^{L}_{\lambda} 
a_{k}^\dag f_\lambda
+
u^{R}_{k} w^{R}_{\lambda} 
b_{k}^\dag f_\lambda + {\rm (h.c.)}
\right)
\label{many levelH}\ ,
\end{eqnarray}
where $u^\nu_{k}w^{\nu}_{\lambda}\ (\nu=L,R,\ \lambda\in{\bf N},\ k\in{\bf R}^2)$ is a tunneling between system and reservoirs, $f_\lambda$ is an annihilation operator of an fermion in finite system with energy $\epsilon_\lambda$, and $a_{k}$/$b_{k}$ is an annihilation operator of left/right reservoir fermion with energies $\omega_{k}/\mu_{k}$ and wave number $k$.

\subsection{Particle current and Landauer formula}
Let us study following operators:
\begin{eqnarray}
\alpha_k=a_k + \sum_\lambda h^k_\lambda f_\lambda
+ \int dk'\ \left(m^k_{k'} a_{k'} + n^k_{k'} b_{k'} \right)
\nonumber
\\
\beta_k=a_k + \sum_\lambda \bar{h}^k_\lambda f_\lambda 
+ \int dk'\ \left(\bar{m}^k_{k'} a_{k'} + \bar{n}^k_{k'} b_{k'} \right)
\label{incoming}
\end{eqnarray}

Suppose that this set of operators~$\{\alpha_k,\beta_k\}$ is complete (It will be studied in \S~\ref{sec_uniqueness}), then, $a_k$ for instance should be written as a summation of this operator, i.e.,
\begin{eqnarray}
a_k=\int dk'\ 
\left( A^k_{k'} \alpha_{k'} + B^k_{k'} \beta_{k'} \right)
\end{eqnarray}
Then, $A^k_{k'}, B^k_{k'}$ should be given by
\begin{eqnarray*}
A^k_{k'}&=&\left\{
a_{k'}^\dag + \sum_\lambda h^{k'*}_\lambda f^\dag_\lambda
+ \int dk''\ \left(m^{k'*}_{k''} a_{k''}^\dag + n^{k'*}_{k''} b_{k''}^\dag \right)
,a_k\right\}
\\
&=& \delta(k-k')+m^{k'*}_{k}
\\
B^k_{k'}&=&\left\{
b_{k'}^\dag + \sum_\lambda \bar{h}^{k'*}_\lambda f^\dag_\lambda
+ \int dk''\ \left(\bar{m}^{k'*}_{k''} a_{k''}^\dag + \bar{n}^{k'*}_{k''} b_{k''}^\dag \right)
,a_k\right\}
\\
&=& \bar{m}^{k'*}_{k}
\ ,
\end{eqnarray*}
where we have used
\begin{eqnarray*}
\{\alpha_{k'}^\dag , a_k\} &=&
\int dk''\ \{\alpha_{k'},\alpha_{k''}\} A^k_{k''} =A^k_{k'}
\\
\{\beta_{k'}^\dag , a_k\} &=&
\int dk''\ \{\beta_{k'},\beta_{k''}\} B^k_{k''} =B^k_{k'}
\ .
\end{eqnarray*}
Similarly, inverse formula for $f_\lambda$ and $b_k$ can be obtained.
As a result, if $\{\alpha_k,\ \beta_k\}$ is complete, then, the inverse formula reads
\begin{eqnarray*}
a_k &=&\alpha_k+
\int dk'\ \left(m^{k'*}_{k} \alpha_{k'} + \bar{m}^{k'*}_{k} \beta_{k'} \right)
\\
b_k &=&
\beta_k+
\int dk'\ \left(n^{k'*}_{k} \alpha_{k'} + \bar{n}^{k'*}_{k} \beta_{k'} \right)
\\
f_\lambda &=&
\int dk\ \left( h^{k*}_{\lambda} \alpha_{k} + \bar{h}^{k*}_{\lambda} \beta_{k} \right)
\ .
\end{eqnarray*}
Next, let us introduce an incoming field. Incoming field is an elements of $C^*$ algebra which formally satisfies
\begin{eqnarray*}
&& [\alpha_{k}, H ]=\omega_{k} \alpha_{k}
\ ,\ \ \ 
e^{iHt} a_{k} e^{-iHt} \ e^{i\omega_{k} t} 
\to \alpha_{k}
\
(t\to -\infty)
\nonumber 
\\
&&[\beta_{k}, H ]=\mu_{k}\beta_{k}
\ ,\ \ \ 
e^{iHt}  b_{k} e^{-iHt} \ e^{i\mu_k t} 
\to \beta_{k} \
(t\to -\infty) \
\ .
\label{incoming_def}
\end{eqnarray*}
For bilinear hamiltonian~(\ref{many levelH}), it should be written in the form~(\ref{incoming}), and $[\alpha_k,H]=\omega_k\alpha_k$ reads
\begin{eqnarray}
&& 
u^L_k w^L_\lambda
+ h^k_\lambda\epsilon_\lambda 
+
\int dk'\ m^k_{k'} u^L_{k'} w^L_\lambda
+
\int dk'\ n^k_{k'} u^R_{k'} w^R_\lambda
=
\omega_k h^k_\lambda
\ ,
\label{incomingTf}
\\
&&
u^{L*}_{k'} A^L_k  + m^k_{k'}
\omega_{k'}
=\omega_k m^k_{k'} 
\ ,
\label{incomingTa}
\\
&& 
u^{R*}_{k'} A^R_k 
+
n^k_{k'}\mu_{k'}
=
\omega_k n^k_{k'}
\ .
\label{incomingTb}
\end{eqnarray}
From (\ref{incomingTa}) and (\ref{incomingTb}), we obtain
\begin{eqnarray*}
 m^k_{k'} =
\frac{u^{L*}_{k'} A^L_k}{\omega_k-\omega_{k'}\pm i0}
\ ,
\\
 n^k_{k'}
=
\frac{u^{R*}_{k'} A^R_k }{\omega_k-\mu_{k'}\pm i0}
\ .
\end{eqnarray*}
By substituting the above equality into (\ref{incomingTf}), we have
\begin{eqnarray}
(\omega_k- \epsilon_\lambda ) h^k_\lambda 
&=&
u^L_k w^L_\lambda
+
\int dk'\ \frac{u^{L*}_{k'} A^L_k}{\omega_k-\omega_{k'}\pm i0}
 u^L_{k'} w^L_\lambda
+
\int dk'\ 
\frac{u^{R*}_{k'} A^R_k }{\omega_k-\mu_{k'}\pm i0}
 u^R_{k'} w^R_\lambda
\nonumber
\\
&=&
u^L_k w^L_\lambda
+
A^L_k w^L_\lambda
\xi_\pm^L(\omega_k)
+
A^R_k w^R_\lambda
\eta_\pm^R(\omega_k)
\ ,
\label{forA}
\end{eqnarray}
where $\xi_\pm(\omega)$ and $\eta_\pm(\omega)$ are defined by
\begin{eqnarray*}
\xi_\pm(\omega)&\equiv& \integ{k'}\frac{|u^L_{k'}|^2}{\omega-\omega_{k'}\pm i0}
\ ,\ \ \ 
\eta_\pm(\omega)\equiv \integ{k'}\frac{|u^R_{ k'}|^2}{\omega-\mu_{k'}\pm i0}
\ \ .
\end{eqnarray*}
We only need to determine $A^\sigma_k$. By substituting (\ref{forA}) into definition of $A^\sigma_k$, we obtain
\begin{eqnarray*}
A_k^\sigma&=&
\sum_\lambda \left(
\frac{u^L_k w^L_\lambda}{\omega_k-\epsilon_\lambda \pm i0}
+
\frac{A^L_k w^L_\lambda
\xi_\pm^L(\omega_k)}{\omega_k-\epsilon_\lambda \pm i0}
+
\frac{A^R_k w^R_\lambda
\eta_\pm^R(\omega_k)}{{\omega_k-\epsilon_\lambda \pm i0}}
\right)
w_\lambda^{\sigma*}
\\
&=&
 u^L_k S_{L\sigma}(\omega_k)
+\xi^L_\pm(\omega_k)S_{L\sigma}(\omega_k)A_k^L
+\eta^R_\pm(\omega_k)S_{R\sigma}(\omega_k)A_k^R
\ .
\end{eqnarray*}
Thus, we have two equations for $A^\sigma_k$:
\begin{eqnarray*}
A^L_k &=&
 u^L_k S_{LL}(\omega_{k})
+\xi_\pm(\omega_{k})S_{LL}(\omega_{k})A_k^L
+\eta_\pm(\omega_{k})S_{RL}(\omega_{k})A_k^R
\\
A_k^R&=& u^L_k S_{LR}(\omega_{k})+\xi_\pm(\omega_{k})S_{LR}(\omega_{k})A_k^L
+\eta_\pm(\omega_{k})S_{RR}(\omega_{k})A_k^R\ ,
\end{eqnarray*}
where $S_{\nu\nu'}(\omega)$ is defined by
\begin{eqnarray*}
S_{\nu\nu'}(\omega)\equiv\sum_\lambda
\frac{w_\lambda^\nu w_\lambda^{\nu'*}}
{\omega-\epsilon_\lambda+\pm i0}\ .
\end{eqnarray*}
The solution is
\begin{eqnarray*}
\mat{ A^L_k }{ A^R_k } 
&=&
\frac{u^L_k }{\Lambda_\pm(\omega)}
\mat{ S_{LL}(\omega) +\eta_\pm(\omega) \left\{ |S_{LR}(\omega)|^2-S_{RR}(\omega)  S_{LL}(\omega)\right\} }
{ S_{LR} (\omega)}\Bigg|_{\omega=\omega_k}
\\
\Lambda_\pm(\omega) &\equiv&
1-\left\{\eta_\pm(\omega) S_{RR}(\omega) + \xi_\pm(\omega) S_{LL}(\omega)\right\}
\\&&\mskip 30 mu+
\xi_\pm(\omega) \eta_\pm(\omega) \left\{S_{LL}(\omega)S_{RR}(\omega)-|S_{LR}(\omega)|^2 \right\}
\end{eqnarray*}
Similarly, we have the condition for 
$\bar{h}^k_\lambda,\ \bar{m}^k_{k'},\ \bar{n}^k_{k'}$ to satisfy $[\beta_k,H]=\mu_k\beta_k$.
In short, the conditions for 
$[\alpha_k,H]=\omega_k \alpha_k,\ [\beta_k,H]=\mu_k \beta_k$, are summarized as follows.
\begin{eqnarray}
 m^k_{k'} &=&
\frac{u^{L*}_{k'} A^L_k}{\omega_k-\omega_{k'}\pm i0}
\nonumber \\
 n^k_{k'}
&=&
\frac{u^{R*}_{k'} A^R_k }{\omega_k-\mu_{k'}\pm i0}
\nonumber \\
h^k_\lambda 
&=&
\frac{u^L_k w^L_\lambda}{\omega_k- \epsilon_\lambda }
+
\frac{A^L_k w^L_\lambda}{\omega_k- \epsilon_\lambda }
\xi_\pm(\omega_k)
+
\frac{A^R_k w^R_\lambda
}{\omega_k- \epsilon_\lambda }\eta_\pm(\omega_k)
\nonumber \\
\bar{ m}^k_{k'} &=&
\frac{u^{L*}_{k'} \bar{A}^L_k}{\mu_k-\omega_{k'}\pm i0}
\nonumber \\
\bar{ n}^k_{k'}
&=&
\frac{u^{R*}_{k'} \bar{A}^R_k }{\mu_k-\mu_{k'}\pm i0}
\nonumber \\
\bar{h}^k_\lambda 
&=&
\frac{u^R_k w^R_\lambda}{\mu_k- \epsilon_\lambda }
+
\frac{\bar{A}^L_k w^L_\lambda}{\mu_k- \epsilon_\lambda }
\xi_\pm(\mu_k)
+
\frac{\bar{A}^R_k w^R_\lambda
}{\mu_k- \epsilon_\lambda }\eta_\pm(\mu_k)
\ ,
\label{incoming_formula}
\end{eqnarray}
where $\bar{A}^\nu_k$ is defined by
\begin{eqnarray*}
\mat{ \bar{A}^L_k }{ \bar{A}^R_k } &=&
\frac{u^R_k }{\Lambda_\pm(\omega)}
\mat{ S_{RL} (\omega)}
{ S_{RR}(\omega) +\xi_\pm(\omega) \left\{ |S_{LR}(\omega)|^2-S_{RR}(\omega)  S_{LL}(\omega)\right\} }
\Bigg|_{\omega=\mu_k}
\ .
\end{eqnarray*}
Moreover, if this field is complete, then, the inverse formula reads
\begin{eqnarray*}
a_k &=&\alpha_k+
\int dk'\ \left(m^{k'*}_{k} \alpha_{k'} + \bar{m}^{k'*}_{k} \beta_{k'} \right)
\\
b_k &=&
\beta_k+
\int dk'\ \left(n^{k'*}_{k} \alpha_{k'} + \bar{n}^{k'*}_{k} \beta_{k'} \right)
\\
f_\lambda &=&
\int dk\ \left( h^{k*}_{\lambda} \alpha_{k} + \bar{h}^{k*}_{\lambda} \beta_{k} \right)
\ .
\end{eqnarray*}
Sign of the denominator is determined by 
the condition~$e^{iHt} a_{k} e^{-iHt} 
e^{i\omega_{k} t}\to
\alpha_{k}\ (t\to -\infty)$:
\begin{eqnarray}
&&e^{iHt}a_{k}e^{-iHt} e^{i\omega_{k} t}-\alpha_{k} 
\nonumber
\\
&&~~~~=\integ{k'} \Big(
\frac{u^{L*}_{k} A^L_{k'}\alpha_{k'}e^{-i(\omega_{k'}-\omega_{k})}}{\omega_{k'}-\omega_{k}\pm i0} 
+\frac{u^{R*}_{k} A^R_{k'}\beta_{k'}e^{-i(\mu_{k'}-\omega_{k})}}{\omega_{k'}-\mu_{k}\pm i0}
\Big) \ ,
\label{vanishing term}
\end{eqnarray}
and its counterpart for $\beta_k$.
Since the term~(\ref{vanishing term}) should vanish in the limit of $t\to -\infty$, 
the sign in denominator should be taken minus, where we have used 
$\displaystyle\lim_{t\to -\infty} {e^{-ixt}\over x+i0}=0$.
The same applies to the right reservoir ($\beta_k$).

Following the idea of Ruelle\cite{Rue00}, if the reservoirs are 
initially
set to be in different equilibria and a set of incoming fields~$\{\alpha_{k},\ \beta_{k}\}$ is complete, then, the whole system is shown to 
approach a NESS in the long time limit\cite{TT05,TT06a,TT06c}.
The NESS so obtained
can be characterized as a state satisfying Wick's theorem with respect to $\alpha_{k}$ 
and $\beta_{k}$ and having the two-point functions:
\begin{equation}
\langle \alpha_{k}^\dag \alpha_{k'} \rangle_\infty
= f_L(\omega_{k}) \delta(k-k') \ ,
\ \
\langle \beta_{k}^\dag \beta_{k'} \rangle_\infty
= f_R(\mu_{k}) \delta(k-k') \ ,
\label{ChNESS}
\end{equation}
where 
$\langle \cdot\rangle_\infty$ represent a NESS average~$\langle \cdot\rangle_\infty\equiv \omega_+(\cdot)$ in a sense of (\ref{natural steady state}),
$f_\nu(x)\equiv 1/(e^{(x-\mu_\nu)/T_\nu}+1)$ is the Fermi distribution function, $T_\nu$ 
is the initial temperature and $\mu_\nu$ is the initial chemical potential of the reservoir $\nu=L,R$.
Formally, this can be understood as follows:\footnote{The very proof of the existence of the limits
requires rigorous and careful arguments\cite{TT06c}.}
Let $\rho_0$ be an initial density matrix, where the two reservoirs are in distinct equilibria, and
$\rho_{\infty}$ be that of the NESS, 
then, $\lim_{t\to +\infty}e^{-iHt}\rho_0e^{iHt}=\rho_\infty$ and,
e.g., ${\rm Tr}\{a_{k}^\dag a_{k'}\rho_0\}
=f_L(\omega_{k}) \delta(k-k')$. 
As the incoming field $\alpha_{k}$ is given by $\lim_{t\to +\infty}
e^{i\omega_{k}(-t)}e^{iH(-t)}a_{k}e^{-iH(-t)}=\alpha_{k}$,
one obtains the desired relation:
\begin{eqnarray}
&&f_L(\omega_{k}) \delta(k-k')
={\rm Tr}\{a_{k}^\dag a_{k'}\rho_0\}e^{i(\omega_{k}-\omega_{k'})t}
\nonumber\\
&&
={\rm Tr}\Big\{
\{e^{iH(-t)}a_{k}e^{-iH(-t)}e^{i\omega_{k}(-t)}\}^\dag 
e^{iH(-t)}a_{k'}e^{-iH(-t)}e^{i\omega_{k'}(-t)}
e^{-iHt}\rho_0 e^{iHt}
\Big\}
\nonumber\\
&&\to 
{\rm Tr}\{\alpha_{k}^\dag \alpha_{k'}\rho_\infty\}
\equiv \langle \alpha_{k}^\dag \alpha_{k'}\rangle_\infty
\quad ({\rm as \ } t\to +\infty) \ .
\end{eqnarray}
Finally, if the incoming field is complete, then, the particle current~$J$ reads
\begin{eqnarray*}
J &\equiv& 
\Big\langle
\integ{k}
\frac{\partial }{\partial t}
\left(a_{k}^\dag a_{k}\right)
\Big\rangle_\infty
\\
&=&
-i
\sum_\lambda \integ{k}
\ave{
u^L_{k} w_\lambda^L 
a_k^\dag f_\lambda
-
u^{L*}_{k} w_\lambda^{L*} 
f_\lambda^\dag a_k
}_\infty
\\
&=&
2\ {\rm Im}
\sum_\lambda \left[
\integ{k}
u^L_{k} w_\lambda^L 
h^{k*}_{\lambda} f_L(k)
+
\int dk\ dk'\ 
u^L_{k} w_\lambda^L 
\left\{
m^{k'}_{k}h^{k'*}_{\lambda}f_L(k')
+
\bar{m}^{k'}_{k} \bar{h}^{k'*}_{\lambda}
f_R(k')
\right\}
\right]
\\
&=&
2{\rm Im}
\integ{k}
u^L_{k}
A^{R*}_k S_{LR}(\omega_k)
\eta_+(\omega_k)
f_L(k)
\\
&&+
2{\rm Im}
\int dk\ 
\xi_-(\omega_{k}) A^L_{k}
A^{R*}_k S_{LR}(\omega_k)
\eta_+(\omega_k)
f_L(k)
\\ &&+
2{\rm Im}
\int dk\
\xi_-(\mu_{k})
\bar{A}^L_{k}
\left\{
u^{R*}_k S_{LR}(\mu_k)
+
\bar{A}^{R*}_k S_{LR}(\mu_k)
\eta_+(\mu_k)
\right\}
f_R(k)
\\ &=&
-2\int dk\integ{\omega}\delta(\omega-\omega_k)
\frac{|u^{L}_k|^2 |S_{LR}(\omega)|^2{\rm Im}\ \eta_-(\omega)}{|\Lambda_-(\omega)|^2}
f_L(\omega)
\\
&&+2
\int dk\integ{\omega}\delta(\omega-\mu_k)
\frac{|u^R_k|^2 |S_{LR}(\omega)|^2}{|\Lambda_-(\omega)|^2}
{\rm Im}\ \xi_-(\omega)
f_R(\omega)
\\
&=&
2\integ{\omega}
\frac{ |S_{LR}(\omega)|^2}{|\Lambda_-(\omega)|^2}
{\rm Im}\ \xi_-(\omega)
{\rm Im}\ \eta_-(\omega)
\left\{ f_R(\omega)-f_L(\omega) \right\}
\ ,
\end{eqnarray*}
where we have used ${\displaystyle {\rm Im}\ \xi_-(\omega)=\pi\integ{k} 
|u^L_k|^2\delta(\omega-\omega_k)}$
,\ 
${\displaystyle {\rm Im}\ \eta_-(\omega)=\pi\integ{k} 
|u^R_k|^2\delta(\omega-\mu_k)}$, 
and equalities (\ref{incoming_formula}) (\ref{ChNESS}).
It is nothing but a Landauer formula, and it gives a tunneling transition 
probability from left to right with energy $\omega$:
\begin{eqnarray*}
T(\omega)=2
\frac{ |S_{LR}(\omega)|^2}{|\Lambda_-(\omega)|^2}
{\rm Im}\ \xi_-(\omega)
{\rm Im}\ \eta_-(\omega)
\end{eqnarray*}

\underline{periodic chain}\\
To be more concrete, let us take the following hamiltonian:
\begin{eqnarray*}
H &=& H_S+H_L+H_R+V
\\
H_S &=& \sum_{j=1}^{n-1} \Big(t_j c_j^\dag c_{j+1} + (h.c.) \Big)
+\sum_{j=1}^{n} U c_j^\dag c_j
\\
H_L &=& \int dk \epsilon_k a_k^\dag a_k
,\quad 
H_R = \int dk \epsilon_k b_k^\dag b_k
,\quad 
\epsilon_k \equiv \theta_F (|k|-k_0)
\\
V &=& \int dk \left( v_L^k a_k^\dag c_1 + v_R^k b_k^\dag c_N\right) + (h.c.)
\end{eqnarray*}
By taking
\begin{eqnarray*}
f_\lambda=\sqrt{\frac{2}{N+1}} \sum_{n=1}^{N}
\sin 
\left( \frac{\pi \lambda}{N+1}n \right)c_n
\ ,
\end{eqnarray*}
we can apply the result discussed.
Since, inverse formula reads
\begin{eqnarray*}
c_n=\sqrt{\frac{2}{N+1}} \sum_{\lambda=1}^{N}\sin
\left( \frac{\pi n}{N+1}\lambda \right)f_\lambda
\ ,
\end{eqnarray*}
we have
\begin{eqnarray*}
S_{LL} &=& \frac{2}{N+1}\sum_\lambda
\frac{
\sin
\left( \frac{\pi }{N+1}\lambda \right)
\sin
\left( \frac{\pi }{N+1}\lambda \right)
}{\omega-\epsilon_\lambda}
\\
S_{LR} &=& \frac{2}{N+1}\sum_\lambda
\frac{
\sin
\left( \frac{\pi N}{N+1}\lambda \right)
\sin
\left( \frac{\pi }{N+1}\lambda \right)
}{\omega-\epsilon_\lambda}
\\
S_{RR} &=& \frac{2}{N+1}\sum_\lambda
\frac{
\sin
\left( \frac{\pi N}{N+1}\lambda \right)
\sin
\left( \frac{\pi N}{N+1}\lambda \right)
}{\omega-\epsilon_\lambda}
\\
\epsilon_\lambda &=& 2t\cos \left( \frac{\pi \lambda}{N+1} \right)+U
\ ,
\end{eqnarray*}
and
\begin{eqnarray*}
v_L^k a_k^\dag c_1 &=&
v_L^k\sqrt{\frac{2}{N+1}} \sum_{\lambda=1}^{N} \sin
\left( \frac{\pi}{N+1}\lambda \right)
a_k^\dag f_\lambda
\\
v_R^k b_k^\dag c_N &=&
v_R^k\sqrt{\frac{2}{N+1}} \sum_{\lambda=1}^{N} \sin
\left( \frac{\pi N}{N+1}\lambda \right)
b_k^\dag f_\lambda
\ .
\end{eqnarray*}
Hereafter, we further assume
\begin{eqnarray*}
u^L_k = v_L|k|^\alpha,\ \qquad
u^R_k = v_R|k|^\alpha
\ ,
\end{eqnarray*}
with some exponent $\alpha\in {\bf R}$.
Then, ${\rm Im}\xi_-(\omega)$ and
${\rm Im}\eta_-(\omega)$ is expressed by very simple form:
\begin{eqnarray*}
{\rm Im}\ \xi_-(\omega)&=& \pi v_L^{2}
\int dk\ |k|^{2\alpha}\delta \big(\omega-\theta_f(|k|-k_0) \big)
\\
&=& 2\pi^2 v_L^{2}
\int dx\ x^{2\alpha+1}\delta \big(\omega-\theta_f(x-k_0) \big)
,\qquad x=|k| 
\\
&=&
2 \frac{\pi^2 v_L^2}{\theta_f}
\left(\frac{\omega-\theta_f k_0}{\theta_f}\right)^{2\alpha+1}
\\
{\rm Im}\ \eta_-(\omega)&=& 
2 \frac{\pi^2 v_R^2}{\theta_f}
\left(\frac{\omega-\theta_f k_0}{\theta_f}\right)^{2\alpha+1}
\end{eqnarray*}
It follows
\begin{eqnarray*}
T(\omega) &=& 8\left(\frac{\pi^2 v_L v_R}{\theta_f}\right)^2
\left(\frac{\omega-\theta_f k_0}{\theta_f}\right)^{4\alpha+2}
\frac{ |S_{LR}(\omega)|^2}{|\Lambda_-(\omega)|^2}
\\
\Lambda_-(\omega) &=&
1-2\frac{\pi^2}{\theta_f} 
\left(\frac{\omega-\theta_f k_0}{\theta_f}\right)^{2\alpha+1}
\left\{v_L^2 S_{RR}(\omega) + v_R^2 S_{LL}(\omega)\right\}
\\
&&\mskip 50 mu
+4\left( \frac{\pi^2 v_L v_R}{\theta_f} \right)^2
\left(\frac{\omega-\theta_f k_0}{\theta_f}\right)^{4\alpha+2}
\left\{S_{LL}(\omega)S_{RR}(\omega)-|S_{LR}(\omega)|^2 \right\}
\ .
\end{eqnarray*}

\subsection{Condition for the existence of unique steady state\label{sec_uniqueness}}
We are going to study the completeness condition for the transformation from
$\{a_k, b_k,\ f_\lambda\}$ to the incoming field $\{\alpha_k,\ \beta_k\}$.

In this subsection, we will give a sufficient condition with which, unique NESS exists.
First, let us suppose $\{\alpha_k^\dag , \beta_{k'}\}={\bf 0},\ 
\{\alpha_k^\dag , \alpha_{k'}\}=\delta(k-k'){\bf 1},\ 
\{\beta_k^\dag , \beta_{k'}\}=\delta(k-k'){\bf 1}$ are satisfied, then, 
only the following operator might match with
$a_k$, $b_k$, and $f_\lambda$.
\begin{eqnarray*}
\widetilde{a}_k &=& \alpha_k +
\integ{k'}
\left( 
m^{k'*}_k \alpha_{k'}
+
\bar{m}^{k'*}_k \beta_{k'}
\right)
\\
\widetilde{b}_k &=& \beta_k +
\integ{k'}
\left( 
n^{k'*}_k \alpha_{k'}
+
\bar{n}^{k'*}_k \beta_{k'}
\right)
\\
\widetilde{f}_\lambda &=&
\integ{k} h^{k*}_\lambda\alpha_k
+\integ{k} \bar{h}^{k*}_\lambda\beta_k
\end{eqnarray*}
Let us start from $\widetilde{a}_k$
\begin{eqnarray*}
\widetilde{a}_k &=& 
a_k  
+\sum_\lambda h^k_\lambda f_\lambda
+\integ{k_1}
\left( m^{k}_{k_1} a_{k_1} + n^{k}_{k_1} b_{k_1} \right)
\\
&&+\integ{k_1}
\Bigg[ 
m^{k_1 *}_k 
\left\{
a_{k_1} 
 +\sum_\lambda h^{k_1}_\lambda f_\lambda
+ \integ{k_2}
\left( m^{k_1}_{k_2} a_{k_2} + n^{k_1}_{k_2} b_{k_2} \right)
\right\}
\\
&&\mskip 80 mu+
\bar{m}^{k_1 *}_k
\left\{
b_{k_1} 
+\sum_\lambda \bar{h}^k_\lambda f_\lambda
+ \integ{k_2}
\left( \bar{m}^{k_1}_{k_2} a_{k_2} + \bar{n}^{k_1}_{k_2} b_{k_2} \right)
\right\}
\Bigg]
\\
 &=& 
a_k + \integ{k_1}
\left( m^{k}_{k_1} +  m^{k1*}_{k}  
+\integ{k_2} m^{k_2*}_{k}  m^{k_2}_{k_1}
+\integ{k_2} \bar{m}^{k_2*}_{k}  \bar{m}^{k_2}_{k_1}
\right)a_{k_1}
\\
&& + \integ{k_1}
\left( n^{k}_{k_1} +  \bar{m}^{k1*}_{k}  
+\integ{k_2} m^{k_2*}_{k}  n^{k_2}_{k_1}
+\integ{k_2} \bar{m}^{k_2*}_{k}  \bar{n}^{k_2}_{k_1}
\right)b_{k_1}
\\
&&
+
\sum_\lambda
\left(
h^k_\lambda
+\integ{k_1} m^{k_1 *}_k h^{k_1}_\lambda
+\integ{k_1} \bar{m}^{k_1 *}_k \bar{h}^{k_1}_\lambda
\right)f_\lambda
\end{eqnarray*}
Thus,
\begin{eqnarray*}
&&
m^{k}_{k'} +  m^{k'*}_{k}  
+\integ{k_1} 
\left(
m^{k_1 *}_{k}  m^{k_1}_{k'}
+
\bar{m}^{k_1*}_{k}  \bar{m}^{k_1}_{k'}
\right)
=0,\ \ \forall k, k'
\\
&&
n^{k}_{k'} +  \bar{m}^{k'*}_{k}  
+\integ{k_1} 
\left(
m^{k_1*}_{k}  n^{k_1}_{k'}
+
\bar{m}^{k_1*}_{k}  \bar{n}^{k_1}_{k'}
\right)
=0,\ \ \forall k, k'
\\
&&
h^k_\lambda
+\integ{k_1} 
\left(
m^{k_1 *}_k h^{k_1}_\lambda
+\bar{m}^{k_1 *}_k \bar{h}^{k_1}_\lambda
\right)
=0,\ \ \forall k, \lambda
\end{eqnarray*}
Similarly, we have the following relation for $\beta$:
\begin{eqnarray*}
&&
\bar{m}^{k}_{k'} +  n^{k'*}_{k}  
+\integ{k_1} 
\left(
n^{k_1 *}_{k}  m^{k_1}_{k'}
+
\bar{n}^{k_1*}_{k}  \bar{m}^{k_1}_{k'}
\right)
=0,\ \ \forall k, k'
\\
&&
\bar{n}^{k}_{k'} +  \bar{n}^{k'*}_{k}  
+\integ{k_1} 
\left(
n^{k_1*}_{k}  n^{k_1}_{k'}
+
\bar{n}^{k_1*}_{k}  \bar{n}^{k_1}_{k'}
\right)
=0,\ \ \forall k, k'
\\
&&
\bar{h}^k_\lambda
+\integ{k_1} 
\left(
n^{k_1 *}_k h^{k_1}_\lambda
+\bar{n}^{k_1 *}_k \bar{h}^{k_1}_\lambda
\right)
=0,\ \ \forall k, \lambda
\end{eqnarray*}
At last, we get the similar relation for $\widetilde{f}_\lambda$:
\begin{eqnarray*}
&&
h^{k*}_\lambda+
\integ{k_1}
\left(
h^{k_1 *}_\lambda m^{k_1}_{k}
+\bar{h}^{k_1 *}_\lambda \bar{m}^{k_1}_{k}
\right)
=0,\ \ \forall k,\lambda
\\
&&
\bar{h}^{k*}_\lambda+
\integ{k_1}
\left(
h^{k_1 *}_\lambda n^{k_1}_{k}
+
\bar{h}^{k_1 *}_\lambda \bar{n}^{k_1}_{k}
\right)
=0,\ \ \forall k,\lambda
\\
&&
\integ{k_1}
\left(
h^{k_1*}_\lambda h^{k_1}_{\lambda'}
+ \bar{h}^{k_1*}_\lambda \bar{h}^{k_1}_{\lambda'}
\right)
=\delta_{\lambda,\lambda'}
,\ \ \forall \lambda,\lambda'
\end{eqnarray*}

If the fermionic anti-commutation is satisfied for $\alpha_k,\ \beta_k$, the following equations are equivalent to completeness of the field.
\begin{eqnarray*}
&&
m^{k}_{k'} +  m^{k'*}_{k}  
+\integ{k_1} 
\left(
m^{k_1 *}_{k}  m^{k_1}_{k'}
+
\bar{m}^{k_1*}_{k}  \bar{m}^{k_1}_{k'}
\right)
=0,\ \ \forall k, k'
\\
&&
n^{k}_{k'} +  \bar{m}^{k'*}_{k}  
+\integ{k_1} 
\left(
m^{k_1*}_{k}  n^{k_1}_{k'}
+
\bar{m}^{k_1*}_{k}  \bar{n}^{k_1}_{k'}
\right)
=0,\ \ \forall k, k'
\\
&&
\bar{n}^{k}_{k'} +  \bar{n}^{k'*}_{k}  
+\integ{k_1} 
\left(
n^{k_1*}_{k}  n^{k_1}_{k'}
+
\bar{n}^{k_1*}_{k}  \bar{n}^{k_1}_{k'}
\right)
=0,\ \ \forall k, k'
\\
&&
h^k_\lambda
+\integ{k_1} 
\left(
m^{k_1 *}_k h^{k_1}_\lambda
+\bar{m}^{k_1 *}_k \bar{h}^{k_1}_\lambda
\right)
=0,\ \ \forall k, \lambda
\\
&&
\bar{h}^k_\lambda
+\integ{k_1} 
\left(
n^{k_1 *}_k h^{k_1}_\lambda
+\bar{n}^{k_1 *}_k \bar{h}^{k_1}_\lambda
\right)
=0,\ \ \forall k, \lambda
\\
&&
\integ{k_1}
\left(
h^{k_1*}_\lambda h^{k_1}_{\lambda'}
+ \bar{h}^{k_1*}_\lambda \bar{h}^{k_1}_{\lambda'}
\right)
=\delta_{\lambda,\lambda'}
,\ \ \forall \lambda,\lambda'
\end{eqnarray*}

Next, we are going to check anti-commutation relation.
Let us start from $\{\alpha_k^\dag, \beta_{k'}\}={\bf 0}$
\begin{eqnarray*}
&&\left\{
a_k^\dag +\sum_{\lambda_1} h^{k*}_{\lambda_1} f^\dag_{\lambda_1} 
+\integ{k_1}
\left( m^{k*}_{k_1} a^\dag_{k_1} + n^{k*}_{k_1} b^\dag_{k_1} \right)
,
b_{k'} + \sum_{\lambda_2} \bar{h}^{k'}_{\lambda_2} f_{\lambda_2}
+\integ{k_2}
\left( \bar{m}^{k'}_{k_2} a_{k_2} + \bar{n}^{k'}_{k_2} b_{k_2} \right)
\right\}
\\
&&=\bar{m}^{k'}_k+n^{k*}_{k'}
 +\sum_{\lambda_1} h^{k*}_{\lambda_1} \bar{h}^{k'}_{\lambda_1}
+\integ{k_1} 
\left(
m^{k*}_{k_1}\bar{m}^{k'}_{k_1}
+
n^{k*}_{k_1}\bar{n}^{k'}_{k_1}
\right)
\\
&&={\bf 0}
\end{eqnarray*}

Then, the other two commutations, $\{\alpha_k^\dag, \alpha_{k'}\}={\bf 1}\delta(k-k')$ and $\{\beta_k^\dag, \beta_{k'}\}={\bf 1}\delta(k-k')$ read

\begin{eqnarray*}
&&\delta(k-k')
+m^{k'}_k+m^{k*}_{k'}
 +\sum_{\lambda_1} h^{k*}_{\lambda_1} h^{k'}_{\lambda_1}
+\integ{k_1} 
\left(
m^{k*}_{k_1}m^{k'}_{k_1}
+
n^{k*}_{k_1}n^{k'}_{k_1}
\right)
=\delta(k-k') 
\\
&&\delta(k-k')
+\bar{n}^{k'}_k+\bar{n}^{k*}_{k'}
 +\sum_{\lambda_1} \bar{h}^{k*}_{\lambda_1} \bar{h}^{k'}_{\lambda_1}
+\integ{k_1} 
\left(
\bar{m}^{k*}_{k_1} \bar{m}^{k'}_{k_1}
+
\bar{n}^{k*}_{k_1} \bar{n}^{k'}_{k_1}
\right)
=\delta(k-k')
\ .
\end{eqnarray*}

Thus, the three commutation relations are satisfied if the following three conditions are satisfied:
\begin{eqnarray*}
&&\bar{m}^{k'}_k+n^{k*}_{k'}
 +\sum_{\lambda_1} h^{k*}_{\lambda_1} \bar{h}^{k'}_{\lambda_1}
+\integ{k_1} 
\left(
m^{k*}_{k_1}\bar{m}^{k'}_{k_1}
+
n^{k*}_{k_1}\bar{n}^{k'}_{k_1}
\right)
=0,\ \ \forall k, k'
\\
&&
m^{k'}_k+m^{k*}_{k'}
 +\sum_{\lambda_1} h^{k*}_{\lambda_1} h^{k'}_{\lambda_1}
+\integ{k_1} 
\left(
m^{k*}_{k_1}m^{k'}_{k_1}
+
n^{k*}_{k_1}n^{k'}_{k_1}
\right)
=0,\ \ \forall k, k'
\\
&& 
\bar{n}^{k'}_k+\bar{n}^{k*}_{k'}
 +\sum_{\lambda_1} \bar{h}^{k*}_{\lambda_1} \bar{h}^{k'}_{\lambda_1}
+\integ{k_1} 
\left(
\bar{m}^{k*}_{k_1} \bar{m}^{k'}_{k_1}
+
\bar{n}^{k*}_{k_1} \bar{n}^{k'}_{k_1}
\right)
=0,\ \ \forall k, k'
\end{eqnarray*}

In summary, we have nine equations to guarantee that the system reaches
{\it unique} steady states
\begin{eqnarray*}
&&
m^{k}_{k'} +  m^{k'*}_{k}  
+\integ{k_1} 
\left(
m^{k_1 *}_{k}  m^{k_1}_{k'}
+
\bar{m}^{k_1*}_{k}  \bar{m}^{k_1}_{k'}
\right)
=0,\ \ \forall k, k'
\\
&&
m^{k}_{k'}+m^{k'*}_{k}
 +\sum_{\lambda_1} h^{k'*}_{\lambda_1} h^{k}_{\lambda_1}
+\integ{k_1} 
\left(
m^{k'*}_{k_1}m^{k}_{k_1}
+
n^{k'*}_{k_1}n^{k}_{k_1}
\right)
=0,\ \ \forall k, k'
\\
&& %3
\bar{m}^{k'*}_{k}  + n^{k}_{k'}
+\integ{k_1} 
\left(
m^{k_1*}_{k}  n^{k_1}_{k'}
+
\bar{m}^{k_1*}_{k}  \bar{n}^{k_1}_{k'}
\right)
=0,\ \ \forall k, k'
\\
&& %4
\bar{m}^{k'*}_{k}  + n^{k}_{k'}
 +\sum_{\lambda_1} h^{k}_{\lambda_1} \bar{h}^{k'*}_{\lambda_1}
+\integ{k_1} 
\left(
m^{k}_{k_1}\bar{m}^{k'*}_{k_1}
+
n^{k}_{k_1} \bar{n}^{k'*}_{k_1}
\right)
=0,\ \ \forall k, k'
\\
&& %5
\bar{n}^{k}_{k'} +  \bar{n}^{k'*}_{k}  
+\integ{k_1} 
\left(
n^{k_1*}_{k}  n^{k_1}_{k'}
+
\bar{n}^{k_1*}_{k}  \bar{n}^{k_1}_{k'}
\right)
=0,\ \ \forall k, k'
\\
&& %6
\bar{n}^{k'*}_k+\bar{n}^{k}_{k'}
 +\sum_{\lambda_1} \bar{h}^{k}_{\lambda_1} \bar{h}^{k'*}_{\lambda_1}
+\integ{k_1} 
\left(
\bar{m}^{k}_{k_1} \bar{m}^{k'*}_{k_1}
+
\bar{n}^{k}_{k_1} \bar{n}^{k'*}_{k_1}
\right)
=0,\ \ \forall k, k'
\\
&&
h^k_\lambda
+\integ{k_1} 
\left(
m^{k_1 *}_k h^{k_1}_\lambda
+\bar{m}^{k_1 *}_k \bar{h}^{k_1}_\lambda
\right)
=0,\ \ \forall k, \lambda
\\
&&
\bar{h}^k_\lambda
+\integ{k_1} 
\left(
n^{k_1 *}_k h^{k_1}_\lambda
+\bar{n}^{k_1 *}_k \bar{h}^{k_1}_\lambda
\right)
=0,\ \ \forall k, \lambda
\\
&&
\integ{k_1}
\left(
h^{k_1*}_\lambda h^{k_1}_{\lambda'}
+ \bar{h}^{k_1*}_\lambda \bar{h}^{k_1}_{\lambda'}
\right)
=\delta_{\lambda,\lambda'}
,\ \ \forall \lambda,\lambda'
\end{eqnarray*}

We have derived a Landauer formula for a quadratic system described by hamiltonian coupled to reservoirs. 
In our calculation there is an explicit expression for the transmission probability in terms of the parameters of the hamiltonian. We have also derived the sufficient conditions under which the system has a unique NESS. These conditions are difficult to interpret, and although they are explicit in terms of the hamiltonian parameters it is not clear how restrictive they are.  We conjecture that if the range of reservoir energy goes from $-\infty$ to $+\infty$ then is not possible to satisfy the sufficient conditions here derived. We leave these problems for future research. 

\section{conclusions}

The derivation of macroscopic irreversible dynamics of nonequilibrium systems from microscopic equations 
was recently revisited from the point of view of infinitely extended quantum systems.  
Here we have briefly reviewed the $C^*$ algebra and its application to equilibrium systems 
as well as introduced some recent results on NESS.  We were too slow to finish the nonequilibrium part of the lectures by Professor Tasaki. We plan to upload to the Arxiv that part in the future.  
In addition to the lecture part we have demonstrated the derivation of Landauer formula rigorously for quadratic systems but using a more physical presentation in the spirit of Professor Tasaki's work. We hope it helps physicists to use these techniques in their work. 

\section*{Acknowledgment}
This work was supported by Fondecyt Grants 1110144 (FB), 3120254 (SA).
We also acknowledge Anillo grant ACT 127.

It can be unusual to thank one of the authors, but SA would like to thank Professor Tasaki for supervising me from 2003 April until his death.
I learned not only mathematics and physics, but also the philosophy and attitude to science.
I would also like to thank that he always takes care of my personal issues.
I pray for the repose of his soul.
His death is still a big shock for me, but his memory will survive forever in my mind.

\bibliographystyle{ieeetr}

\bibliography{ajisaka-preprint}

\begin{thebibliography}{10}

\bibitem{Bussei_Tasaki}
S.~Tasaki, ``Nonequilibrium Statistical Mechanics in infinitely extended systems (Lecture note),'' {\em Bussei Kenkyu},
  vol.~96(3), p.~289, 2011.

\bibitem{Seg47}
I.~E. Segal {\em The Annals of Mathematics Second ser.}, vol.~48, No.~4,
  p.~930, 1947.

\bibitem{Cornfeld00}
I.~Cornfeld, S.~Fomin, and Y.~Sinai, {\em Ergodic Theory}.
\newblock Springer, New York, 1982.

\bibitem{Bunimovich82}
A.~Bunimovich {\em et~al.}, {\em Dynamical Systems, Ergodic Theory and
  Applications}.
\newblock Encyclopaedia of Mathematical Sciences (Spinger), 1982.

\bibitem{Rue69}
D.~Ruelle, {\em tatistical Mechanics: Rigorous Results}.
\newblock Benjamin, reading, 1969.

\bibitem{Sinai82}
Y.~G. Sinai, {\em Theory of Phase Transitions: Rigorous Results}.
\newblock Oxford; New York: Pergamon Press, 1982.

\bibitem{BR02}
O.~Bratteli and D.~Robinson, {\em Operator Algebras and Quantum Statistical
  Mechanics vol.1, vol.2}.
\newblock Springer, Berlin-Heidelberg-New York, 2002.

\bibitem{Rue00}
D.~Ruelle {\em J. Stat. Phys.}, vol.~98, p.~57, 2000.

\bibitem{Rue01}
D.~Ruelle {\em Comm. Math. Phys.}, vol.~224, p.~3, 2001.

\bibitem{JP01}
V.~Jak\v{s}i\'c and C.-A. Pillet {\em Commun. Math. Phys.}, vol.~217, p.~285,
  2001.

\bibitem{JP02a}
V.~Jak\v{s}i\'c and C.-A. Pillet {\em Commun. Math. Phys.}, vol.~226, p.~131,
  2002.

\bibitem{JP02b}
V.~Jak\v{s}i\'c and C.-A. Pillet {\em J. Stat. Phys.}, vol.~108, p.~787, 2002.

\bibitem{LecMath1880}
S.~Attal, A.~Joye, and C.-A. Pillet, {\em Open Quantum Systems I, II, III
  (Lecture Notes in Mathematics, 1880, 1881, 1882)}.
\newblock Springer, Berlin-Heidelberg-New York, 2006.

\bibitem{FMSU03}
J.~Fr\"ohlich, M.~Merkli, S.~Schwarz, and D.~Ueltschi, {\em A garden of
  quanta}, p.~345.
\newblock World Scientific, River Edge, 2003.

\bibitem{TM03}
S.~Tasaki and T.~Matsui, {\em Fundamental Aspects of Quantum Physics}, p.~100.
\newblock World Scientific, Singapore, 2003.

\bibitem{TT05}
J.~Takahashi and S.~Tasaki {\em J. Phys. Soc. Jpn. Suppl.}, vol.~74, p.~261,
  2005.

\bibitem{TM06}
S.~Tasaki and T.~Matsui {\em RIMS Kohkyuroku}, vol.~1507, p.~118, 2006.

\bibitem{TT06a}
J.~Takahashi and S.~Tasaki {\em Physica E}, vol.~34, p.~651, 2006.

\bibitem{TT06b}
J.~Takahashi and S.~Tasaki {\em J. Phys. Soc. Jpn.}, vol.~75, No.9, p.~094712,
  2006.

\bibitem{TT06c}
S.~Tasaki and J.~Takahashi {\em Prog. Theor. Phys. Suppl.}, vol.~165, p.~57,
  2006.

\bibitem{Tasaki06}
S.~Tasaki {\em J. Phys. Conf. Ser.}, vol.~31, p.~35, 2006.

\bibitem{JP07}
V.~Jak\v{s}i\'c and C.-A. Pillet {\em Contemporary Mathematics}, vol.~447,
  p.~153, 2007.

\bibitem{Merkli07}
M.~Merkli, M.~Mueck, and I.~Sigal {\em Ann. Henri Poincar\'e}, vol.~8, p.~1539,
  2007.

\bibitem{Merkli08}
M.~Merkli, I.~Sigal, and G.~Berman {\em Ann. Phys.}, vol.~323, p.~373, 2008.

\bibitem{Zagrebnov09}
H.~D. Cornean, H.~Neidhardt, and V.~A. Zagrebnov {\em Ann. Henri Poincar\'e},
  vol.~10, p.~61, 2009.

\bibitem{JOP06a}
V.~Jak\v{s}i\'c, Y.~Ogata, and C.-A. Pillet {\em Comm. Math. Phys.}, vol.~265,
  p.~721, 2006.

\bibitem{JOP06b}
V.~Jak\v{s}i\'c, Y.~Ogata, and C.-A. Pillet {\em Comm. Math. Phys.}, vol.~268,
  p.~369, 2006.

\bibitem{JOP06c}
V.~Jak\v{s}i\'c, Y.~Ogata, and C.-A. Pillet {\em J Stat. Phys.}, vol.~123,
  p.~547, 2006.

\bibitem{JOP07}
V.~Jak\v{s}i\'c, Y.~Ogata, and C.-A. Pillet {\em Ann. Henri Poincar\'e},
  vol.~8, p.~1013, 2007.

\bibitem{SalemFrohlich07}
W.~Salem and J.~Fr\"ohlich {\em J. Stat. Phys.}, vol.~126, p.~1045, 2007.

\bibitem{Salem07}
W.~A. Salem {\em Ann. Henri Poincar\'e}, vol.~8, p.~569, 2007.

\bibitem{Tasaki01}
S.~Tasaki {\em Chaos, Solitons and Fractals}, vol.~12, p.~2657, 2001.

\bibitem{FMU06}
J.~Fr\"ohlich, M.~Merkli, and D.~Ueltschi {\em Ann. Henri. Poincar\'e}, vol.~4,
  p.~897, 2006.

\bibitem{AJPP07}
W.~Aschbacher {\em et~al.} {\em J. Math. Phys.}, vol.~48, p.~032101, 2007.

\bibitem{Nenciu07}
G.~Nenciu {\em J. Math. Phys}, vol.~48, p.~033302, 2007.

\bibitem{AjiTa09}
S.~Ajisaka {\em et~al.} {\em Prog. Theor. Phys.}, vol.~121, p.~1289, 2009.

\bibitem{AjiTa11}
S.~Ajisaka, S.~Tasaki, and I.~Terasaki {\em Phys. Rev. B}, vol.~83, p.~212301,
  2011.

\bibitem{Resibois77}
P.~Resibois {\em et~al.}, {\em Classical Kinetic Theory of Fluids}.
\newblock John Wiley \& Sons Inc, 1977.

\bibitem{ST06}
S.~Sasa and H.~Tasaki {\em J. Stat. Phys.}, vol.~125, p.~125, 2006.

\bibitem{Kubo88}
R.~Kubo, {\em Statistical physics II: Nonequilibrium Statistical Mechanics}.
\newblock Springer, 1988.

\bibitem{Koss76}
V.~Gorini, A.~Kossakowski, and E.~C.~G. Sudarshan {\em J. Math. Phys.},
  vol.~17, p.~821, 1976.

\bibitem{Lindblad76}
G.~Lindblad {\em Commun. Math. Phys.}, vol.~48, p.~119, 1976.

\bibitem{Breuer02}
H.-P. Breuer and F.~Petruccione, {\em Theory of open quantum systems}.
\newblock Oxford: Oxford University Press, 2002.

\bibitem{Kamenev04}
A.~Kamenev, ``Many-body theory of non-equilibrium systems,'' {\em
  arXiv:cond-mat/}, p.~0412296v2, 2004.

\bibitem{Imry99}
Y.~Imry and R.~Landauer, ``Conductance viewed as transmission,'' {\em Rev. Mod.
  Phys.}, vol.~71(2), p.~s306, 1999.

\bibitem{Zubarev95}
D.~Zubarev, {\em Nonequilibrium Statistical Thermodynamics}.
\newblock Springer-Verlag, 1995.

\bibitem{Hayakawa03}
H.-D. Kim and H.~Hisao {\em K\^o ky\^u roku}, vol.~73, p.~1305, 2003.


\bibitem{Arai08}
A.~Arai, {\em Mathematical Principles of Quantum Statistical Mechanics
  (Japanese)}.
\newblock Kyoritsu Shuppan, 2008.

\bibitem{PW05}
W.~Pusz and S.~L. Woronowicz {\em Commun. Math. Phys.}, vol.~58, p.~273, 1978.

\bibitem{Oji89}
I.~Ojima {\em J. Stat. Phys.}, vol.~56, p.~203, 1989.

\bibitem{OHI88}
I.~Ojima, H.~Hasegawa, and M.~Ichiyanagi {\em J. Stat. Phys.}, vol.~50, p.~633,
  1988.

\bibitem{Maclennan63}
J.~M. Jr. {\em Adv. Chem. Phys.}, vol.~5, p.~261, 1963.

\end{thebibliography}

\end{document}